\documentclass[]{iopart}
\usepackage{graphicx} 
\usepackage{graphics} 
\usepackage{amssymb,amsfonts,latexsym}
\usepackage{dcolumn,epsfig}
\usepackage{setspace}
\usepackage{subfigure}
\usepackage{url}
\usepackage{citesort}
\usepackage{rotating}
\usepackage[usenames]{color}

\newcommand\T{\rule{0pt}{2.6ex}}
\newcommand\B{\rule[-1.2ex]{0pt}{0pt}}
\newcommand\TT{\rule{0pt}{4.2ex}}
\newcommand\BB{\rule[-2.4ex]{0pt}{0pt}}
\newcommand\TTT{\rule{0pt}{3.8ex}}

\usepackage{ulem}
\newcommand{\Ytwo}{{{}^{-2}Y}}

\def\IL{\relax{\rm I\kern-.18em L}}

\newcommand{\be}{\begin{equation}}
\newcommand{\ee}{\end{equation}}
\newcommand{\bel}[1]{\begin{equation}\label{#1}}
\newcommand{\ba}{\begin{eqnarray}}
\newcommand{\ea}{\end{eqnarray}}
\newcommand{\bal}[1]{\begin{eqnarray}\label{#1}}

\newcommand{\Mc}{{\cal M}}
\newcommand{\Ms}{M_{\odot}}

\newcommand{\M}{\rangle}

\def\ltsima{$\; \buildrel < \over \sim \;$}
\def\simlt{\lower.5ex\hbox{\ltsima}}
\def\gtsima{$\; \buildrel > \over \sim \;$}
\def\simgt{\lower.5ex\hbox{\gtsima}}

\begin{document}

\title[Results from the first NINJA project]{%
Testing gravitational-wave searches with numerical
relativity waveforms: Results from the first Numerical INJection
Analysis (NINJA) project}

\author{
Benjamin~Aylott${}^{1}$,
John~G.~Baker${}^{2}$,
William~D.~Boggs${}^{3}$,
Michael~Boyle${}^{4}$,
Patrick~R.~Brady${}^{5}$,
Duncan~A.~Brown${}^{6}$,
Bernd~Br\"ugmann${}^{7}$,
Luisa~T.~Buchman${}^{4}$,
Alessandra~Buonanno${}^{3}$,
Laura~Cadonati${}^{8}$,
Jordan~Camp${}^{2}$,
Manuela~Campanelli${}^{9}$,
Joan~Centrella${}^{2}$,
Shourov~Chatterji${}^{10,11}$,
Nelson~Christensen${}^{12}$,
Tony~Chu${}^{4}$,
Peter Diener${}^{13,14}$,
Nils~Dorband${}^{15}$,
Zachariah~B.~Etienne${}^{16}$,
Joshua~Faber${}^{9}$,
Stephen~Fairhurst${}^{17}$,
Benjamin~Farr${}^{9,17}$,
Sebastian~Fischetti${}^{8}$,
Gianluca~Guidi${}^{10,18}$,
Lisa~M.~Goggin${}^{5}$,
Mark~Hannam${}^{19}$,
Frank~Herrmann${}^{20,30}$,
Ian~Hinder${}^{20}$,
Sascha~Husa${}^{21,15}$,
Vicky~Kalogera${}^{22}$,
Drew~Keppel${}^{11}$,
Lawrence~E.~Kidder${}^{23}$,
Bernard~J.~Kelly${}^{2}$,
Badri~Krishnan${}^{15}$,
Pablo~Laguna${}^{24}$,
Carlos~O.~Lousto${}^{9}$,
Ilya~Mandel${}^{22}$,
Pedro~Marronetti${}^{25}$,
Richard~Matzner${}^{29}$,
Sean~T.~McWilliams${}^{2}$,
Keith~D.~Matthews${}^{4}$,
R.~Adam~Mercer${}^{5}$,
Satyanarayan~R.~P.~Mohapatra${}^{8}$,
Abdul H. Mrou{\'e}${}^{23}$,
Hiroyuki~Nakano${}^{9}$,
Evan~Ochsner${}^{3}$,
Yi~Pan${}^{3}$,
Larne~Pekowsky${}^{6}$,
Harald~P.~Pfeiffer${}^{4}$,
Denis~Pollney${}^{15}$,
Frans~Pretorius${}^{26}$,
Vivien~Raymond${}^{22}$,
Christian~Reisswig${}^{15}$,
Luciano~Rezzolla${}^{15}$,
Oliver~Rinne${}^{27}$,
Craig~Robinson${}^{12}$,
Christian~R\"over${}^{28}$,
Luc{\'i}a~Santamar{\'i}a${}^{15}$,
Bangalore~Sathyaprakash${}^{17}$,
Mark~A.~Scheel${}^{4}$,
Erik~Schnetter${}^{13,14}$,
Jennifer~Seiler${}^{15}$,
Stuart~L.~Shapiro${}^{16}$,
Deirdre~Shoemaker${}^{24}$,
Ulrich~Sperhake${}^{7}$,
Alexander~Stroeer${}^{31,2}$,
Riccardo~Sturani${}^{10,18}$,
Wolfgang~Tichy${}^{25}$,
Yuk~Tung~Liu${}^{16}$,
Marc~van~der~Sluys${}^{22}$,
James~R.~van Meter${}^{2}$,
Ruslan~Vaulin${}^{5}$,
Alberto~Vecchio${}^{1}$,
John~Veitch${}^{1}$,
Andrea~Vicer\'e${}^{10,18}$,
John~T.~Whelan${}^{9,15}$,
Yosef~Zlochower${}^{9}$
} 

\address{$^{1}$ School of Physics and Astronomy, University of Birmingham, Edgbaston, Birmingham B15 2TT, UK}
\address{$^{2}$ NASA Goddard Space Flight Center, Greenbelt, MD 20771, USA}
\address{$^{3}$ Maryland Center for Fundamental Physics, Department of Physics,
University of Maryland, College Park, MD 20742, USA}
\address{$^{4}$ Theoretical Astrophysics 130-33, California Institute of Technology, Pasadena, CA 91125}
\address{$^{5}$ University of Wisconsin-Milwaukee, P.O.~Box 413, Milwaukee, WI 53201, USA}
\address{$^{6}$ Department of Physics, Syracuse University, Syracuse, New York, 13254}
\address{$^{7}$ Theoretisch Physikalisches Institut, Friedrich Schiller Universit\"at, 07743 Jena, Germany}
\address{$^{8}$ Department of Physics, University of Massachusetts, Amherst, MA 01003}
\address{$^{9}$ Center for Computational Relativity and Gravitation and School of Mathematical Sciences, Rochester Institute of Technology, 85 Lomb Memorial Drive, Rochester, NY 14623}
\address{$^{10}$ INFN-Sezione Firenze/Urbino, I-50019 Sesto Fiorentino, Italy}
\address{$^{11}$ LIGO -- California Institute of Technology, Pasadena, CA 91125, USA}
\address{$^{12}$ Physics \& Astronomy, Carleton College, Northfield, MN, USA}
\address{$^{13}$ Center for Computation \& Technology, Louisiana State University, Baton Rouge, LA 70803}
\address{$^{14}$ Department of Physics \& Astronomy, Louisiana State University, Baton Rouge, LA 70803}
\address{$^{15}$ Max-Planck-Institut f\"ur Gravitationsphysik, Am M\"uhlenberg 1, D-14476 Potsdam, Germany.}
\address{$^{16}$ Department of Physics, University of Illinois at Urbana-Champaign, Urbana, IL 61801}
\address{$^{17}$ School of Physics and Astronomy, Cardiff University, The Parade, Cardiff, UK}
\address{$^{18}$ Istituto di Fisica, Universit\`a di Urbino, I-61029 Urbino, Italy}
\address{$^{19}$ Physics Department, University College Cork, Cork, Ireland}
\address{$^{20}$ Center for Gravitational Wave Physics, The Pennsylvania State University, University Park, PA 16802}
\address{$^{21}$ Departament de F\'isica, Universitat de les Illes Balears, Palma de Mallorca, Spain}
\address{$^{22}$ Department of Physics and Astronomy, Northwestern University, Evanston, IL, USA}
\address{$^{23}$ Center for Radiophysics and Space Research, Cornell University, Ithaca, New York, 14853}
\address{$^{24}$ Center for Relativistic Astrophysics and School of Physics, Georgia Institute of Technology, Atlanta, GA 30332}
\address{$^{25}$ Department of Physics, Florida Atlantic University, Boca Raton, FL 33431}
\address{$^{26}$ Department of Physics, Princeton University, Princeton, NJ 08540}
\address{$^{27}$ Department of Applied Mathematics and Theoretical Physics, Centre for Mathematical Sciences, Wilberforce Road, Cambridge CB3 0WA, UK, and King's College, Cambridge CB2 1ST, UK}
\address{$^{28}$ Max-Planck-Institut f\"ur Gavitationsphysik, Hannover, Germany}
\address{$^{29}$ University of Texas at Austun, Austin, Texas, 78712}
\address{$^{30}$ Center for Scientific Computation and Mathematical Modeling, University of Maryland, 4121 CSIC Bldg. 406, College Park, Maryland 20742, USA}
\address{$^{31}$ CRESST, University of Maryland, College Park, Maryland 20742, USA}

\begin{abstract} 
The Numerical INJection Analysis (NINJA) project is 
a collaborative effort between members of the numerical
relativity and gravitational-wave data analysis communities. 
The purpose of NINJA is to study the 
sensitivity of existing gravitational-wave search
algorithms using numerically generated waveforms and to foster closer
collaboration between the numerical relativity and data analysis
communities. We describe the results of the first NINJA analysis which 
focused on gravitational waveforms from binary black hole
coalescence.  Ten numerical relativity groups contributed 
numerical data which were used to generate a set of 
gravitational-wave signals. These signals were injected into a 
simulated data set, designed to mimic the
response of the Initial LIGO and Virgo gravitational-wave detectors. 
Nine groups analysed this data using search and parameter-estimation pipelines.
Matched filter algorithms, un-modelled-burst searches and Bayesian
parameter-estimation and model-selection algorithms were applied to the data. 
We report the efficiency of these search methods in detecting the numerical
waveforms and measuring their parameters. We describe preliminary comparisons
between the different search methods and suggest improvements for future NINJA
analyses.
\end{abstract}

\maketitle

\section{Introduction}
Binary systems of compact objects, i.e., black holes and neutron stars,
are among the most important objects for testing general relativity
and studying its astrophysical implications~\cite{thorne.k:1987}. The
general solution of the binary problem 
in Newtonian gravity is given by the Keplerian orbits.  In general
relativity, the Keplerian orbits for a bound system decay due to the
emission of gravitational radiation, leading eventually to the merger
of the two compact objects and to a single final remnant~\cite{Einstein:1916,Peters:1963ux,Peters:1964zz}.  The
decay of the orbits is due to the emission of gravitational waves and
these waves carry important information about the dynamics of the
binary system. In particular, the waves produced during the merger
phase contain important non-perturbative general relativistic
effects potentially observable by gravitational-wave detectors.  Gravitational
waves could be detectable by the current generation of gravitational
wave detectors such as LIGO and
Virgo~\cite{Barish:1999vh,Acernese:2005yh}, and detection is very
likely with future generations of these detectors.

Two important advances have occurred in recent years that have brought
us closer to the goal of observing and interpreting gravitational waves
from coalescing compact objects.  The first is the successful
construction and operation of a world-wide network of large
interferometric gravitational-wave detectors; these include the three
LIGO detectors in the United States, Virgo in Italy, TAMA in
Japan~\cite{Tsubono:2000if} and the GEO600 detector in
Germany~\cite{Willke:2007zz}.  The TAMA detector was the first
interferometric detector to achieve its design goals, and it collected
science data between 1999 and 2003~\cite{Tsubono:2000if}.  The LIGO
detectors started observations in 2002~\cite{Abbott:2003vs}.
From 2005 to 2007 these detectors operated at design sensitivity collecting 
more than a year of coincident data from the three LIGO detectors; these
observations are referred to as the ``fifth science run''
(S5)~\cite{Abbott:2007kva}.  The Virgo detector is
also close to achieving its design goals and collected six 
months of data coincident with the last six months of the LIGO S5
run (referred to as VSR1)~\cite{Acernese:2008zz}.  
The GEO600 detector has been operating
since $2002$ in coincidence with the LIGO instruments~\cite{Willke:2007zz}.  
The two 4km LIGO detectors are currently being upgraded to improve their
sensitivity by a factor of 2--3 (Enhanced LIGO~\cite{LIGOEnhancedLIGO}) and will resume observations
in 2009. Upgrades to the Virgo detectors to yield comparable sensitivity to
Enhanced LIGO are proceeding on a similar schedule.
During this time, the GEO600 and the LIGO Hanford 2km detector continue to
make best-effort observations (called ``astro-watch'') to capture
any possible strong events, such as a galactic supernova. 
Following the Enhanced LIGO and Virgo
observations, the Advanced LIGO~\cite{Fritschel:2003qw} and Virgo~\cite{AdvVirgo} upgrades will improve detector
sensitivities by a factor of $\sim 10$ above the Initial LIGO detectors; these
upgrades are expected to be complete by 2014.
There are also plans to build a second-generation
cryogenic detector in Japan known as LCGT \cite{Uchiyama:2004vr}.
Searching data from these detectors for weak gravitational wave signals
over a vast parameter space is a challenging task. The gravitational-wave
community has invested significant resources in this effort.  
A number of searches on S5/VSR1
data for un-modelled bursts and binary coalescence are in progress and
many results, including those from previous science runs, have already been
reported~\cite{Abbott:2009tt,Abbott:2003pj,Abbott:2005pe,Abbott:2005pf,%
Abbott:2007xi,Abbott:2007ai,Abbott:2004rt,Abbott:2005fb,Abbott:2005at,%
Abbott:2005rt,Abbott:2007wu}.

The second important advance has been the impressive success of
numerical relativity in simulating the merger phase of
binary black hole (BBH) coalescence.  The first breakthroughs occurred in
2005 with simulations by Pretorius \cite{Pretorius:2005gq}, closely
followed by the independent Goddard and Brownsville (now at RIT) results
\cite{Campanelli:2005dd,Baker:2005vv}.  Since then, a number of
numerical relativity groups around the world have successfully evolved
various configurations starting from the inspiral phase all
the way through the merger to the final remnant black hole (for recent
overviews on the field see e.g.~\cite{Pretorius:2007nq,Husa:2007zz,Hannam:2009rd}).  This
has led to important new physical insights in BBH
mergers. These include the prediction of large recoil velocities
produced by asymmetric emission of gravitational radiation during the
merger process
\cite{Herrmann:2006ks,Baker:2006vn,Gonzalez:2006md,Herrmann:2007ac,Koppitz:2007ev,Campanelli:2007ew,Gonzalez:2007hi,Tichy:2007hk,Campanelli:2007cga,Baker:2007gi,Herrmann:2007ex,Brugmann:2007zj,Schnittman:2007ij,Pollney:2007ss,Lousto:2007db,Baker:2008md,Dain:2008ck,Healy:2008js,Gonzalez:2008bi}
 and the prediction of the parameters of the remnant
Kerr hole for a wide class of initial states
\cite{Campanelli:2006fg,Gonzalez:2006md,Campanelli:2006fy,Berti:2007fi,Rezzolla:2007xa,Boyle:2007sz,Rezzolla:2007rd,Marronetti:2007wz,Rezzolla:2007rd,Sperhake:2007gu,Hinder:2007qu,Berti:2007nw,Boyle:2007ru,Tichy:2008du,Rezzolla:2008sd}.  
Since the inspiral, merger and coalescence of black holes 
are also among the most important targets of gravitational-wave
detectors, we expect that the detailed
information provided by numerical simulations can be used to increase
the reach and to quantify the efficacy of data analysis pipelines.
Indeed the driving motivation of research on numerical simulations of
black-hole binaries over the last few decades has been their use in
gravitational-wave observations.  

Thus far, most searches for gravitational waves from 
BBH mergers have relied on post-Newtonian results, which are valid when
the black holes are sufficiently far apart.  Within its range of
validity, post-Newtonian theory provides a convenient analytic
description of the expected signals produced by binary systems.  The
numerical relativity results, on the other hand, have not yet been
synthesised into an analytic model for the merger phase covering a broad
range of parameters, i.e., a wide range of mass ratios, spins and if
necessary, eccentricity; there has however been significant progress for
the non-spinning case~\cite{Buonanno:2006ui,Berti:2007fi,Ajith:2007kx,Pan:2007nw,%
Buonanno:2007pf,Boyle:2007ft,%
Ajith:2007qp,Damour:2007yf,Damour:2007vq,Damour:2008te,Boyle:2008ge,Boyle:2009dg}.
Similarly, despite significant progress, there
is not yet a complete detailed description over the full parameter space
of how post-Newtonian and numerical simulations are to be matched with
each other.  On the data analysis side, many pipelines, especially ones
that rely on a detailed model for the signal waveform, have made a
number of choices based on post-Newtonian results, and it is important
to verify that these choices are sufficiently robust.  More generally,
it is necessary to quantify the performance of these data analysis
pipelines for both detection and parameter estimation.  This is critical
for setting astrophysical upper limits in case no detection has been
made, for following up interesting detection candidates, and of course
for interpreting direct detections. Work on this to date has primarily used 
post-Newtonian waveforms. Numerical relativity now provides
an important avenue for extending this to the merger phase.  

There are significant challenges to be overcome before numerical relativity
results can be fully exploited in data-analysis pipelines.  The Numerical
INJection Analysis (NINJA) project was started in the spring of 2008 with the
aim of addressing these challenges and fostering close collaboration between
numerical relativists and data analysts. Participation in NINJA is open to all
scientists interested in numerical simulations and gravitational-wave data
analysis.  NINJA is the first project of its kind that attempts to form a
close working collaboration between the numerical relativity and data analysis
communities.  Several decisions were made that restrict the scope of the
results reported here: we consider only BBH simulations and have not used
results from supernova simulations or simulations containing neutron stars;
the waveform data comes purely from numerical simulations and we do not
attempt to extend numerical data using post-Newtonian waveforms; the NINJA
data set is constructed using Gaussian noise to model the response of the
Initial LIGO and Virgo detectors -- no attempt has been made to include
non-Gaussian noise transients found in real detector data.  The comparisons
and conclusions reported here are thus necessarily limited, and in many cases
are only the first steps towards fully understanding the sensitivity of
data-analysis pipelines to black hole signals.  Further studies are needed
regarding the accuracy and comparison of numerical waveforms, and of how
systematic errors in these waveforms can affect parameter estimation.  Some
analyses of numerical waveforms with regard to gravitational-wave detection
have already been
performed~\cite{Baumgarte:2006en,Vaishnav:2007nm,Pan:2007nw,Boyle:2009dg},
accuracy standards have been developed for use of numerical waveforms in data
analysis~\cite{Lindblom:2008cm} and a detailed comparison of some of the
waveforms used in the NINJA project was performed in the related Samurai
project~\cite{Hannam:2009hh}.  We expect that subsequent NINJA analyses will
build on these results to address these issues.  

Despite the limited scope of the first NINJA project, we are able to draw the following broad conclusions from this
work.  Our first conclusion is that the current data analysis pipelines used to search LIGO, Virgo
and GEO600 data for black hole coalescence are able to detect numerical waveforms injected into the 
NINJA data set at the expected sensitivities. Indeed, several of these pipelines are able to detect signals
that lie outside the parameter space that they target.  This is a
non-trivial statement since most detectability estimates to date for these
sources have relied on post-Newtonian waveforms, which are valid only when the black
holes are sufficiently far apart. For many of these pipelines, this is the
first time they have been tested against numerical waveforms. It should be
noted, however, that the
NINJA data set does not contain non-stationary noise transients so more work
is needed to understand how detection performance is affected by the noise
artifacts seen in real gravitational-wave detector data. Our second conclusion is that significant
work is required to understand and improve the measurement of signal
parameters.  For instance, among the pipelines used in this first NINJA 
analysis only the Markov-chain Monte-Carlo algorithm attempted
to estimate the spins of the individual black holes, and the estimation of the
component masses by the detection pipelines is poor in most cases.  Improvement
in this area will be crucial for bridging the gap between gravitational wave
observations and astrophysics. NINJA has proven to be extremely valuable at
framing the questions that need to be answered.

This paper is organised as follows: In the next section we describe the
contributed numerical waveforms and Section~\ref{sec:ninja_data}
describes the construction of the simulated gravitational-wave detector
data used in the NINJA analyses.  Descriptions of the search methods and
results are given in Section~\ref{sec:da}.  The results are grouped by
search method into search pipelines using modelled waveforms (Section
\ref{ssec:modeled}), search pipelines using un-modelled waveforms
(Section \ref{ssec:unmodeled}) a comparison of inspiral-burst-ringdown
results (Section \ref{ssec:comparison}), and Bayesian pipelines (Section
\ref{ssec:param_est}). We conclude with a discussion of our results and
future directions for NINJA in Section~\ref{sec:discussion}.

\section{Numerical Waveforms}
\label{sec:nrwaveforms}

\begin{figure}
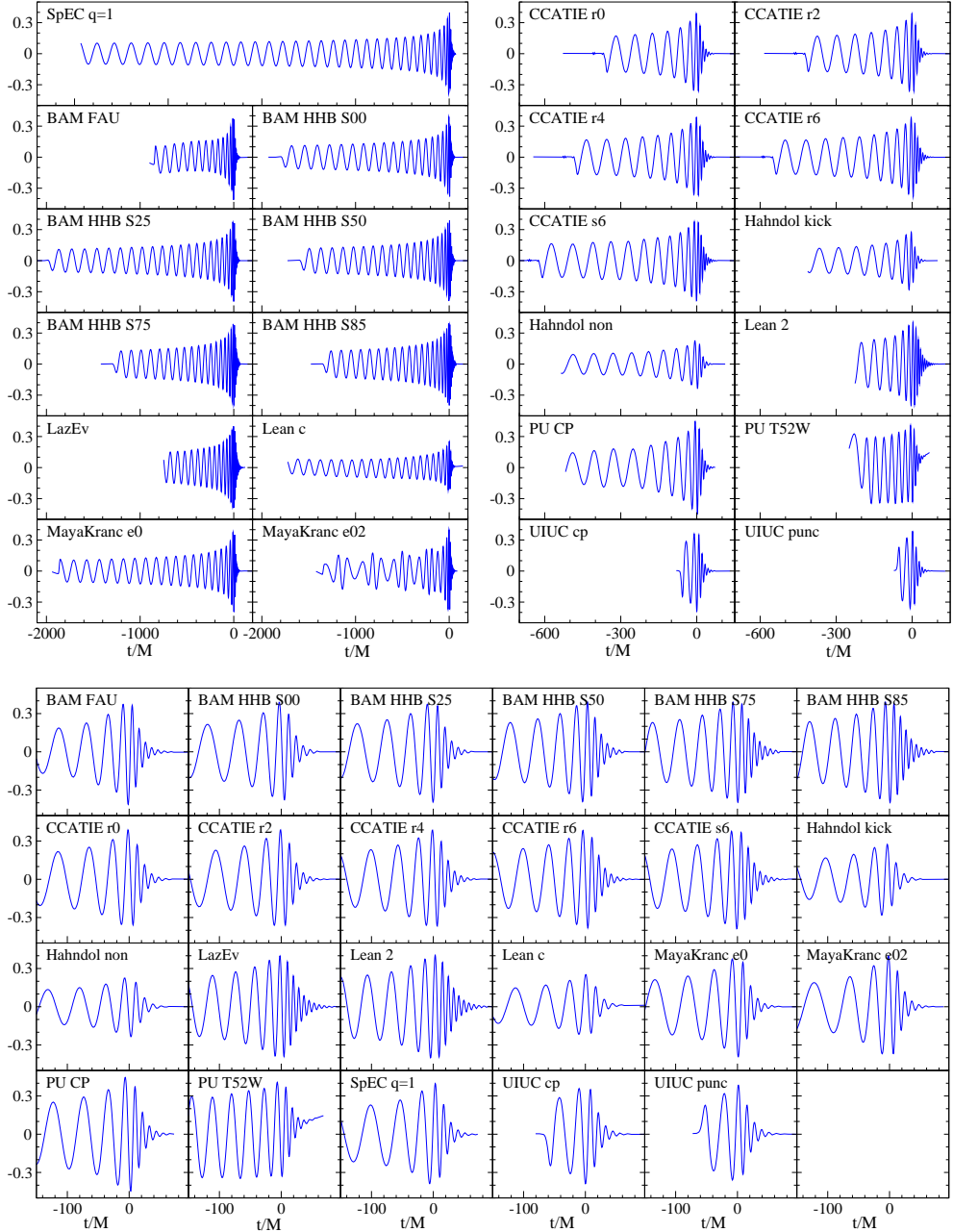

\includegraphics[width=0.49\textwidth]{figures/Prune_Re_h22-A}
$\;$
\includegraphics[width=0.49\textwidth]{figures/Prune_Re_h22-B}\\[-.25em]

\includegraphics[width=\textwidth]{figures/Prune_Re_h22-Merger}

\caption{\label{fig:NR-Reh22} {\bf Summary of all submitted numerical waveforms: \boldmath$r/M\,\mbox{Re}(h_{22})$ }
The $x$-axis shows time in units of $M$ and the $y$-axis shows the 
real part of the $(\ell,m)=(2,2)$
  component of the dimensionless wave strain $r h = r h_+ - i r
  h_\times$.
  The top panels show the complete
  waveforms: the top-left panel includes waveforms that last more
  than about $700M$, and the top-right panel includes waveforms
  shorter than about $700M$. The bottom panel shows an enlargement of
  the merger phase for all waveforms.}
\end{figure}

\begin{figure}
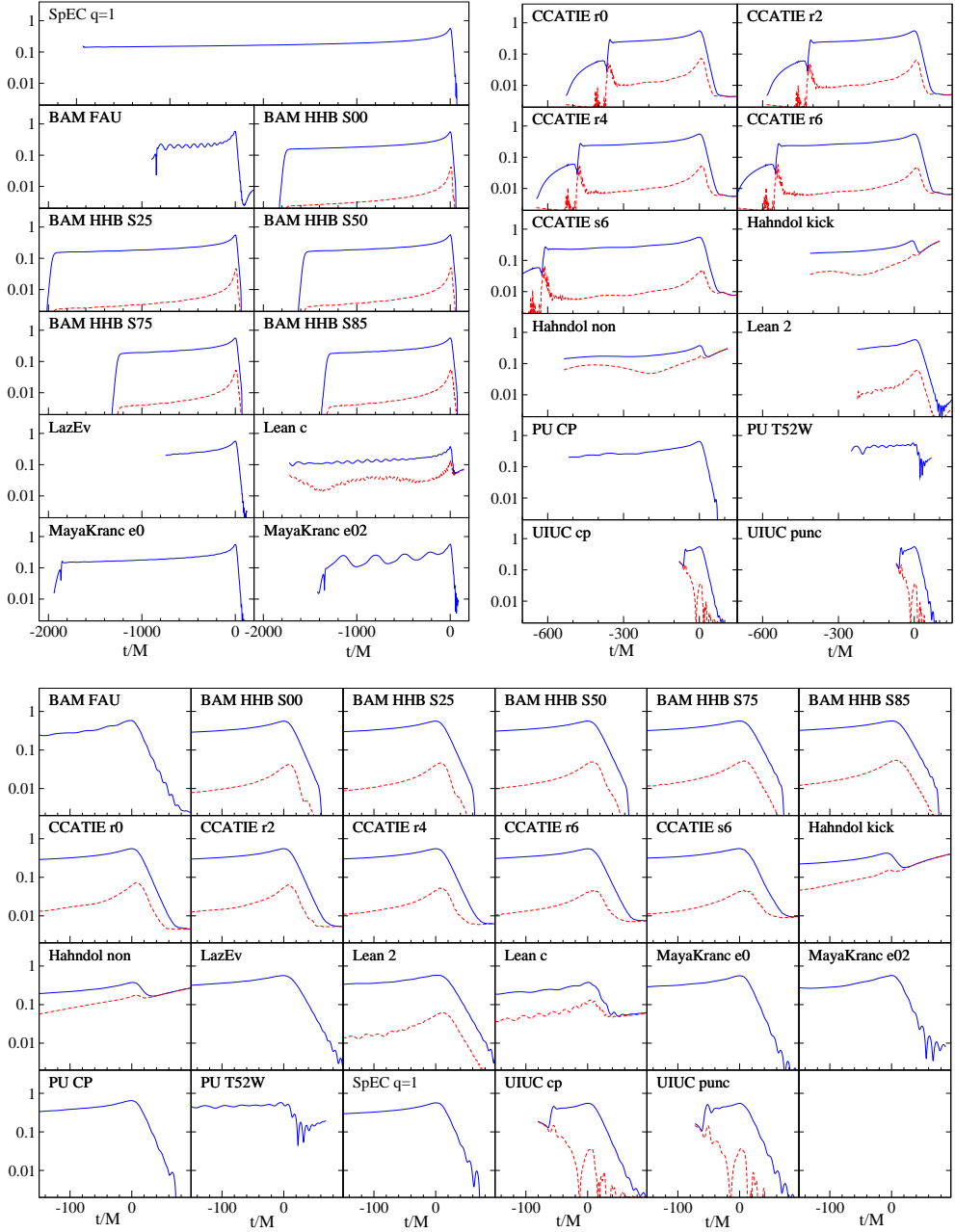

\includegraphics[width=0.49\textwidth]{figures/Prune_SumAllModes-A}
$\;$
\includegraphics[width=0.49\textwidth]{figures/Prune_SumAllModes-B}\\[-0.25em]

\includegraphics[width=\textwidth]{figures/Prune_SumAllModes-Merger}

\caption{\label{fig:NR-SumAllModes} {\bf Distribution of power into
  different spherical harmonics.}  The blue line shows
  $\left(\Sigma_{\ell,m}|h_{\ell m}\,r/M|^2\right)^{1/2}$.  A dashed red line, if
  present, shows the same sum, but {\em excluding} the $(\ell,m)=(2,\pm
  2)$ modes. 
  The separation between the two lines gives the relative importance
  of non $(2,\pm 2)$ modes.  If no red line is present for a certain
  run, then only the $(2,\pm 2)$ modes were supplied.  The layout is
  as in Fig.~\ref{fig:NR-Reh22}: The top panels show the complete
  waveforms, whereas the bottom panel shows an enlargement of the
  merger phase. The $x$-axis shows time in units of $M$.
}
\end{figure}

The NINJA project has studied BBH coalescence
waveforms submitted by ten individuals and teams.
Participation in NINJA was open to anyone and the only restrictions were
that each contribution: (i)
was a numerical solution of the full Einstein equations, (ii)
consisted of only two waveforms, or up to five waveforms if they were part of a one-parameter
  family.
No restrictions were placed on the accuracy of each waveform. All
contributions followed the format specified
in~\cite{Brown:2007jx}. The waveforms are plotted in
Figures~\ref{fig:NR-Reh22} and \ref{fig:NR-SumAllModes}.
The contributed waveforms cover a variety of physical and numerical
parameters. Most simulations model low-eccentricity inspiral, the mass
ratio $q = m_1/m_2$ ranges from 1 to 4, and the simulations cover a range
of spin configurations.  The initial angular frequency of the $\ell=m=2$ mode
ranges from $0.033/M$ to $0.203/M$ (where $M$ denotes the 
sum of the initial black-hole masses). This initial angular frequency
marks where contributors consider the waveform sufficiently clean to represent 
the physical system (e.g. this will be chosen after initial unphysical radiation content, often referred to as ``junk radiation'' in numerical relativity, 
is radiated away).   The length of the waveforms 
varies between a few 100M to over 4000M. 
The contributions naturally differ in accuracy, both
regarding how well they capture the black-hole dynamics and in the
extraction of the gravitational-wave
signal. 
Table~\ref{tab:allwaveforms} lists a few key parameters that
distinguish the waveforms, and introduces the following tags for the
different contributions and codes:
%
%
BAM~HHB~\cite{Brugmann:2008zz,Husa:2007hp,Hannam:2007ik,Hannam:2007wf,Bruegmann:2003aw} 
and BAM~FAU~\cite{Brugmann:2008zz,Husa:2007hp,Tichy:2008du,Bruegmann:2003aw}
are contributions using the {\tt BAM} code,
{\tt CCATIE} is the AEI/LSU code~\cite{Alcubierre:2000xu,Alcubierre:2002kk,Koppitz:2007ev,Pollney:2007ss,Rezzolla:2007xa},
{\tt Hahndol} is the Goddard Space Flight Center's code~\cite{Imbiriba:2004tp,vanMeter:2006vi}, 
{\tt LazEv} is the RIT code~\cite{Zlochower:2005bj,Campanelli:2005dd,Dain:2008ck}, 
{\tt Lean} is Ulrich Sperhake's code~\cite{Sperhake:2006cy,Sperhake:2007gu,Sperhake:2008ga}, 
{\tt MayaKranc} is the Georgia Tech/Penn State code~\cite{Vaishnav:2007nm,Hinder:2007qu}, 
PU stands for the Princeton University code~\cite{Pretorius:2004jg,Pretorius:2005gq,Buonanno:2006ui,Pretorius:2007jn},
{\tt SpEC} for the Cornell/Caltech collaboration 
code~\cite{Scheel:2006gg,Pfeiffer:2007yz,Boyle:2007ft,Scheel:2008rj},
and
UIUC stands for the University of Illinois at
Urbana-Champaign team~\cite{Etienne:2007hr}.

The codes listed above use different formulations
of the Einstein equations, gauge conditions, mesh structures, initial
data and wave extraction methods;
we will attempt to give a unified presentation of common features
first, and then list further details of the approaches separately for
each contribution. Full details of each code are given in the
references.

The numerical codes follow either of two approaches to solving the
Einstein equations: (1) the generalised harmonic formulation, which
was the basis of Pretorius' initial breakthrough simulation of
coalescing black holes \cite{Pretorius:2005gq}, or 
(2) the moving-puncture approach, following
\cite{Campanelli:2005dd,Baker:2005vv}.  
Both approaches result in canonical choices for the construction of
initial data, the evolution system for the Einstein equations, and 
the treatment of the singularity inside the black-hole horizons.

\begin{table}
\begin{center}
\begin{tabular}{|l|l|l|l|l|l|c|c|}\hline
Code & Run & $q$ & $\vec S_i/m_i^2$ & $e$ & $\omega_{22}\, M$ & $D/M$ &
eccentricity
\\ 
\hfill Ref.     &     &   &                        &   &          &       &
     removal \\ \hline

BAM~FAU      & \cite{Tichy:2008du}       & 1 & 
see caption
& qc  & 0.06 & $9.58\,\hat y$ & T-PN \cite{Marronetti:2007ya,Marronetti:2007wz} \\
\hfill\cite{Brugmann:2008zz,Husa:2007hp}              &        &   &
&    &      & &   \\[1em]

BAM~HHB      & S00 \cite{Hannam:2007ik}   & 1 & $0$       & $< 0.002$   &
0.045 & $12\,\hat y$ & TR-PN \cite{Husa:2007rh} \\ 
\hfill\cite{Brugmann:2008zz,Husa:2007hp}              & S25 \cite{Hannam:2007wf}   & 1 & $0.25\,\hat z$    & $\approx 0.006$  &
              0.045 & $12\,\hat y$ & T-PN \cite{Brugmann:2007zj} \\
              & S50 \cite{Hannam:2007wf}   & 1 & $0.50\,\hat z$    & $\approx 0.006$  &
              0.052 & $11\,\hat y$ & -- '' -- \\ 
              & S75 \cite{Hannam:2007wf}   & 1 & $0.75\,\hat z$    & $\approx 0.006$ &
              0.06 &  $10\,\hat y$ & -- '' -- \\ 
              & S85 \cite{Hannam:2007wf}   & 1 & $0.85\,\hat z$    & $\approx 0.006$  &
              0.06 &  $10\,\hat y$ & -- '' -- \\[1em] 

{\tt CCATIE}
              & r0~\cite{Pollney:2007ss}    & 1 & $0.6\,\hat z$, $-0.6\,\hat z$  & qc &
              0.079 & $8\,\hat x$ & TR-PN \cite{Husa:2007rh} \\
\hfill\cite{Alcubierre:2000xu,Alcubierre:2002kk,Koppitz:2007ev,Pollney:2007ss}
	      & r2~\cite{Pollney:2007ss}     & 1 & $0.6\,\hat z$, $-0.3\,\hat z$    & qc  &
              0.078 & $8\,\hat x$ & -- '' -- \\ 
              & r4~\cite{Pollney:2007ss}     & 1 & $0.6\,\hat z$, $0$       & qc  &
              0.076 & $8\,\hat x$ & -- '' -- \\ 
              & r6~\cite{Pollney:2007ss}     & 1 & $0.6\,\hat z$, $0.3\,\hat z$    & qc  &
              0.075 & $8\,\hat x$ & -- '' -- \\
              & s6~\cite{Rezzolla:2007xa}     & 1 & $0.6\,\hat z$     & qc  &
              0.074 & $8\,\hat x$ & -- '' -- \\[1em] 

{\tt Hahndol}      & kick   & 3 & $0.2\,\hat x$, $0.022\,\hat x$ & qc  &0.078 &  $8.007\,\hat{y}$ & T-PN \cite{Kidder:1995zr}\\ 
\hfill\cite{Imbiriba:2004tp,vanMeter:2006vi}            & non    & 4 & $0$     & qc
              &0.070 & $8.470\,\hat{y}$ & -- '' -- \\ 

{\tt LazEv}$\,$\cite{Zlochower:2005bj,Campanelli:2005dd}$\!$    & MH~\cite{Dain:2008ck} & 1 & $0.92\,\hat z$   & qc  &
0.07  & $8.16\,\hat x$ & T-PN \cite{Kidder:1995zr,Blanchet:1999pm} \\[1em]

{\tt Lean}\hfill\cite{Sperhake:2006cy}      &  c     & 4 & $0$       & qc  &
0.05  & $10.93\,\hat x$ & T-PN \cite{Brugmann:2008zz} \\ 
             &  2     & 1 & $0.926\,\hat z$   & qc  &
              0.11  & $6.02\,\hat x$ & T-PN \cite{Kidder:1995zr} \\[1em]

{\tt MayaKranc}     & e0  \cite{Hinder:2007qu}   & 1 & $0$       &  qc &
0.05 & $12\,\hat x$ & TR-PN \cite{Husa:2007rh} \\
  \hfill\cite{Vaishnav:2007nm}            & e02 \cite{Hinder:2007qu}   & 1 & $0$       & 0.2&
              0.05 & $15.26\,\hat x$ & n/a \\[1em]

PU\hfill\cite{Pretorius:2004jg,Pretorius:2005gq}  & CP \cite{Buonanno:2006ui}    & 1 & $0.063\,\hat z$      & qc  &
0.07  & $9.5\,\hat x$ & T-ID \cite{Cook:2004kt} \\
              & T52W \cite{Pretorius:2007jn}  & 1 & $0$       & $\ge
              0.5$ & 0.07 & & n/a \\[1em]

{\tt SpEC}\hfill  \cite{Scheel:2006gg}        & q=1 \cite{Boyle:2007ft,Scheel:2008rj}       & 1 & $0$       & $5\times 10^{-5}$  &
0.033 & $15\,\hat x$ & TR-it \cite{Pfeiffer:2007yz} \\[1em]

UIUC \hfill\cite{Etienne:2007hr}          & cp \cite{Etienne:2007hr}     & 1 & $0$       & qc  &
0.194 & $4.790\,\hat x$ & T-ID \cite{Cook:2004kt} \\ 
             & punc \cite{Etienne:2007hr}   & 1 & $0$     & qc  &
              0.203 & $4.369\,\hat y$ & T-ID \cite{Tichy:2003qi} \\  \hline
\end{tabular}
\end{center}
\caption{{\bf Initial conditions for numerical waveforms.}
The columns list, in order from left
to right, the name of the contribution or code, the name of the run
where appropriate, 
the mass ratio $q=m_1/m_2$ where $m_1\ge m_2$, the
spins of the black holes in vector form (if only one spin is given, both spins are equal), an estimate of the initial
eccentricity of the orbit (the entry qc denotes cases where quasi-circular inspiral, i.e.~zero eccentricity
is modelled, but a value of the eccentricity has not been reported), the initial frequency of the $(\ell,m)=(2,2)$
mode (rounded to three digits), the initial coordinate separation
of either the black-hole punctures or the excision surfaces, and where
appropriate the method of eccentricity removal.  All binaries start
out in the $xy$-plane with initial momenta tangent to the $xy$-plane.  See text for the
identification of each contribution, and a description of the notation
in the last column. 
The dimensionless spins of the BAM~FAU run are $(-0.634,-0.223, 0.333)$ and 
$(-0.517,-0.542,0.034)$.}
\label{tab:allwaveforms}
\end{table}

\begin{table}
\begin{center}
\begin{tabular}{|l|l|c||c|c|c|}\hline
Code & Run & q 
& $\Delta T_{\rm 100}$ [s] & $f_{i,\rm 100}$ [Hz] & $M_{30 Hz} [M_\odot]$ \\ 
 \hline

BAM~HHB    & S00  & 1 
& 1.03 & 15 & 48  \\ 
            & S25  & 1 
& 1.15 & 15 & 48  \\ 
            & S50  & 1 
& 1.03 & 17 & 56  \\ 
            & S75  & 1 
& 0.81 & 19 & 65  \\ 
            & S85  & 1 
& 0.87 & 19 & 65  \\ 

BAM~FAU    &      & 1 
& 0.54 & 19 & 65  \\ 

{\tt CCATIE}& r0   & 1 
& 0.34 & 26 & 85  \\ 
            & r2   & 1 
& 0.37 & 25 & 84  \\ 
            & r4   & 1 
& 0.40 & 25 & 82  \\ 
            & r6   & 1 
& 0.45 & 24 & 81  \\ 
            & s6   & 1 
& 0.59 & 24 & 80  \\ 

{\tt Hahndol}&kick & 3 
& 0.25 & 25 & 84  \\ 
            & non  & 4 
& 0.32 & 23 & 75  \\ 

{\tt LazEv} & MH   & 1 
& 0.43 & 23 & 75  \\ 

{\tt Lean}  &  c   & 4 
& 0.92 & 16 & 54  \\ 
            &  2   & 1 
& 0.20 & 36 & 118\,\,\,  \\ 

{\tt MayaKranc}& e0& 1 
& 1.23 & 16 & 54  \\ 
            & e02  & 1 
& 0.74 & 16 & 54  \\ 

PU          & CP   & 1 
& 0.29 & 23 & 75  \\ 
            & T52W & 1 
& 0.16 & 23 & 75  \\ 

{\tt SpEC}  & q=1  & 1 
& 1.96 & 11 & 36  \\ 

UIUC        & cp   & 1 
& 0.10 & 63 & 209\,\,\,  \\ 
            & punc & 1 
& 0.10 & 66 & 219\,\,\,  \\ 

 \hline
\end{tabular}
\end{center}
\caption{\label{tab:allwaveforms-SI}%
{\bf Characteristic duration, mass and frequencies} of
the waveforms summarised in table~\ref{tab:allwaveforms}.
The columns $\Delta T_{\rm 100}$ and $f_{i, 100}$ give the duration and initial
frequency of the waveform when scaled to total mass $M=100M_{\odot}$.
$M_{30Hz}$ is the total mass of the waveform when it is scaled so that
the initial frequency is 30Hz (this sets the lowest mass at which each waveform  can be injected into the NINJA data).  }
\end{table}


\subsection{Summary of the simulation algorithms}
\label{ssec:sumalg}

\subsubsection{Initial Data}
\label{ssec:id}

Due to the presence of constraint equations, specifying initial data 
in numerical relativity is far from trivial, 
for a general overview see e.g.~\cite{Cook:2000vr}.
All of the results presented here make the simplifying assumption of
conformal flatness for the spatial metric of the initial slice, which
leads to some spurious gravitational
radiation in the initial data.  All contributions attempt to
model non-eccentric inspiral, except for the two data sets
 PU--T52W and {\tt MayaKranc}--e02.  However, the degree of
``quasi-circularity'' varies, and in general one should bear in mind
that the definition of eccentricity for fully general-relativistic
orbits is not unique (see for
example~\cite{Sperhake:2007gu,Hinder:2007qu}).   
The data set PU--T52W is notable for the fact
that the BBH
was constructed via scalar field collapse. Specifically,
the initial data consists of two, compact, dense distributions of 
scalar field energy, separated by some distance and Lorentz boosted in opposite
directions orthogonal to the line between them. Upon subsequent
evolution, each scalar field pulse quickly collapses to form a
black hole, with all remnant scalar
field energy radiating away from the domain on the order
of the light-crossing time of the orbit. This is the same time scale on
which spurious gravitational radiation present in all current
initial-data sets leaves the domain of the inspiral, and hence for
practical purposes this can be considered a vacuum merger.  All other
runs start from vacuum initial data. 

Most codes ({\tt BAM}, {\tt CCATIE}, {\tt Hahndol}, {\tt LazEv}, {\tt
  Lean}, {\tt MayaKranc} and the UIUC 
code) adopt the ``moving puncture'' approach, following
\cite{Campanelli:2005dd,Baker:2005vv}.  These codes use 
puncture initial data
\cite{Bowen:1980yu,Beig:1993gt,Brandt:1997tf} to model 
black holes, resulting in initial data that contain a separate 
asymptotically flat end within each black hole.
Constructing such initial data is mathematically 
well understood~\cite{Beig:1993gt,Dain:2001ry}. 
The codes {\tt CCATIE}, {\tt LazEv}, {\tt Lean}
and {\tt MayaKranc} all use the same
pseudo-spectral solver for the Einstein constraint
equations~\cite{Ansorg:2004ds}, and {\tt BAM} uses a variant thereof~\cite{Husa:2007hp}.  {\tt UIUC-punc} initial data is generated via the
  \textsc{Lorene} \cite{Lorene} multi-domain spectral libraries.  The {\tt Hahndol} code uses the second-order-accurate multi-grid solver \textsc{amrmg}
\cite{Brown:2004ma}, which is however tuned to give truncation errors
typically much smaller than those produced by the evolution code.

The generalised harmonic codes use conformal thin sandwich initial
data~\cite{York:1998hy}.  PU-CP and {\tt SpEC} use quasi-equilibrium
excision initial data  
 where the interior 
 of the black-hole horizons has been excised from
the numerical grid. The presence of black holes with desired linear
momenta and spins is enforced through the boundary conditions on the
excision surfaces and the numerical outer boundary during the solution
of the initial-value equations
\cite{Cook:2001wi,Cook:2004kt,Caudill:2006hw,Pfeiffer:2007yz}. This 
``excision technique'' is based on the 
defining property of black holes --- the horizons act as causal
membranes and information cannot escape from the inside.  The {\tt UIUC-cp}
simulation uses the same excised initial data, but fills the BH
interior with ``smooth junk'', as described in~\cite{Etienne:2007hr},
before evolving with the moving puncture technique.   

All codes take input parameters that ultimately determine the
individual black-hole masses $m_i$, spins $\vec{S}_i$, 
momenta $\vec{P}_i$ and coordinate separation $D$ of
the black holes (one should however be aware that in the strong field
regime of general relativity various subtleties 
are associated with the definition of all of these quantities). In
addition, the black-hole masses and dimensionless spins slowly change 
during the inspiral, which requires additional caution regarding the
definition and accuracy of the values of mass, spin, etc.  
There are two common methods to estimate the instantaneous individual
black-hole masses. One is to calculate the
\emph{apparent-horizon mass}, computed from the irreducible mass
(given by the area of each hole's horizon) and the spin according
to Christodoulou's \cite{Christodoulou:1970wf} relation
$m_i^2 = m_{i{\rm,irr}}^2 + S_i^2/(4 m_{i,{\rm irr}}^2)$. The other,
applicable only to puncture data, is to compute the
Arnowitt-Deser-Misner (ADM) mass~\cite{Arnowitt1962} at each puncture,
which corresponds to spatial infinity in a space that contains only
that black hole~\cite{Brandt:1997tf}. We
generally use the total black-hole mass $M=m_1+m_2$ to scale
dimensionful quantities, although sometimes the total conserved energy
($M_\mathrm{ADM}$) is used for this purpose. 
%
%
Without loss of generality all codes chose the rest frame where $\vec
P_1 = -\vec P_2$ and, thus, the net linear momentum vanishes
initially.     
%

Those simulations that attempt to model non-eccentric inspiral use
initial parameters calculated by a number of different methods. Ideal
initial parameters would produce tangential motion consistent with 
circular orbits, and radial motion consistent with slow inspiral. The
various methods to choose initial parameters can be broadly
characterised as those that attempt to provide only 
tangential motion (so that initially the black holes have no radial
momenta), denoted by ``T'' in the last column of Tab.~\ref{tab:allwaveforms}, 
and those that provide both
tangential and radial motion (denoted by ``TR''). The procedures to
estimate these parameters are based on properties of the
initial-data set (``ID''), post-Newtonian methods (``PN''), or an
iterative procedure following the results of several trial simulations
(``it''). In Tab.~\ref{tab:allwaveforms} we indicate which of
these variants was used, and provide a reference to the specific
algorithm; for the post-Newtonian methods in particular there are
several variants.  Note that the
estimates of the resulting eccentricity range from $e\sim 5\times 10^{-5}$
(for the {\tt SpEC} contribution) up to $e \sim 0.02$. 

The two data sets from the UIUC contribution actually compare
two {\em alternative} sets of non-spinning, equal-mass,
quasi-circular initial data, with initial orbital frequency
$M\Omega=0.0824$: 
(i) Puncture initial data with coordinate separation
$D/M=4.369$ and initial linear momentum of each BH set
according to \cite{Tichy:2003qi}, and (ii) Cook-Pfeiffer initial data
with coordinate separation $D/M=4.790$
\cite{Pfeiffer_data,Cook:2004kt} (measured from the centroids of the
apparent horizons), filling 
the BH interior with data that smoothly connect to the exterior as
described in \cite{Etienne:2007hr}. 
Both data sets yield the same final spin $\vert\vec
S_{BH}\vert/M_{BH}^2 = 0.68$, 
but differ at the level of a few percent in radiated energy and
angular momentum. 

For the eccentric {\tt MayaKranc} simulation (data set e02), the
conservative, third-post-Newtonian-order (3PN) expressions in 
Ref.~\cite{Konigsdorffer:2006zt} have been used to specify initial data.
These expressions require the
specification of the eccentricity $e$ and the mean motion $n =
2\pi/T_r$, where $T_r$ is the radial (pericenter to pericenter)
orbital period.  There are three PN eccentricities, which are the same
to 1PN order, and we choose $e_t$, which appears in the PN Kepler
equation, following Ref.~\cite{Konigsdorffer:2006zt}.  The quantity
$n$ has been chosen as $n = 0.01625/M$ ($T_r \sim 387 M$) and $e=0.2$.
The binary separation, $D/M=15.264$, was determined from equation~(23) in
Ref.~\cite{Konigsdorffer:2006zt}, and the tangential linear momentum,
$P/M$=0.0498, of each black hole at apocenter was obtained from $J = P
D$, where $J$ is the total angular momentum computed as a
post-Newtonian expansion in $n$ and $e$ (equation~(21) in
Ref.~\cite{Konigsdorffer:2006zt}).


\subsubsection{Evolution systems}
\label{ssec:ev}

There is a long history of casting the Einstein equations into systems
of partial differential equations, and in 
particular into the form of a well-posed initial value problem. The
process of writing the covariant Einstein equations 
in the form of three-dimensional tensor quantities that evolve in time
is commonly referred to as a 3+1 split. The fundamental idea
is to choose coordinates $\{x^{i},t\}$ $(i=1,2,3)$ such 
that the spacetime metric can be written in the form
\be\label{3+1_split}
ds^2 = -(\alpha^{2}-\gamma_{ij}\beta^{i}\beta^{j})dt^{2}
   + 2 \gamma_{ij}\beta^{j}dt\,dx^{i}
   + \gamma_{ij}dx^{i}\,dx^{j}, 
\ee
where $\gamma_{ij}$ is a positive-definite metric on the slices of
constant time $t$, and the scalar function $\alpha$ and  
vector field $\beta^i$ are commonly used to encode the freedom of
coordinate choice. They may in principle be freely specified, but in
practice they are judiciously prescribed, usually through further evolution equations.

\begin{table}
\begin{center}
\begin{tabular}{|c|c|c|c|c|c|c|c|}\hline
Code        & $\!$System$\!$ & $\!$Technique$\!$   & shift   &  $M \eta$ & $r_{max}/M$ & $r_{ext}/M$ & \TTT \BB $\displaystyle\frac{h_{min}}{0.001M}$   
\\\hline

BAM~HHB   & BSSN  &  FD--6         & 000 & 2         &  $773$      & $90$        & 56, 19 \\ 

BAM~FAU   & BSSN &  FD--6         & 000 & 2         &  $436$      &  $50$        & $16$\\ 

{\tt CCATIE}   & BSSN   &  FD--4         & 000 & 1         &  $819$    & $160$   & 20      \\

{\tt Hahndol} & BSSN    &  FD--$4,6$         & 000 &  2        &  $> 1000$   & $45$        & 19, 13 \\

{\tt LazEv}   & BSSN &  FD--4         & ttt & 6         &  1281      & $40$            &   3.1 \\ 

{\tt Lean}   & BSSN     &  FD--$4,6$ & 000 &  1.25,1     &  $153.6$, $256$   & $60$, $61$         & 19, 13 \\ 

{\tt MayaKranc} & BSSN  &  FD--4         & 000 & 2         &  $317.4$     & $70$            & 16, 19 \\ 

PU          &  GH &  FD--2         & n/a     &  n/a      &   $\infty$ & $50$           &         \\

{\tt SpEC}     & GH   &  Spectral       & n/a     &  n/a      & $\!450\to 230\!$           & $\!75-225\!$&       $\sim 3$ \\

UIUC       & BSSN  &  FD--4         & 000 & 0.25      &  $409.6$    & $70$            & 25 \\ 

\hline
\end{tabular}
\end{center}
\caption{{\bf Some properties of the NR evolution codes.}  The columns
  list, for each contribution, the employed evolution system, the
  numerical technique (FD-k stands for finite differences using k-th
  order stencils in the bulk), the time derivative and $\eta$ choices
  for the $\tilde{\Gamma}$-driver shift, the approximate location of
  the outer boundary, the radii used for wave extraction, and the
  finest grid--spacing.  If two numbers are given they correspond to
the two runs of the respective code listed in table~\ref{tab:allwaveforms} (for BAM\_HBB, $h_{\rm min}=0.019M$ applies to all runs with spin).
For the {\tt SpEC} run, $r_{\rm max}$ decreases during the run and
the waveform is extrapolated to $r_{\rm ext}=\infty$ based
on extraction at radii in the given interval~\cite{Boyle:2007ft,Scheel:2008rj}. }
\label{tab:numparameters}
\end{table}

The waveforms contributed to NINJA use versions of either of the two
formulations for which successful multi-orbit evolutions of black-hole 
binaries have been published so far: the generalised harmonic and the
BSSN/moving-puncture formulation of the Einstein equations. For
overviews of writing the covariant Einstein equations as a time
evolution problem, see e.g.\ \cite{York1979,Wald84,Friedrich:2000qv}.

The generalised harmonic formulation (see
e.g.\ \cite{Friedrich:2000qv}) writes the evolution equations in   
manifestly hyperbolic form as a set of coupled wave equations for
the space--time metric $g_{\mu\nu}$. The
{\tt SpEC} code uses this formulation in first 
order form \cite{Lindblom:2005qh}, while the PU contribution is based
on a second order version of the equations. 
Gauge conditions are enforced by specification of gauge-source functions $H^\mu$, either as a specified function of time, or through evolution equations
\cite{Pretorius:2005gq,Pretorius:2006tp,Boyle:2007ft,Scheel:2008rj}. 

All other codes use the first-order-in-time, second-order-in-space BSSN
formulation of the Einstein evolution equations
\cite{Nakamura:1987zz,Shibata:1995we,Baumgarte:1998te} in combination with
hyperbolic evolution equations for the lapse and shift. The BSSN
formulation consists of making a conformal decomposition of the spatial
metric, $\gamma_{ij} = \psi^4 \tilde{\gamma}_{ij}$, and all other
variables, and the introduction of $\tilde{\Gamma}^i = \partial_j
\tilde{\gamma}^{ij}$, which is treated as an independent variable. The
moving-puncture treatment of the BSSN system involves evolving not the
conformal factor $\psi$ but either $\phi = \ln\psi$ ({\tt CCATIE}), $W
= \psi^{-2}$ (BAM FAU, {\tt Hahndol}~\cite{Marronetti:2007wz,Baker:2008mj}), or $\chi = \psi^{-4}$
(used by all other BSSN codes); it also consists of the gauge choices
that we will summarise next.

All BSSN-based contributions evolve the lapse according to the 1+log
slicing condition \cite{Bona:1997hp}, \begin{equation}
(\partial_t - \beta^i \partial_i) \alpha = -2 \alpha
K\,. \label{oplwithshift} 
\end{equation}  
The shift vector field $\beta^i$ is evolved according to some variant of the
$\tilde\Gamma$-driver condition 
\cite{Alcubierre:2002kk,vanMeter:2006vi}). 
During the evolution these gauge conditions change the geometry of the
``puncture singularity'' and soften the singularity as discussed in
\cite{Hannam:2006vv,Hannam:2006xw,Brown:2007tb,Hannam:2008sg}. 

The original $\tilde\Gamma$-driver condition introduced in
\cite{Alcubierre:2002kk} is  
\begin{equation}
\label{Gfreezing0}
  \partial_t \beta^i = \frac{3}{4} B^i, \quad
  \partial_t B^i     = \partial_t \tilde \Gamma^i - \eta B^i.
\end{equation} 
The factor of $3/4$ is chosen such that at large distances the
propagation speed of the hyperbolic equation (\ref{Gfreezing0}) equals
the coordinate speed of light~\cite{Alcubierre:2002kk}, and the quantity
$\eta$ is a parameter with the dimensions of the inverse of a mass and
affects coordinate drifts: larger values of
$\eta$ lead to a stronger initial growth of the apparent horizon, and
thus to a magnification effect for the black
holes~\cite{Brugmann:2008zz}.
Variants of this condition
\cite{Campanelli:2005dd,Baker:2005vv,Baker:2006yw,vanMeter:2006vi,Gundlach:2006tw} 
consist of replacing some or all of the $\partial_t$ derivatives with
$\partial_0 = \partial_t - \beta^i \partial_i$. We will label these options
with reference to each of the three time derivatives in
(\ref{Gfreezing0}): ``ttt'' denotes that $\partial_t$ is used for all three
derivatives, ``000'' denotes usage of $\partial_0$. The properties of
the different choices are studied in
\cite{Gundlach:2006tw,vanMeter:2006vi}, and in
\cite{Gundlach:2006tw} it is proven that the combination of the BSSN
equations with the ``1+log'' slicing condition (\ref{oplwithshift})
and the ``000'' shift choice yields a well-posed initial-value problem. 

Small differences in the evolutions also originate in the choice of
initial lapse (all BSSN codes initialise the shift 
quantities $\beta^i$ and $B^i$ to zero). We first define a
Brill-Lindquist-like conformal factor, $\psi_{BL} = 1 + m_{1,p}/2 r_1 +
m_{2,p}/2 r_2$, where $r_A$ is the distance to the $A$th
puncture, and $m_{1,p}$ and $m_{2,p}$ parametrise the masses of the black
holes, although they are not in general equal to $m_1$ and $m_2$. 
The RIT contributions choose $\alpha(t=0) = 2/(1+\psi_{BL}^{4})$,
as does the {\tt Hahndol}--non contribution, while the {\tt Hahndol}--kick
contribution uses an approximate $\alpha(t=0)$ derived from the
late-time ``1+log'' Schwarzschild slicing~\cite{Hannam:2006xw}.
BAM~HHB, {\tt MayaKranc} and the UIUC group use $\alpha(t=0) =
\psi_{BL}^{-2}$, and 
BAM~FAU choose $\alpha(t=0) = \left[ (\psi_{BL} - 1)/2 + 1
  \right]^{-4}$. 

The generalised harmonic codes (PU and {\tt SpEC}) employ black-hole
excision, i.e., they excise from the computational grid a region around
the singularities inside each black hole.  

\subsubsection{Radiation Extraction}
\label{ssec:rad}

All groups use one of two popular methods to estimate the
gravitational-wave signal at a finite distance from the source: The {\tt
  SpEC} and {\tt CCATIE} contributions use the Zerilli-Moncrief/Sarbach-Tiglio
perturbative formalism~\cite{Moncrief:1974am,Nagar:2005ea,Sarbach:2001qq} (with
{\tt SpEC} following a version restricted to a Minkowski background in standard coordinates~\cite{Rinne:2008vn}), all other contributions use the
Newman-Penrose curvature scalar $\psi_4$. Both methods are implemented
in the {\tt CCATIE} code, and have been shown to give similar 
results~\cite{Koppitz:2007ev,Pollney:2007ss}). Summaries and details on
the implementations within particular codes can be found, for instance
in the references listed in Table~\ref{tab:allwaveforms}. 
Since the gravitational-wave signal can only be defined
unambiguously at null infinity, one typically considers several
extraction radii and performs some form of convergence test, although for
the present purpose most groups only report results for a single
extraction radius.  At finite radius both methods depend on the
coordinate gauge, and the Newman-Penrose method additionally requires the
choice of a tetrad, which is obtained by Gram-Schmidt orthonormalisation
of a tetrad of coordinate vectors.

For this work, all waveforms have been contributed as spherical harmonic
modes of spin-weight $-2$ of the strain, according to the specification
in \cite{Brown:2007jx}.  Computation of the strain from the
Zerilli-Moncrief odd- and even-parity
multipoles of the metric perturbation requires one time
integration~\cite{Nagar:2005ea,Pollney:2007ss}, 
in the Sarbach-Tiglio formalism the strain is algebraically related to 
the invariants at leading order in the inverse 
radius~\cite{Ruiz:2007yx,Nagar:2005ea}, and computation of
the strain from the Newman-Penrose curvature scalar $\psi_4$ requires two
time integrations. Time integration requires the proper choice of integration 
constants, and may require further ``cleaning procedures'' to get rid
of artifacts resulting from the finite extraction radii. For example, for
the BAM~HHB contribution unphysical linear drifts were removed by a
variant of the method described in \cite{Damour:2008te}, where higher
order than linear polynomials were used to remove unphysical drifts from
higher modes to further improve the properties of the derived strain. In
the RIT contribution, the strain was computed by taking the Fourier
transform of $\psi_4$, removing modes in a small region around $\omega =
0$, then dividing by $- \omega^2$ and taking the inverse Fourier
transform.

\subsubsection{Numerical Methods and Computational Infrastructure}
\label{ssec:num}

There are large overlaps regarding the numerical methods in the
present waveform contributions. With the exception of the {\tt SpEC} code,
which uses a multi-domain pseudo-spectral method, all codes use
finite-difference methods to discretise the equations. With the
exception of 
the PU contribution, which uses a second-order-accurate implicit
evolution scheme, all other codes use an explicit algorithm based on
method of lines: Usually standard fourth-order-accurate Runge-Kutta time stepping,
except for the {\tt SpEC} code which uses a fifth order Cash--Karp time-stepper with adaptive step--size.

The moving-puncture/BSSN-based codes use standard centred finite
differencing stencils; however the terms corresponding to the
Lie-derivative with respect to the shift vector are off-centred
(up-winded) by one grid-point. The {\tt CCATIE}, {\tt MayaKranc}, {\tt LazEv} and UIUC codes use
fourth-order-accurate stencils, the \texttt{BAM} 
code uses sixth-order stencils, the {\tt Hahndol} code uses 
sixth-order stencils combined with fifth-order up-winded stencils \cite{Baker:2005xe},
and the \texttt{Lean} code uses fourth-order for equal-mass and
sixth-order for unequal-mass data sets. All of these codes add standard
fifth-order Kreiss-Oliger dissipation~\cite{Kreiss73,Gustafsson95} to the
right-hand-sides of the evolution equations. The finite-difference orders
described here apply to the bulk of the computational domain. There are
contributions at other orders in different parts of the codes, which we
will describe below. However, the finite-difference order in the bulk
plays the dominant role in defining the accuracy of the present
simulations (and indeed the spatial finite-differencing order seems to
dominate over the order of time integration when sufficiently small
time steps are used), and for that reason we list in 
Tab.~\ref{tab:numparameters} the bulk spatial finite-difference order.

All codes except the {\tt SpEC} code use variants of Berger-Oliger mesh-refinement.
The PU and {\tt Hahndol} codes employ full adaptive mesh refinement, while
the other codes use a hierarchy of fixed refinement boxes which follow
the motion of the black holes. Several of the codes are based on the
\texttt{Cactus} computational toolkit~\cite{Goodale02a,cactus}  
and the \texttt{Carpet} mesh-refinement
code~\cite{Schnetter:2003rb,carpet} 
({\tt CCATIE}, {\tt Lean}, {\tt MayaKranc}, {\tt LazEv}, UIUC). The BAM~HHB and BAM~FAU
contributions both use the {\tt BAM} mesh refinement code. The
{\tt Hahndol} code  
uses the \texttt{PARAMESH} infrastructure~\cite{MacNeice00} with 
a uniform time step; all other mesh refinement
codes use a time step
that depends on the grid spacing, and for these codes time interpolation
at mesh-refinement boundaries introduces second-order errors. 

For interpolation between meshes of different spacing, the groups that
used fourth- or higher-order methods all use fifth-order-accurate ({\tt CCATIE},
UIUC, {\tt LazEv}, {\tt Lean}, {\tt MayaKranc} and {\tt Hahndol}'s 4:1 ``non'' data)
or sixth-order-accurate ({\tt BAM} and {\tt Hahndol}'s 3:1 ``kick'' data)
polynomial interpolation in space between different refinement levels so
that all spatial operations of the AMR method (i.e., restriction and
prolongation) are sixth-order accurate and the second derivatives of
interpolated values are at least fourth-order accurate.

A proper numerical treatment of gravitational waves in asymptotically
flat spacetimes would include null infinity and not require
boundary conditions at some finite distance from the source.  Most codes
circumvent this problem in essentially heuristic ways. The PU code
uses spatial compactification combined with numerical dissipation, all
BSSN codes use heuristic outgoing wave boundary conditions (which will
in general violate constraint preservation and potentially
well-posedness and will result in reflections of the outgoing
radiation).  The {\tt SpEC} code, in contrast, uses
constraint-preserving outer boundary conditions which are nearly
transparent to outgoing gravitational radiation and gauge
modes~\cite{Rinne:2007ui}. 

Note that several of the groups use the same apparent horizon finder 
code (\textsc{AHFinderDirect})
\cite{Thornburg:1995cp,Thornburg:2003sf}
({\tt Hahndol}, UIUC, {\tt CCATIE}, {\tt LazEv}, {\tt MayaKranc}, {\tt Lean}). 


\subsection{Accuracy}
\label{ssec:accuracy}

Estimates on accuracy are reported for the BAM~HHB
and {\tt SpEC} contributions. 
For the BAM~HHB simulations reasonably clean sixth-order convergence was
observed, as reported in \cite{Hannam:2007ik,Hannam:2007wf}. In the
waveform $r\Psi_4$, extracted at $R_{ex} = 90M$, the
uncertainty due to numerical errors and the use of finite extraction
radii is estimated as 0.25 radians in the phase and less than 3\% in the amplitude
of the $l=2,m=2$ mode. Modes up to $l = 8$ were calculated; the {\it relative} phase
uncertainty is the same for all of them (the {\it absolute} phase uncertainty is
proportional to $m$), but we estimate that the
amplitude uncertainty increases to as much as $10\%$ for the highest
modes. 
The {\tt SpEC} contribution is the only one that extrapolates the gravitational
wave signal to infinite
extraction radius (using third-order polynomial
extrapolation~\cite{Boyle:2007ft}). Various convergence tests indicate
that the resulting extrapolated waveform is accurate to $0.02$ radians
in phase and $0.5$ percent in amplitude~\cite{Boyle:2007ft}.

\section{Construction of the NINJA data set}
\label{sec:ninja_data}
The data provided by the numerical relativity groups follows the
format outlined in~\cite{Brown:2007jx}, which is based on the
mode decomposition of the gravitational radiation field at large
distances from the source. If we specify a gravitational
waveform $h_{\mu\nu}$ in the Transverse-Traceless (TT) gauge, we only
need the spatial components $h_{ij}$. We assume that we are
sufficiently far away from the source so that the $1/r$ piece dominates:
\begin{equation}
  h_{ij} = A_{ij}\frac{M}{r} + \mathcal{O}\left(r^{-2}\right)\,,
\end{equation}
where $M$ is the total mass of the system, $r$ is the distance from
the source, and $A_{ij}$ is a time-dependent TT tensor.  In the TT
gauge, $h_{ij}$ has two independent polarisations denoted $h_+$ and
$h_\times$ and the complex function
$h_+-ih_\times$ can be decomposed into modes using spin-weighted spherical
harmonics $\Ytwo_{lm}$ of weight -2:
\begin{equation}
\label{eq:mode-decomposition}
  h_+ - ih_\times = \frac{M}{r}\sum_{\ell=2}^{\infty}\sum_{m=-\ell}^\ell H_{\ell m}(t)\,
  \Ytwo_{\ell m}(\iota,\phi)\,.
\end{equation}
The expansion parameters $H_{lm}$ are complex functions of the
retarded time $t-r$, however if we fix $r$ to be the radius of the sphere
at which we extract waves then $H_{lm}$ are functions of $t$
only. The angles $\iota$ and $\phi$ are respectively the polar and
azimuthal angles in a suitable coordinate system centred on the
source. This decomposition is directly applicable to non-precessing
binaries. Otherwise, a comparison of the waveforms requires a careful
treatment of mode-mixing effects due to rotations of the frame;
see for instance \cite{Gualtieri:2008ux}.
The numerical data contributed to NINJA is given in the form of an
ASCII data file for each mode $(\ell,m)$, with accompanying meta-data describing the simulation~\cite{Brown:2007jx}. Only modes that contribute appreciably
to the final waveform are included, at the discretion of the contributing
group.  Each data file consists of three columns: time
in units of the total mass, and the real and imaginary parts of the
mode coefficients $H_{\ell m}$ as a function of time. Note that the
total mass $M$ scales both the time and the amplitude; thus the
BBH waveforms for each simulation can be scaled to an
arbitrary value of the mass.  (This is not true in the case of simulations which
include matter fields, but we do not consider such waveforms here.)

To model the signal seen by a gravitational-wave detector, we need to
calculate the detector strain $h(t)$ from the above mode decomposition. To do
this, we must choose particular values of the total mass, orientation and
distance from the detector.  Given the $H_{\ell m}$, the total
mass, the distance to the source, and the angles $(\iota,\phi)$, we
calculate $h_{+,\times}$ using Equation.~(\ref{eq:mode-decomposition}), and
use the detector response functions $F_{+,\times}$ (see, for example, Ref.~\cite{thorne.k:1987}) to calculate the observed
strain
\begin{equation}
  h(t) = h_+(t) F_+(\alpha,\delta,\psi) + h_\times(t) F_\times(\alpha,\delta,\psi)\,.
\end{equation}
Here $(\alpha,\delta)$ are sky-angles in the detector frame, $\psi$ is
the polarisation angle and the time $t$ is measured in seconds. In this
analysis, we wish to simulate signals that might be observed by the Initial
LIGO and Virgo detectors. There are three LIGO detectors: a 4~km detector and
a 2~km detector at the LIGO Hanford Observatory (called H1 and H2,
respectively) and a 4~km detector at the
LIGO Livingston Observatory (called L1). The Virgo detector is a 3~km detector in Cascina,
Italy (called V1). We used the same two-letter codes for the simulated NINJA detectors.
Since the location and alignment of the three observatories 
differ, we must use the appropriate detector response and arrival time to compute
the strain waveform $h(t)$ seen at each observatory. This ensures that the
waveforms are coherent between the detectors and simulate a true signal.

\begin{figure}
  \begin{center}
  \includegraphics[width=\textwidth]{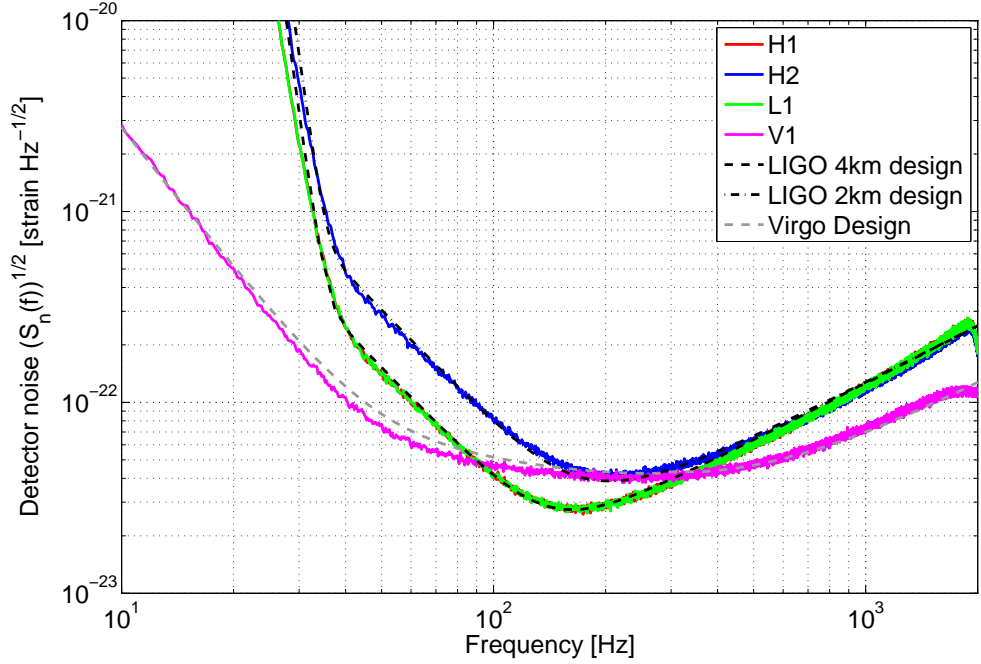}
  \end{center}
  \caption{The NINJA data noise curves and the design spectra of the
    first generation LIGO and Virgo detectors.}
  \label{fig:ninjapsd}
\end{figure}

To model the detector noise, we generated independent Gaussian noise time
series $n(t)$, sampled at $4096$~Hz, for each detector. This sample rate was
chosen to mimic that used in LSC-Virgo searches and assures a tolerable loss
in signal-to-noise ratio due to the discrete time steps. Stationary white noise time series
are generated and coloured by a number of time-domain filters designed to mimic  the design response of each of the LIGO and Virgo
detectors. Figure~\ref{fig:ninjapsd} shows the one-sided amplitude spectral 
density $\sqrt{S_n(f)}$ of each time detector's time series, where $S_n(f)$ is 
defined by
\begin{equation}
\left\langle \tilde{n}(f) \tilde{n}(f') \right\rangle = \frac{1}{2} S_n(|f|)
\delta(f-f').
\end{equation}
$\tilde{n}(f)$ denotes the Fourier transform of $n(t)$ and angle brackets
denote averaging over many realisations of the noise.
We see from Figure~\ref{fig:ninjapsd} that the noise power spectrum of 
the NINJA data set
closely approximates the Initial LIGO design sensitivity in the frequency
range of interest ($30$-$10^3\,$~Hz).
There is a
slight discrepancy with the Virgo design curve at low frequencies
(between approximately $20$ and $150\,$Hz), which is an
artefact of the Virgo noise generation procedure. Narrow-band
features such as the violin and mirror modes were removed from the detector
response used to compute the NINJA data, but were included in the calculation of
the Virgo design curve~\cite{SIESTA}. The $1/f$ tails of
these narrow-band features are responsible for the small discrepancy. 

Having generated the simulated detector data, we then generated a population of
simulated signals using the numerical relativity data.  This population was
constructed to cover a broad range of masses and signal amplitudes.  We
required that the starting frequency of the dominant $\ell=m=2$ mode of the
signal was not more than $30\,$Hz, an appropriate threshold given the
sensitivity curve of the Initial LIGO and Virgo detectors.  This sets a
minimum mass at which each waveform can be injected, which is given in Table
\ref{tab:allwaveforms-SI}. The minimum possible injection mass is therefore
$36 M_{\odot}$. The maximum mass was chosen as $350 M_{\odot}$.  To
get a good sample of long injected waveforms, we systematically chose a lower
range of masses for the longer waveforms. No restrictions were placed on the
other simulation parameters, i.e., the spins, mass-ratios and eccentricities. 
We ensured that waveforms from all the participating groups were equitably
represented by generating approximately 12 signals from the waveforms supplied
by each group. The time interval between adjacent injected signals was chosen
to be a random number in the range $700\pm 100\,$~s. 

Given these constraints, we generated the parameters of the signal population.
The logarithm of the distance to the binary was drawn from a uniform
distribution ranging from $50\,$Mpc to $500\,$Mpc, and the source 
locations and orientations were drawn from an isotropic distribution of angles. 
We then computed
waveforms corresponding to this population and at the appropriate sampling
rate. We required that the optimal matched filter signal-to-noise ratio of any
injection be greater than five in at least one of the four simulated detectors.
Any waveform that did not satisfy this constraint was discarded from the
population.  Subject to this condition, the distances of injected signals varied
from $52\,$Mpc to $480\,$Mpc (median at $145\,$Mpc), the injected total mass
range was $36M_\odot \le M \le346M_\odot$ (median at $155 M_\odot$), with
individual component masses in the range $11 M_\odot \le m_i \le193M_\odot$.  

Finally, the waveforms $h(t)$ were added to the simulated detector noise
$n(t)$ to generate the NINJA data set $s(t) = n(t) + h(t)$. As described
above, care was taken to ensure that signals were coherently injected in the
data streams from the four detectors.  The software for carrying out this
procedure is freely available as part of the LSC Algorithm Library
(LAL)~\cite{lal}.

\begin{figure}
  \begin{center}
  \includegraphics[width=0.9\textwidth]{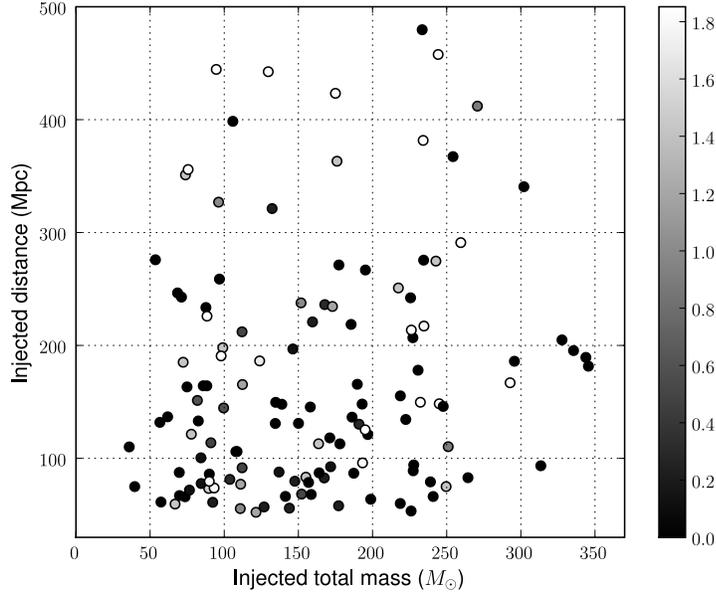}
  \end{center}
  \caption{{\bf The total mass and distance of the 126 NINJA injections.} The
    grey scale encodes the sum of the dimensionless spins of the black holes, 
$|\vec{S_1}/m_1^2 + \vec{S_2}/m_2^2|$.}
  \label{fig:distvsmass}
\end{figure}

The data set used in this analysis consisted of a total of 126 signals
injected in a total of 106 contiguous segments of noise each $1024\,$ s
long, thus spanning a duration of a little over $30\,$hours.
Figure~\ref{fig:distvsmass} shows the mass, spin and distance of the waveforms
contained in the NINJA data set.

\section{Data Analysis Results}
\label{sec:da}
Analysis of the NINJA data was open to all and nine groups submitted
contributions using a variety of analysis techniques.  Participating groups
were provided with the NINJA data set containing signals embedded in noise
and the parameters of the injected signals.
Analysts were not given access to the raw numerical-relativity waveforms or
noiseless injection data.

Methods used to analyse the NINJA data include: matched-filter based searches,
un-modelled waveform searches using excess-power techniques, and 
Bayesian model-selection and parameter-estimation techniques.  
Where possible, the
performance of different searches is compared. The limited scope of
the NINJA data set makes detailed comparisons difficult, however. 
A list of the data-analysis contributions is shown in Table~\ref{tab:allda}. 
\begin{table}
\begin{center}
\begin{tabular}{|l|l|l|}\hline
Group & Analysis & Section\\\hline
AEI & Phenomenological Waveforms in CBC pipeline & \ref{sssec:imr}\\
Birmingham & Bayesian Model Selection & \ref{ssec:bayesian} \\
Cardiff & Post-Newtonian (PN) Templates in CBC pipeline & \ref{sssec:insp} \\
Cardiff, Maryland & EOBNR waveforms in CBC pipeline & \ref{sssec:imr}\\
Goddard & Hilbert Huang Transform & \ref{sssec:hht} \\
Northwestern & Markov Chain Monte Carlo & \ref{ssec:mcmc} \\
Syracuse & Extended $\eta$ PN Templates in CBC pipeline & \ref{sssec:insp}\\
UMass, Urbino & Q-pipeline analysis & \ref{ssec:q_pipeline} \\
UWM & PN templates in CBC pipeline, Neyman-Pearson criteria & 
\ref{sssec:neyman}\\
UWM, UMass, Urbino & Ringdown analysis & \ref{sssec:ring} \\
UWM, UMass, Urbino & Inspiral, Merger, Ringdown combined search & 
\ref{ssec:comparison}\\
\hline
\end{tabular}
\end{center}
\caption{{\bf The data-analysis contributions to the NINJA project.}  ``CBC pipeline'' refers to the LSC-Virgo Compact Binary Coalescence group's analysis pipeline, described in section \ref{ssec:modeled}.}
\label{tab:allda}
\end{table}

In sections~\ref{ssec:modeled} and \ref{ssec:unmodeled} we describe results of
analyses using modelled (matched-filter) and un-modelled waveforms,
respectively.  Comparisons between these analyses are given in
Section~\ref{ssec:comparison} and Section~\ref{ssec:param_est} presents the
results of Bayesian model-selection and parameter-estimation analyses.

\subsection{Search pipelines using modelled waveforms}
\label{ssec:modeled}
When the waveform of the target signal is known, matched filtering is the
optimal search technique for recovering signals buried in stationary
noise~\cite{Wainstein:1962,Helstrom:1968}. This section describes the results
of filtering the NINJA data with matched-filter based analysis pipelines.
Results are given for waveforms that span only the inspiral signal, the
ringdown alone, and the full inspiral, merger and ringdown.  Although the
morphologies of these waveforms differ, the underlying analysis techniques are
similar in all cases.  All the contributions in this section use a pipeline
developed by the LSC and Virgo Collaboration to search for gravitational waves
from binary neutron stars and black holes in a network of
detectors~\cite{LIGOS3S4Tuning,Abbott:2007xi}.   We first describe the features
of this pipeline common to all the contributed matched-filter analyses before
presenting the results of searching the NINJA data using different
matched-filter templates.

The LSC-Virgo search pipeline performs a series of hierarchical
operations in order to search for real signals buried in the detector
noise: Given a desired search parameter space and waveform model, a
``bank'' of templates is created to cover the parameter space such that
the fractional loss in signal-to-noise ratio (SNR) between any signal
and the nearest template is less than a specified value (typically
$3\%$). All the NINJA inspiral searches use a non-spinning template bank
parametrised by the two component masses of the
binary~\cite{Owen:1998dk,Babak:2006ty,Cokelaer:2007kx}.  It has been
found that inspiral searches for spinning binaries using waveforms which
neglect the effect of spin are reasonably effective in most
cases~\cite{LIGOS3S4Tuning,VanDenBroeck:2009}. Ringdown searches use a
two parameter template bank parametrised by the frequency and quality
factor of the signal constructed to cover the desired range of mass and
spin~\cite{Creighton:1999pm}. Data from each of the detectors is
separately match filtered against this bank of
waveforms~\cite{Allen:2005fk,Creighton:1999pm} and a ``trigger'' is
produced whenever the SNR exceeds the desired threshold. All the
analyses used a threshold of $5.5$. A test is then performed which
discards triggers which do not have coincident parameters in two or more
detectors (time and masses for inspiral searches, and time, mass and
spin for ringdown searches)~\cite{Robinson:2008un,Goggin:2008}. These
coincident triggers provide the gravitational-wave candidates for the
ringdown analysis.  The triggers are ranked by a detection statistic
$\rho_\mathrm{c}$ constructed from the SNRs of the $N \ge 2$ individual
triggers in a coincidence by $\rho_\mathrm{c} = ({\sum_{i=1}^N
\rho_{i}^2})^{1/2}$. Coincident inspiral triggers are subject to a
second stage of filtering in which ``signal-based vetoes'' are also
calculated, which aim to separate true signals from noise fluctuations.
These include the $\chi^{2}$~\cite{Allen:2004gu} and
$r^{2}$~\cite{Rodriguez:2007} tests. Signal-based vetoes could also be
employed for ringdown searches, but at present they are not implemented
in the pipeline. For each trigger, we construct an effective SNR
$\rho_\mathrm{eff}$, which combines the matched-filter SNR and the value
of the $\chi^{2}$ signal based veto~\cite{Allen:2004gu}.  Explicitly,
the effective SNR is defined as~\cite{LIGOS3S4Tuning,Abbott:2007xi}
\begin{equation}\label{eq:eff_snr}
  \rho_{\mathrm{eff}}^{2} = \rho^{2} /
  \sqrt{\left(\frac{\chi^{2}}{\mathrm{DOF}}\right) \left( 1 +
  \frac{\rho^{2}}{250} \right)} \, .  
\end{equation}
where DOF signifies the number of degrees of freedom in the $\chi^{2}$
test.  For signals of moderate SNR, which are a good match to the
template waveform, the expected value of the $\chi^{2}$ is unity per
degree of freedom and consequently the effective SNR is approximately
equal to the SNR.  Non-stationarities in the data typically have large
values of $\chi^{2}$ and consequently the effective SNR is significantly
lower than the SNR.  A second test is then performed to discard
coincidences in which signal-based vetoes reduce the number of triggers
to less than two.  These coincidences provide the candidate
gravitational wave signals for the inspiral-based pipelines and they are
ranked by the combined effective SNR $\rho_{\mathrm{eff}} =
(\sum_{i=1}^N \rho_{\mathrm{eff}\ i}^2)^{1/2}$. To evaluate the
sensitivity of the analyses, we compare the list of gravitational-wave
candidates generated by filtering the NINJA data to the parameters of
the inject numerical relativity signals.

Six groups contributed matched-filter results to this analysis and the
results can be roughly divided into three categories based on the
waveform templates used: (i) searches based on the stationary-phase
approximation to the inspiral signal, which are designed to capture
various stages of the inspiral, merger and ringdown, (ii) searches which
use waveforms designed to model the full inspiral-merger-ringdown signal,
(iii) searches using ringdown-only waveforms obtained from black hole
perturbation theory.  Within these categories, different parameter
choices were made in order to investigate the ability of the pipeline to
detect the numerical relativity simulations.  Each of these three
approaches is described independently in the following sections.  A
comparison between these results is given in
Section~\ref{ssec:comparison}.

\subsubsection{Stationary Phase Inspiral Templates}
\label{sssec:insp} 
The workhorse template of the LSC-Virgo search pipeline is based on the
stationary-phase approximation to the Fourier transform of the
non-spinning post-Newtonian inspiral~\cite{Droz:1999qx,Allen:2005fk}.
This waveform (referred to as SPA or TaylorF2) has been used in the
search for binary neutron
stars~\cite{Abbott:2003pj,Abbott:2005pe,Abbott:2007xi,Abbott:2009tt}, 
sub-solar mass black holes~\cite{Abbott:2005pf,Abbott:2007xi,Abbott:2009tt}
and stellar mass black holes~\cite{Abbott:2009tt}. The TaylorF2 waveform
is parametrised by the binary's component masses $m_1$ and $m_2$ (or
equivalently the total mass $M = m_1 + m_2$ and the symmetric mass ratio
$\eta = m_1 m_2 / M^2$) and an upper frequency cutoff $f_\mathrm{c}$.
Amplitude evolution is modelled to leading order and phase evolution is
modelled to a specified post-Newtonian order. In this section we investigate
the performance of TaylorF2-based searches on the three simulated LIGO
detectors. Results which include the simulated Virgo detector are described in
the next section.  Several analyses were performed 
which test the ability of TaylorF2 waveforms to detect numerical relativity
signals. The analyses differed in the way the TaylorF2 waveforms or the
template bank were constructed.  The results of these searches are summarised
in Table~\ref{tab:inspiral_results}, each column giving the results from a
different search with a summary of the chosen parameters.  We first describe
the parameters varied between these analyses and then present a more detailed
discussion of the results.

\begin{figure}
  \begin{center}
  \includegraphics[width=0.8\textwidth]{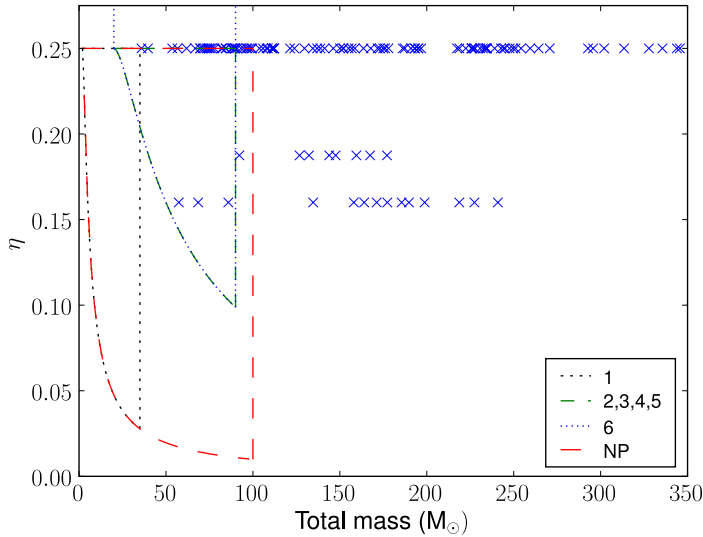}
  \end{center}
  \caption{{\bf Boundaries of the template banks used in inspiral searches} as a
  function of total mass $M$ and symmetric mass ratio $\eta$. The crosses show
  the location of the injections in the NINJA data set. The numbers in the
  legend correspond to entries in table~\ref{tab:inspiral_results}. Bank 6
  extends in a rectangle up to $\eta = 1.00$, as indicated by the arrows. NP
  is the bank used in the Neyman-Pearson analysis described in 
  Section~\ref{sssec:neyman}.}
  \label{f:ninjaBanks}
\end{figure}

All TaylorF2 NINJA analyses used restricted templates (i.e.~the amplitude is
calculated to leading order), however the phase was calculated to various
different post-Newtonian orders~\cite{Blanchet:2002av}. Phases were computed to
either two~\cite{Blanchet:1996pi,Blanchet:1995ez} or three point five
post-Newtonian order~\cite{Blanchet:2001ax,PhysRevD.71.129902,Blanchet:2004ek}
since these are, respectively, the order currently used in LSC-Virgo
searches~\cite{Abbott:2009tt} and the highest order at which post-Newtonian
corrections are known. After choosing a post-Newtonian order, one chooses a
region of mass-parameter space to cover with the template bank.
Figure~\ref{f:ninjaBanks} shows the boundaries of the template banks used in
the analyses. One search used the range used by the LSC-Virgo ``low-mass''
search \cite{Abbott:2009tt} ($m_1,m_2 \ge  1 M_\odot, M \le 35 M_{\odot}$) and
all other searches used templates with total masses in the range $20 M_\odot
\le M \le 90 M_\odot$.  These boundaries were chosen since there were no
signals in the NINJA data with mass smaller than $36 M_\odot$ and there is
little, if any, inspiral power in the sensitive band of the NINJA data for
signals with $M \gtrsim 100 M_\odot$.  The standard LSC-Virgo template bank
generation code~\cite{Babak:2006ty} restricts template generation to signals
with $\eta \le 0.25$, since it is not possible to invert $M$ and $\eta$ to
obtain real-valued component masses for $\eta > 0.25$. All but one of the
searches enforced this constraint, with the $0.03 \le \eta \le 0.25$ for the
low-mass CBC search and $0.1 \le \eta \le 0.25$ for the other
``physical-$\eta$'' searches. It is, however, possible to generate TaylorF2
waveforms with ``unphysical'' values of $\eta > 0.25$.  In two separate studies
using Goddard and Pretorius waveforms~\cite{Pan:2007nw}, and Caltech-Cornell
waveforms~\cite{Boyle:2009dg} it was observed that match between numerical
signals and TaylorF2 templates could be increased by relaxing the condition
$\eta \le 0.25$. One NINJA contribution uses a template bank with $0.1 \le \eta
\le 1.0$ to explore this.

Finally, it is necessary to specify a cutoff frequency at which to terminate the
TaylorF2 waveform. In the LSC-Virgo analyses, this is chosen to be the
innermost stable circular orbit (ISCO) frequency for a test mass in a
Schwarzschild spacetime 
\begin{equation}
\label{f_ISCO}
f_\mathrm{ISCO} = \frac{c^3}{6\sqrt{6}\pi GM}.
\end{equation}
This cutoff was chosen as the point beyond which the TaylorF2 waveforms
diverge significantly from the true evolution of the
binary~\cite{Blanchet:2002av}.  More recently, comparisons with numerical
relativity waveforms have shown that extending the waveforms up to higher
frequencies improves the sensitivity of TaylorF2 templates to higher mass
signals~\cite{Pan:2007nw,Boyle:2009dg}. The NINJA TaylorF2 analyses use
templates terminated at the ISCO frequency and two additional cut-off
frequencies: the effective ringdown (ERD) frequency and a weighted
ringdown ending (WRD) frequency. The ERD frequency was obtained by comparing
post-Newtonian models to the Pretorius and Goddard
waveforms~\cite{Pan:2007nw}. The ERD almost coincides with the fundamental
quasi-normal mode frequency of the black hole formed by the merger of an
equal-mass non-spinning black-hole binary. The weighted ringdown ending (WRD)
frequency lies between ISCO and ERD, and was obtained by comparing
TaylorF2 waveforms to the Caltech-Cornell numerical
signals~\cite{Boyle:2009dg}.

\begin{table}
\begin{tabular}{| l || c | c | c | c | c | c | c |}
\hline
\bf{Analysis} \T \B & $(1)$ & $(2)$ & $(3)$ & $(4)$ & $(5)$ & $(6)$ \\ \hline
\bf{Freq. Cutoff} \T \B & ISCO & ISCO & ERD & ERD &  WRD & WRD  \\ 
\hline
\bf{PN Order} & 2 PN & 2 PN & 2 PN & 3.5 PN &  3.5 PN& 3.5 PN  \\
\hline
\bf{Total Mass $M_{\odot}$} \T \B & 2--35 & 20--90 & 20--90 & 20--90 & 20--90 & 20--90  \\ 
\hline
\bf{$\eta$ range} \T \B & 0.03--0.25 & 0.10--0.25 & 0.10--0.25 & 0.10--0.25 & 0.10--0.25 & 0.10--1  \\ 
\hline
\parbox{2.3cm}{
\bf{Found Single\\ (H1, H2, L1)}}\TT \BB  & 69, 66, 75 & 72, 43, 66 & 83, 51, 81 & 91, 56, 87 & 90, 55, 88 & 90, 56, 88 \\ 
\hline
\parbox{2.3cm}{
\bf{Found \\Coincidence }} \TT \BB & 49 & 59 & 79 & 82 &  82 & 84 \\ 
\hline
\parbox{2.5cm}{
\bf{Found Second\\Coincidence}} \TT \BB & 48 & 59 & 77 & 81 &  81 & 81 \\ 
\hline 
\end{tabular}
\caption{{\bf Results of inspiral searches using TaylorF2 templates.}  There were 126
injections performed into the data.  The table above shows the number of
injections which were recovered from the three simulated LIGO detectors (H1, H2 and L1) using various different waveform families,
termination frequencies $f_\mathrm{ISCO}$, $f_\mathrm{ERD}$ and $f_\mathrm{WRD}$ 
(as described in the text), and post-Newtonian orders.} 
\label {tab:inspiral_results}
\end{table}

The results of these searches are reported in
Table~\ref{tab:inspiral_results}.  The principal result is the number of
injected signals detected by the search.  For simplicity, we define a
detected signal as one for which there is a candidate gravitational-wave
signal observed within $50$~ms of the coalescence time of the injection,
determined by the maximum gravitational-wave strain of the injected
signal.  We do not impose any additional threshold on the measured SNR or
effective SNR of the candidate.  For a single detector, this will lead
to a small number of falsely identified injections, but for coincidence
results the false alarm rate is so low that we can be confident that the
triggers are associated with the injection. We now describe these
results in the order that they appear in
Table~\ref{tab:inspiral_results}.

Search~$(1)$ used second order post-Newtonian templates terminated at
$f_\mathrm{ISCO}$ with a maximum mass of $M \le 35 M_{\odot}$.  Despite the
fact that no NINJA injections had a mass within the range of this search, a
significant number of signals were still recovered in coincidence both before
and after signal consistency tests.  Although the templates are not a
particularly good match to the injected signals, they are still similar enough
to produce triggers at the time of the injections.  Search~$(2)$ changed the
boundary of the template bank to $20 M_\odot \le M \le 90 M_{\odot}$, but left
all other parameters unchanged.  The number of detected signals increases
significantly as more signals now lie within the mass range searched. 

Search~$(3)$ extended the upper cutoff frequency of the waveforms to
$f_\mathrm{ERD}$. The number of signals detected increased from 59 to 77, as
expected since these waveforms can detect some of the power contained in the
late inspiral or early merger part of the
signal~\cite{Pan:2007nw,Boyle:2009dg}. Search~$(4)$ extends the post-Newtonian
order to 3.5~PN, slightly increasing the number of detected signals to 81.
With the limited number of simulations performed in this first NINJA analysis,
it is difficult to draw a strong conclusion, although there does seem to be
evidence that the higher post-Newtonian order waveforms perform better,
consistent with previous comparisons of post-Newtonian and numerical
relativity waveforms
\cite{Pan:2007nw,Baker:2006ha,Hannam:2007ik,Boyle:2008ge,Boyle:2009dg}.
Search~$(5)$ uses an upper-frequency cutoff of $f_\mathrm{WRD}$ for the
templates. The number of injections found in coincidence for this search is
the same as the search using $3.5$ order templates with a cutoff of
$f_\mathrm{ERD}$, although there are slight differences in the number of found
injections at the single detector level.

\begin{figure}
  \includegraphics[width=0.50\textwidth]{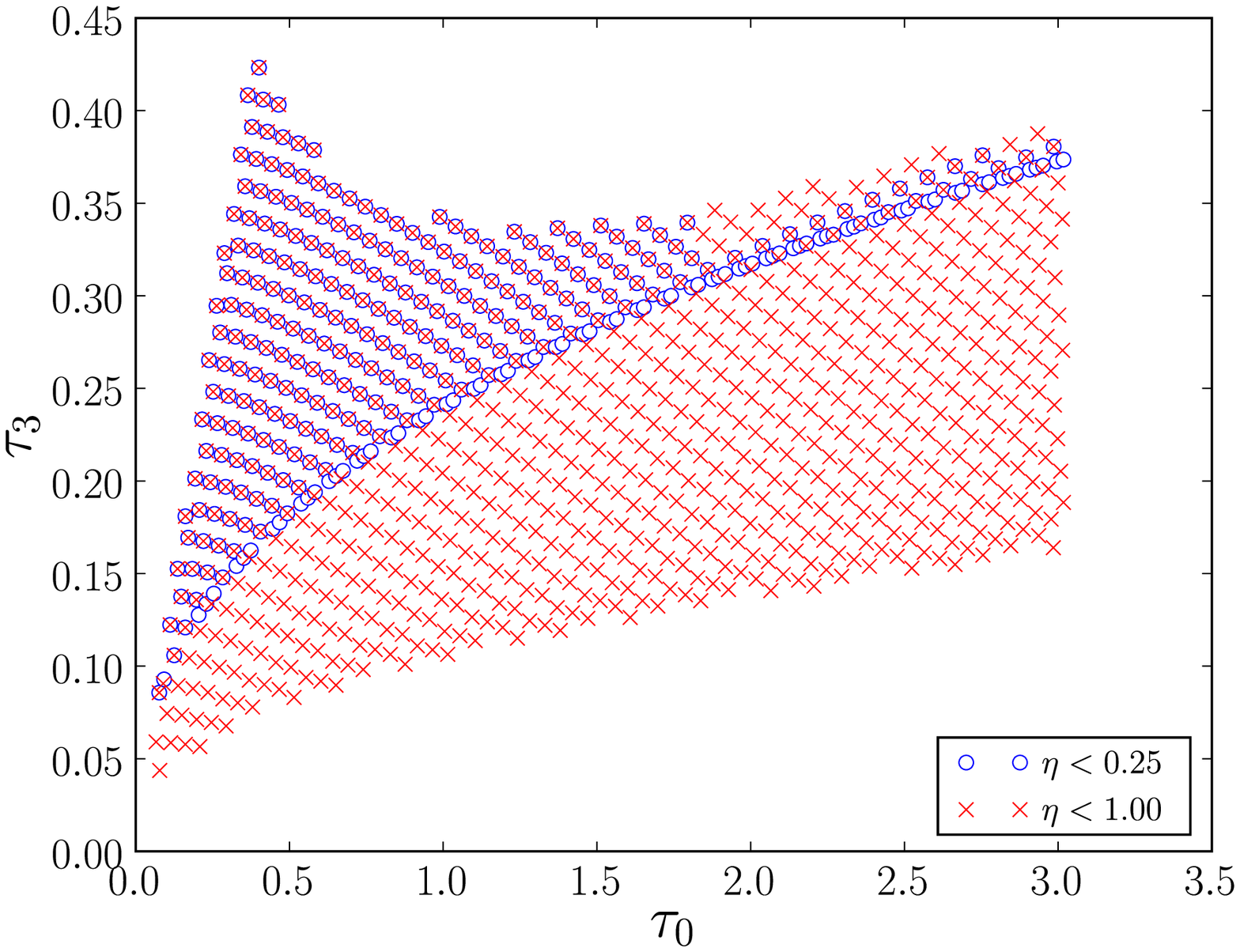}
  \includegraphics[width=0.50\textwidth]{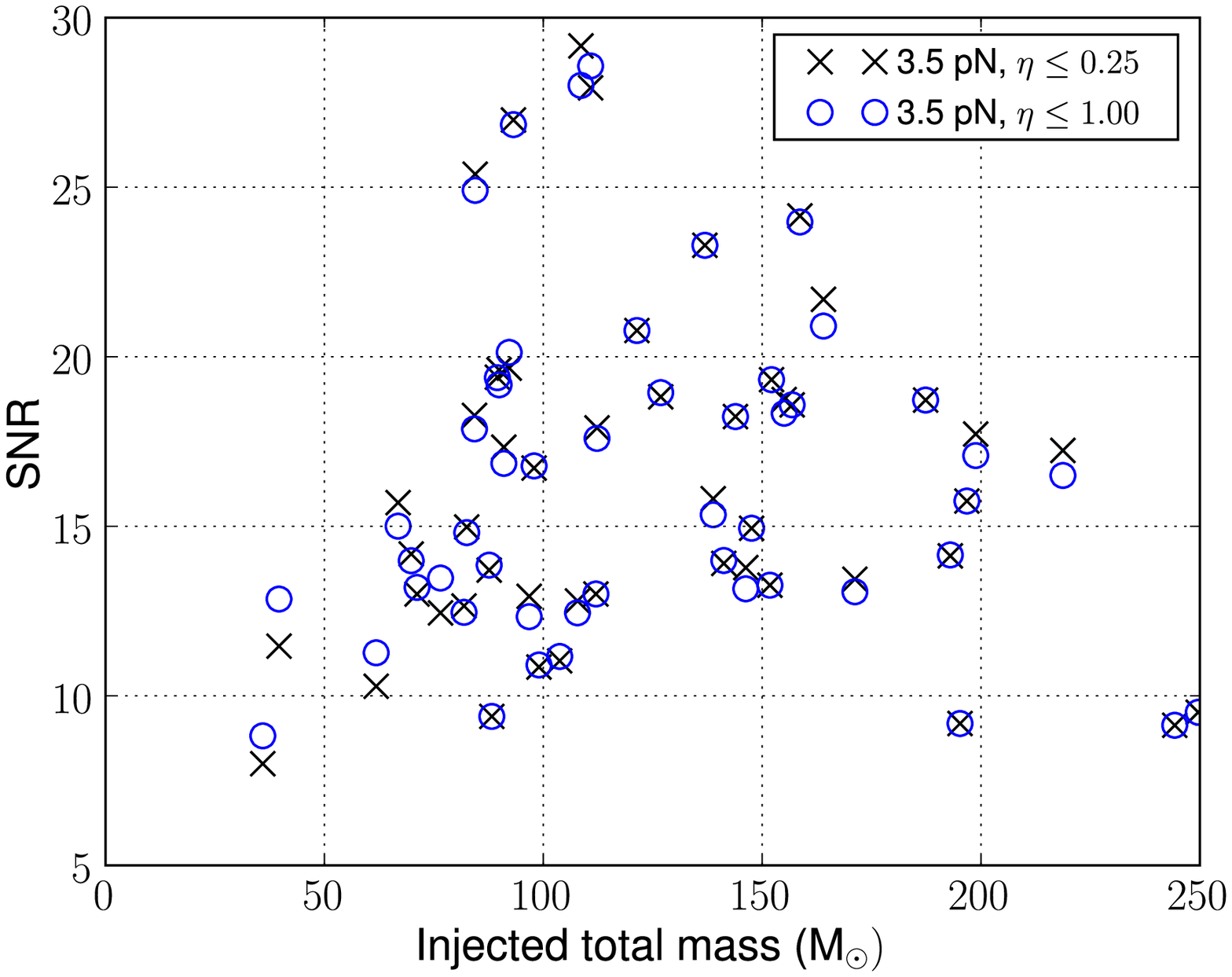}
  \caption{{\bf Results from the extended template bank.} 
  {\bf Left:} The template bank generated by the LSC-Virgo
  search pipeline (circles) and the bank obtained by extending to
  $\eta \leq 1.00$ (crosses). In this figure the bank is parametrised
  by $\tau_0$ and $\tau_3$ which are related to the binary masses by
  $\tau_0 = 5M/(256\eta v_0^8)$ and $\tau_3 = \pi M/(8\eta v_0^5)$,
  where $v_0 = (\pi M f_0)^{1/3}$ is a fiducial velocity parameter
  corresponding to a fiducial frequency $f_0 = 40.0 Hz$.
  {\bf Right:} The
  signal-to-noise (SNR) ratio at which NINJA injections were recovered using
  the $\eta \le 0.25$ bank (squares) and the $\eta \le 1$ extended bank
  (circles) in the Hanford detectors, given by $\rho =
  (\rho_\mathrm{H1}^2 + \rho_\mathrm{H2}^2)^{1/2}$. The SNR of the signal
  recovered using the extended bank shows with significant ($> 10\%$) 
  increases over the standard bank for certain injections.}
  \label{f:templateBanks}
\end{figure}

Search~$(6)$ extends the template bank of search~$(5)$ to unphysical values of
the symmetric mass ratio. Extending the bank to $\eta\le 1$ increases the
number of templates in the bank by a factor of $\sim 2$. The original and
modified template banks are shown in Figure~\ref{f:templateBanks}. With the
extended template bank the number of injections found in coincidence remains
the same as search~$(5)$ after signal-based vetoes are applied.  However, many
of the injections are recovered at a higher SNR, particular the low-mass
signals, as shown in Figure~\ref{f:templateBanks}.  Some injections show a
reduction in SNR; more work is needed to understand this effect.

Finally, we note that the majority of signals passed the $\chi^2$
signal-based veto with the thresholds used in the LSC-Virgo pipeline.  The
last two lines of Table~\ref{tab:inspiral_results} show the number of
recovered signals before and after these signal-based vetoes are
performed. The post-Newtonian templates and numerical relativity signals are
similar enough that virtually all of the injected signals survive the signal
based vetoes. 

\begin{figure}
\begin{center}
  \includegraphics[width=0.49\textwidth]{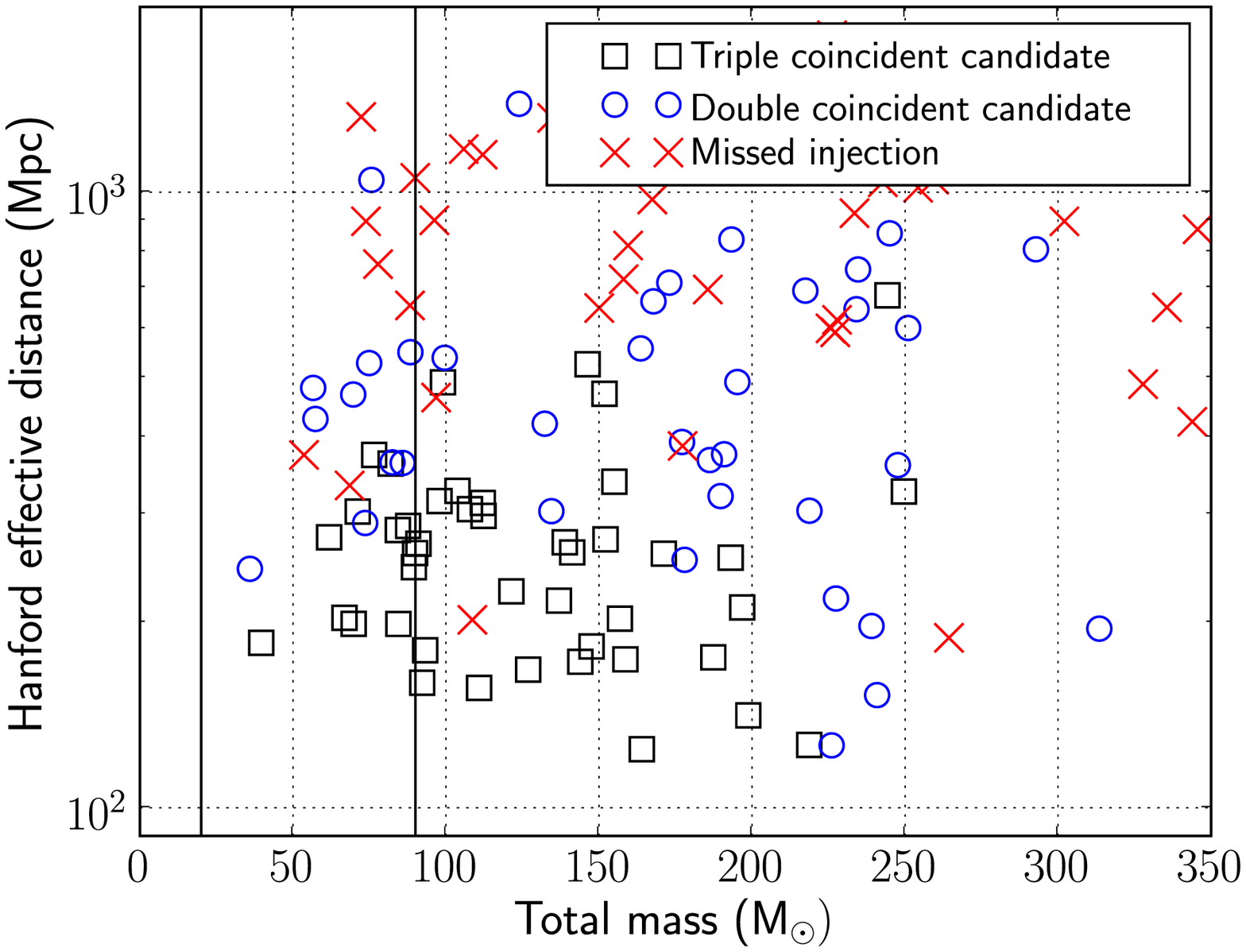}
  \includegraphics[width=0.49\textwidth]{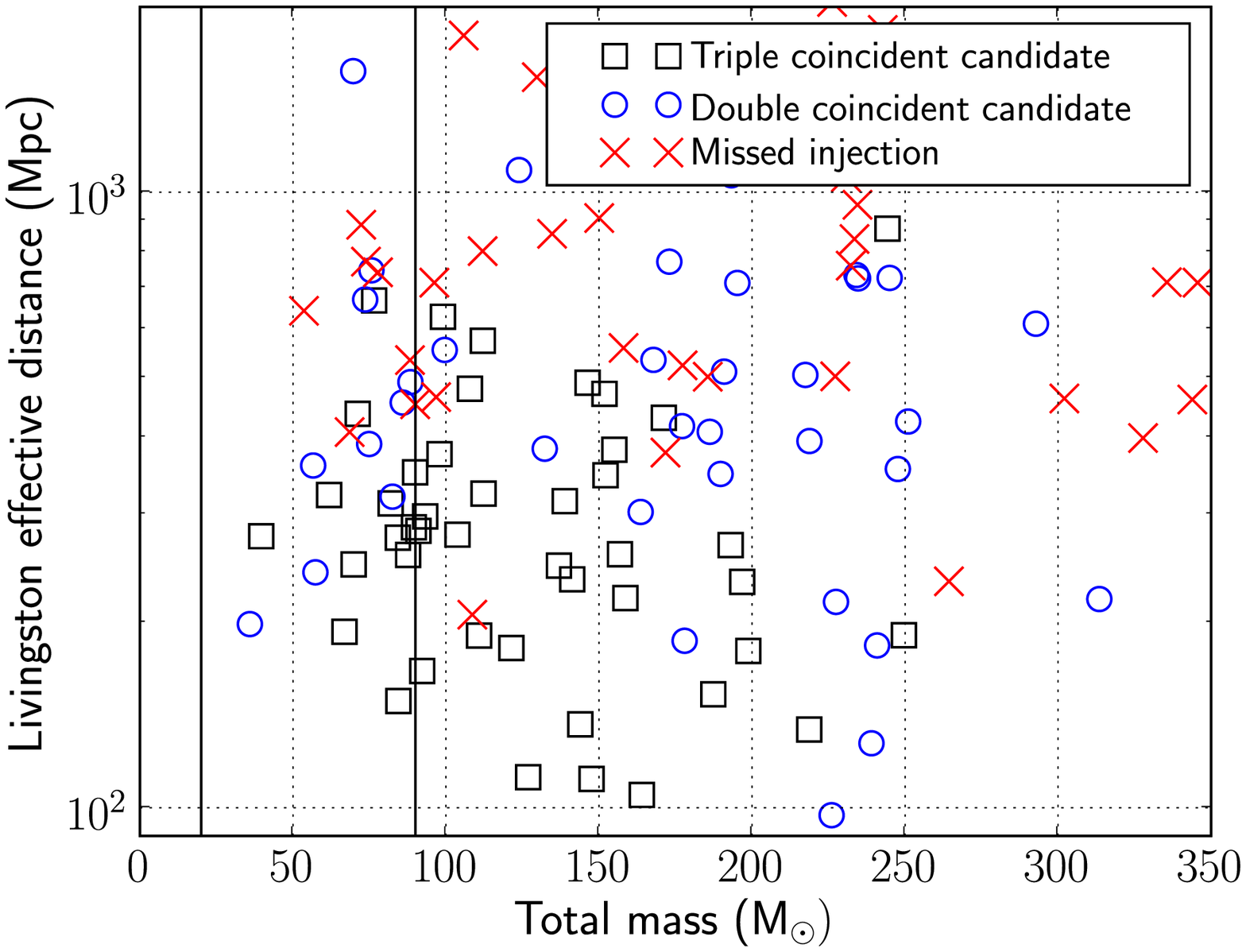}
\end{center}
\caption{\textbf{Found and missed injections using TaylorF2 templates terminated at ERD}, plotted as a function
of the injected effective distance in Hanford (left) and Livingston (right) and the total mass of the injection. Since the LIGO Observatories are not exactly aligned, the effective distance of a signal can differ, depending on the sky location of the signal.
The vertical bars mark the limits of the template bank used in the search.  For
the lower masses, we see that the majority of the closer injections 
are found in coincidence in all three of
the detectors.  There is then a band of injections which are found only
in two detectors -- H1 and L1 and not the less sensitive H2 detector.
For higher masses, the results are less meaningful as the template bank
was only taken to a total mass of $90 M_{\odot}$.}
\label{fig:3_5pn_found_missed}
\end{figure}

To illustrate the results of these analyses in more detail, 
Figure~\ref{fig:3_5pn_found_missed} shows which signals were detected and which were
missed by the 3.5 order post-Newtonian TaylorF2 templates terminated at
$f_\mathrm{ERD}$, as a function of injected
total mass and effective distance of the binary (a measure of the
amplitude of the signal in the detector), defined by~\cite{Allen:2005fk}
\begin{equation}
D_\mathrm{eff} = d \left/ \sqrt{F_+^2 (1 + \cos^2 \iota)^2/4 + F_\times^2 \cos^2 \iota}\right.,
\label{eq:effdist}
\end{equation}
where $d$ is the luminosity distance of the binary.

One signal, with total mass of $110 M_{\odot}$ and effective distance $\sim
200$ Mpc, was missed while others with similar parameters were found.  This
signal was one of the Princeton waveforms (labelled \verb|PU-e0.5| in
Figure \ref{fig:NR-Reh22}) for which the maximum amplitude occurs at the start
of the waveform rather than at coalescence\footnote{That the maximum 
occurs at the start of the waveform is in part an ``artifact'' of the 
double-time integration from the Newman-Penrose scalar $\psi_4$ to the 
metric perturbation $h$, and in part a coordinate artifact.
The two integration constants were chosen to remove a 
constant and linear-in-time piece for $h$, however, there is still 
a non-negligible quadratic component; we {\em suspect} this is purely gauge, 
though lacking a better understanding of this it was not removed from the 
waveform.}, rendering our simple coincidence
test invalid.  The injection finding algorithm compares the peak time to the
trigger time and, even though triggers are found at the time of the simulation,
there are no triggers within the $50$~ms window used to locate detected
signals.

\begin{figure}
    \includegraphics[width=0.50\textwidth]{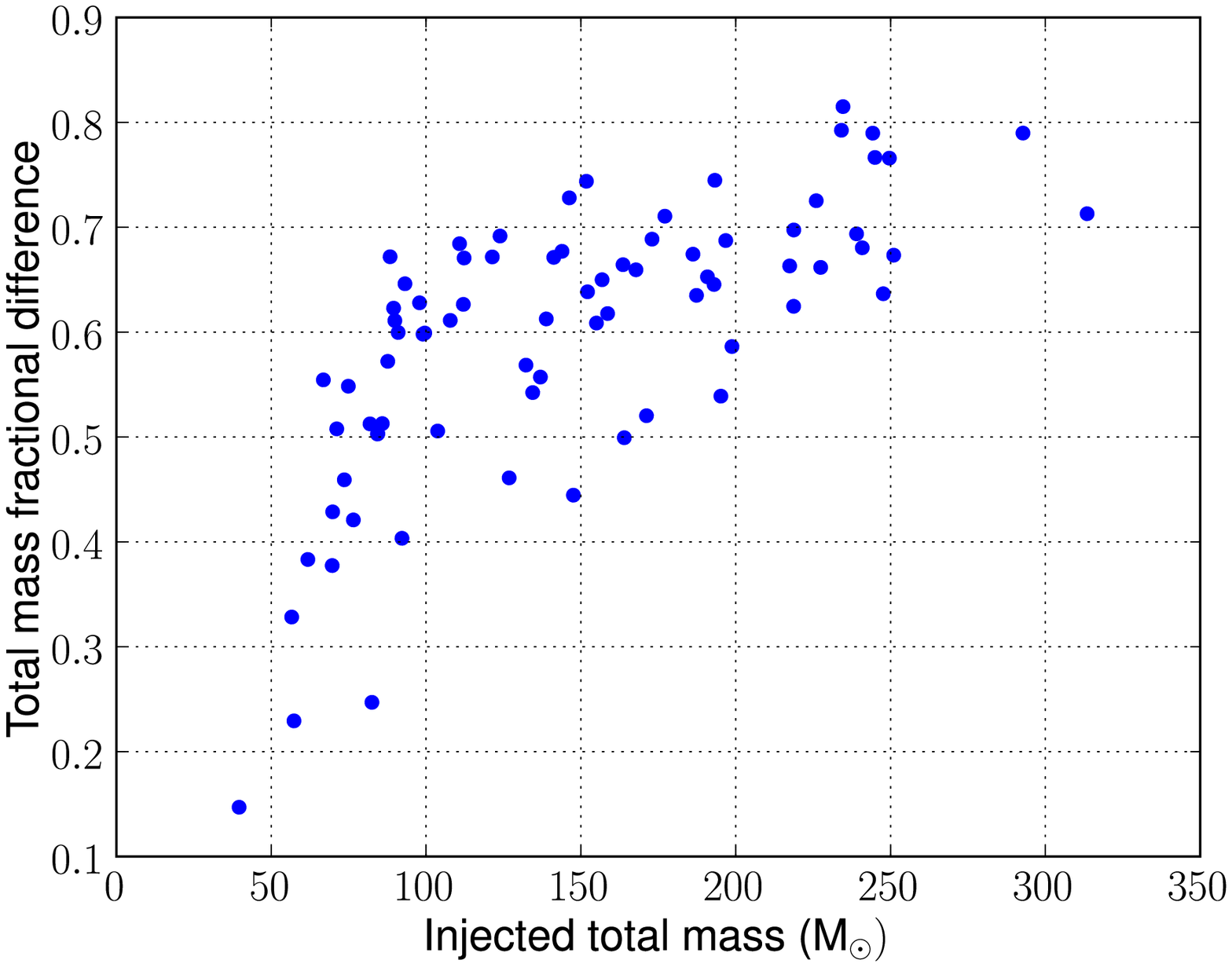}
    \includegraphics[width=0.50\textwidth]{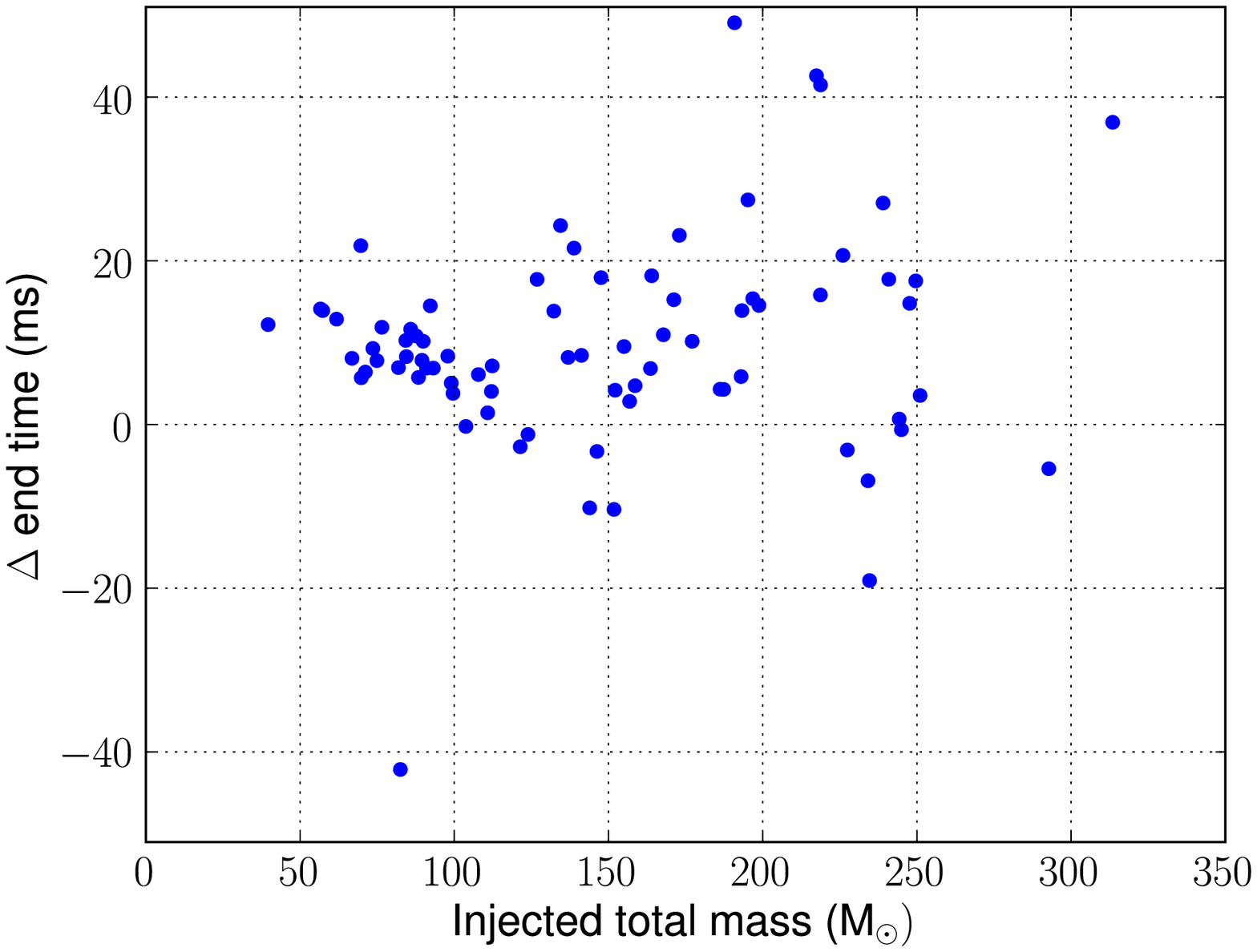}
\caption{\textbf{Parameter accuracy using TaylorF2 templates terminated at
ERD}.\textbf{Left:} Accuracy with which the total mass is recovered. The
template bank covers the region $20 M_\odot \le M \le 90 M_\odot$, hence
the mass of injections with $M > 90 M_\odot$ are always underestimated.
Even within the region covered by the bank, the TaylorF2 templates
systematically underestimate the mass of the injected signals and the total
mass is recovered accurately only for a few injections.  The vast majority of
recoverd signals have an error of $40\%$ or greater. \textbf{Right:} Accuracy
of determining the coalescence time of the injections.  The end time is not
recovered accurately, the timing error can become as large as $50
\mathrm{ms}$, the limits of the injection window.  }
\label{fig:3_5pn_params}
\end{figure}

Figure~\ref{fig:3_5pn_params} shows the accuracy with which the total mass and
coalescence time of the binary are recovered when using the 3.5 post-Newtonian
order Taylor F2 templates. The total mass fraction difference is computed as
$(M_\mathrm{injected} - M_\mathrm{detected})/ M_\mathrm{injected}$. For lower
mass signals, the end time is recovered reasonably accurately, with accuracy
decreasing for the high mass systems. The total mass recovery is poor for the
majority of signals, with good parameter estimation for only a few of the
lowest mass simulations.

\subsubsection{Four-detector Inspiral Search}  
\label{sssec:neyman}
The inspiral analysis described in Section~\ref{sssec:insp} considered 
data from the three simulated LIGO detectors. We now
extend the analysis to
include data from the simulated Virgo detector.  
In addition, we impose an alternative
criterion, based on the Neyman-Pearson formalism~\cite{Helstrom:1968},
to determine those injections which were detected by the pipeline.  In
the previous section an injection was classified as found by the search
if gravitational-wave candidate existed within 50~ms of the
peak time of the numerical data.  Here, we consider a signal to
be found is there is an associated candidate whose significance exceeds
a pre-determined threshold.  Specifically, we require the candidate
to have a significance greater than \textit{any} candidate arising due
to noise alone.  This allows us to probe in more detail the effect
of signal-based vetoes and the efficaciousness of the effective SNR 
statistic in analysis of the NINJA data. 

Data from all four simulated NINJA detectors was 
analysed using the CBC pipeline as described in
column~$(1)$ of Table~\ref{tab:inspiral_results}.  
In addition, a second analysis was performed with that the
template bank extended to cover the region from $2 M_\odot \le M \le 100
M_\odot$, with all other parameters unchanged.  The search can therefore
be though of as the simplest extension of the standard LSC-Virgo ``low
mass'' CBC search~\cite{Abbott:2009tt}. The boundary of the template
bank used is shown in Figure~\ref{f:ninjaBanks}. 

In this analysis, we choose a detection statistic and claim that a
gravitational-wave candidate is present if the value of this
statistic exceeds a pre-determined threshold. All candidates are considered
detections. The threshold is chosen so
that the false alarm probability---the probability that a noise event
will be mistaken for a real signal---is tolerable.  The efficiency of
this method depends on how close the chosen statistic is to the optimal
detection statistic. It is well known that the matched filter SNR is the
optimal statistic for known signals in a single detector \emph{if the
noise is stationary}~\cite{Wainstein:1962,Helstrom:1968}.  For a network
of detectors containing stationary noise, the optimal statistic is the
coherent signal-to-noise ratio
$\rho_\mathrm{coherent}$~\cite{Bose:1999pj}. At the time of this
analysis, calculation of $\rho_\mathrm{coherent}$ was not available in
the CBC pipeline, so we instead compute a combined SNR from the $i$
detectors, $\rho_\mathrm{c} = ({\sum_{i=1}^N \rho_{i}^2})^{1/2}$, as a
simple alternative.  In the presence of non-Gaussian noise, the
effective SNR, described in Section~\ref{ssec:modeled}, has shown to be
an effective detection statistic~\cite{Abbott:2007xi}. In this analysis,
we also consider the combined effective SNR $\rho_{\mathrm{eff}} =
(\sum_{i=1}^N \rho_{\mathrm{eff}\ i}^2)^{1/2}$.

We investigate three choices of detection statistic: (i) the combined
matched filter SNR of coincident candidates before signal-based
vetoes are applied 
($\rho_\mathrm{c}^\mathrm{first}$), (ii) the combined matched
filter signal-to-noise ratio \emph{after} the $\chi^2$ signal-based veto has
been applied applied to coincidences
($\rho_\mathrm{c}^\mathrm{second}$), (iii) the combined effective SNR
($\rho_\mathrm{eff}^{\mathrm{second}}$). This statistic is only available after
the second coincidence stage, since it is a function of matched filter SNR and
the $\chi^2$ statistic for a candidate.  To set a threshold for each statistic
we
choose the highest value of that statistic NINJA data containing only noise.
To do this, we discard all triggers within $5$~s of an injected signal;
the remaining triggers will be due to the simulated noise alone (we note that
this approach is not possible in real data where the locations of the signals
are unknown). This crude method of background estimation should
provide us with consistent criteria for elimination of spurious
detections. Therefore, we mark an injection as found only if it resulted
in a trigger with statistic higher then any background trigger found in
the data.  

\begin{table}
\begin{center}
\begin{tabular}{|l|c|c|c|c|}
\hline
Bank mass range& \multicolumn{2}{c|}{$2M_\odot\le M\le 35 M_\odot$} & \multicolumn{2}{c|}{$2M_\odot\le M\le 100 M_\odot$} \\
\hline 
Statistic & Statistic  & Found & 
Statistic & Found \\
 & Threshold& Injections& 
Threshold& Injections \\
\hline
$\rho_\mathrm{c}^\mathrm{first}$  & 9.18 & 73  & 9.8 & 91 \\
$\rho_\mathrm{c}^\mathrm{second}$ & 9.18  & 69 & 9.8 & 93\\
$\rho_\mathrm{eff}^\mathrm{second}$  & 10.05 & 27  & 10.05 & 85 \\
 \hline
\end{tabular}
\end{center}
\caption{{\bf Number of injections found as determined by the Neyman-Pearson
criteria} for different choices of detection statistic $\Lambda$ and
threshold $\Lambda^\ast$. The mass range of the template bank is shown in the
first row, all other parameters of the search as the same as those described
in column $(1)$ of Table~\ref{tab:inspiral_results}.}
\label{tab:threshold_found}
\end{table}

Table~\ref{tab:threshold_found} shows the threshold and the number of triggers
found for each choice of statistic. It is interesting to compare the results
for the low-mass search when we threshold on
$\rho_\mathrm{c}^\mathrm{second}$, rather than using a $50$~ms time window to
determine detected signals. When using the time-window method, the number of
injections found by the low-mass search is $51$, but this increase to $69$
when using the the threshold method. Since all the injected signals lie
outside the boundary of the low-mass bank, the coalescence time of the signals
will be poorly estimated. This will result in triggers outside the $50$~ms
window, which are nevertheless are loud enough to lie above the background.

Signal-based vetoes are applied at the second stage of the inspiral pipeline
and are used to compute $\rho_\mathrm{eff}$. By comparing the number of triggers
found before and after signal-based vetoes are applied, we can evaluate their
effect on the sensitivity of the search. Note that we observe the same
threshold for both $\rho_\mathrm{c}^\mathrm{first}$ and
$\rho_\mathrm{c}^\mathrm{second}$. However, the number of detected signals in
the low-mass search is reduced by 4 as the the $\chi^2$ veto has removed
triggers where the templates are not a good match for the signals.  More
intriguing is a slight increase in the number of detected signals after the
$\chi^2$ veto in the bank with the extended mass range (from $91$ to $93$). 
Additional investigations revealed that, despite having fewer triggers in each
detector after the $\chi^2$ test has been applied, the total number of
coincident triggers actually increases. This is due to the fact that the
signal-based vetoes cause the time of the signal to be measured more accurately
in the detectors; more triggers therefore survive the coincidence test.
We do not observe this in the case of the low mass search.  

Finally, we turn our attention to the effective SNR statistic, defined
in equation~(\ref{eq:eff_snr}).  Since the NINJA detector noise is stationary and
Gaussian, the expected value of the $\chi^{2}$ is one per degree of
freedom.  Therefore, we do not expect that the effective SNR will be
useful in reducing the significance of loud background triggers.  This
is borne out by the fact that the statistic threshold actually increases
slightly when using effective SNR.  For the low mass
search the number of signals found by thresholding on $\rho_\mathrm{eff}$ is
significantly less than when using the combined SNR statistic.  This is to be
expected as the simulated signals do not match well with the templates.
Although the low mass templates produce candidates, these will have
large values of $\chi^{2}$ since signal and template do not match well.
Thus, the effective SNR will be smaller than the original SNR and fewer
signals will be recovered above the threshold.  This effect is less
significant for the second search with a larger mass range as the
templates provide a better match to the simulated signals. Since effective SNR
has been a powerful statistic in real detector data, this highlights the need
for further NINJA studies using data containing non-stationary noise
transients.

\subsubsection{Inspiral-Merger-Ringdown Templates}
\label{sssec:imr}
The calculation of the full binary binary black hole coalescence waveform
accessible to ground-based detectors requires numerical methods. At the moment, it
is not possible to accurately model a coalescing binary over hundreds of
orbits due to the computational cost of evolutions.  Furthermore, it is not
necessary to model the entire waveform, since post-Newtonian gives a valid
description of the system when the black holes are sufficiently separated.
During their final orbits before merger the black holes' velocities increase
and the post-Newtonian expansion becomes less reliable. At this stage the
non-perturbative information contained in numerical simulations is required.
A successful approach has been to combine analytical and numerical results
to obtain full waveform templates. Two different families of such waveforms have
been used to analyse the NINJA data: the effective one body (EOB)
\cite{Buonanno:1998gg,Buonanno:2000ef,Damour:1997ub,Damour:2000we} and phenomenological
\cite{Ajith:2007kx,Ajith:2007qp} models.    

By combining together results from post-Newtonian theory and
perturbation theory, the EOB model~\cite{Buonanno:1998gg,Buonanno:2000ef} predicts the
full inspiral, merger and ringdown waveform. More recently, the non-spinning
EOB model has been further improved by calibrating it to NR results,
achieving high overlaps without the need to maximise the intrinsic
mass parameters of the binary~\cite{Buonanno:2006ui,Pan:2007nw,Buonanno:2007pf,Damour:2007yf,%
Damour:2007vq,Damour:2008te,Boyle:2008ge}. 
The LSC Algorithm Library (LAL)~\cite{lal} contains two implementations of the 
effective one body template: one (called EOB) which only evolves the waveform to the light-ring frequency
\begin{equation}
f_\mathrm{LR}=\frac{c^3}{3\sqrt{3}\pi GM},
\end{equation}
and a second (called EOBNR) which implements the full 
EOB waveform described in~\cite{Buonanno:2007pf}. This template
which was constructed
to match the NASA-Goddard binary black hole simulations with 
mass ratios $m_1$:$m_2$~$=$~1:1, 3:2, 2:1 and 4:1,
however LAL waveforms do not yet implement higher harmonics of the signal.
Both of these implementations were used to search for
black hole binary signals in NINJA data. 

Another approach for constructing the full waveform is to ``stitch''
together the results of post-Newtonian and numerical relativity calculations.  The model presented in
\cite{Ajith:2007kx,Ajith:2007qp,Ajith:2007xh} consists of matching the
post-Newtonian and numerical
waveforms in an appropriate matching regime (where both are
sufficiently accurate) to obtain a ``hybrid'' waveform. This hybrid is then fit
by a phenomenological model in the frequency
domain determined entirely by
the physical parameters of the system. This procedure has been carried
out for non-spinning black holes and a two-dimensional template family
of waveforms that attempts to model the inspiral, merger and ringdown
stages for non-spinning binary black holes has been obtained. Each
waveform is parametrised by the physical parameters of the system, i.e.,
the masses $m_1$ and $m_2$ of the black holes.  

Since the EOBNR and phenomenological models provide complete waveforms,
the search was performed to higher masses ($200 M_{\odot}$ and $160
M_{\odot}$ respectively) than for inspiral only searches.  In principle,
the search could be extended to even higher masses, but technical issues
with the current waveform generation procedures prevent this.  The
minimum component mass was also increased, in an effort to reduce the
size of the template bank by limiting the number of highly asymmetric
signals.  Finally, the template bank for all these searches was
constructed using the standard second order post-Newtonian metric, and
hexagonal placement algorithm~\cite{Cokelaer:2007kx}.  At high masses,
the parameter space metric for the full waveforms will differ
significantly from the standard second-order post-Newtonian metric.
However, the current template bank placement suffices for detection
purposes, although probably not for good parameter estimation.

\begin{table}
\small
\begin{tabular}{| l || c | c | c |}
\hline
\bf{Template} & EOB & EOBNR & Phenom \\ 
\hline
\bf{Freq. Cutoff} & Light ring & Full waveform & Full waveform\\ 
\hline
\bf{Filter Start Freq.} & 40 Hz & 30 Hz & 30 Hz \\
\hline
\bf{Component Mass $M_{\odot}$} & 10-60 & 15-160 & 20-80 \\ 
\hline
\bf{Total Mass $M_{\odot}$} & 20-90 & 30-200 & 40-160 \\ 
\hline
\bf{Minimal Match} & 0.97 & 0.99 & 0.99 \\ 
\hline
\bf{Found Single (H1, H2, L1, V1)} & 91, 64, 82, - & 97, 68, 92, 102 & 
92, 61, 87, - \\ 
\hline
\bf{Found Coincidence (LIGO, LV) } & 83, - & 88, 106 & 81, - \\ 
\hline
\bf{Found Second Coincidence (LIGO, LV)} & 80, - & 85, 102 & 80, - \\ 
\hline 
\end{tabular}
\caption{{\bf Results of the search for NINJA signals using IMR template
banks.} There were 126 injections performed into the analysed data.  The
signal-based vetoes have little influence in the rejection of triggers,
confirming their efficiency in separating inspiral-like signals from
other kind of glitches.}
\label{tab:imr_results} 
\end{table}

The parameters of the NINJA analyses using the EOB,
EOBNR and phenomenological waveforms are also given in Table
\ref{tab:imr_results}. Again, the primary result is the number of
gravitational-wave candidates found to be coincident with an injected signal.
For the EOB model truncated at light ring, the
parameters were chosen to match the TaylorF2 analyses described in
Section~\ref{sssec:insp}.  Therefore, it is unsurprising that the results
are very similar to the TaylorF2 search extended to ERD (the fourth
column of Table \ref{tab:inspiral_results}).  Further details of the EOB search, and a
comparison to TaylorF2 results are available in  The EOBNR results show
some improvement for detecting the numerical relativity signals over the
usual post-Newtonian or EOB waveforms.  For the phenomenological
waveforms, time windows of $120\mathrm{ms}$ in single detector and
$80\mathrm{ms}$ in coincidence have been used to associate triggers to
injections.  These parameters differ from those employed in other
searches to compensate for a relatively large observed error in the
estimation of the coalescence time.   By comparing the results with the
standard post-Newtonian analyses presented in Section \ref{sssec:insp}, we conclude
that in the present case the phenomenological
waveforms~\cite{Ajith:2007kx,Ajith:2007qp} do not seem to provide a
clear benefit
over the usual post-Newtonian waveforms extended to higher cutoff
frequency and/or to unphysical regions of the parameter
space~\cite{Pan:2007nw,Boyle:2009dg}. For an extended description of
the search with phenomenological waveforms
see~\cite{Santamaria:2009}. In all cases, the signal-based
vetoes have little influence in the rejection of triggers, confirming
their efficiency in separating inspiral-like signals from other kind of
glitches.

\begin{figure}
\subfigure[Search with EOBNR templates.]
{
\label{fig:EOBNRfoundMissed}
\includegraphics[width=0.5\textwidth]{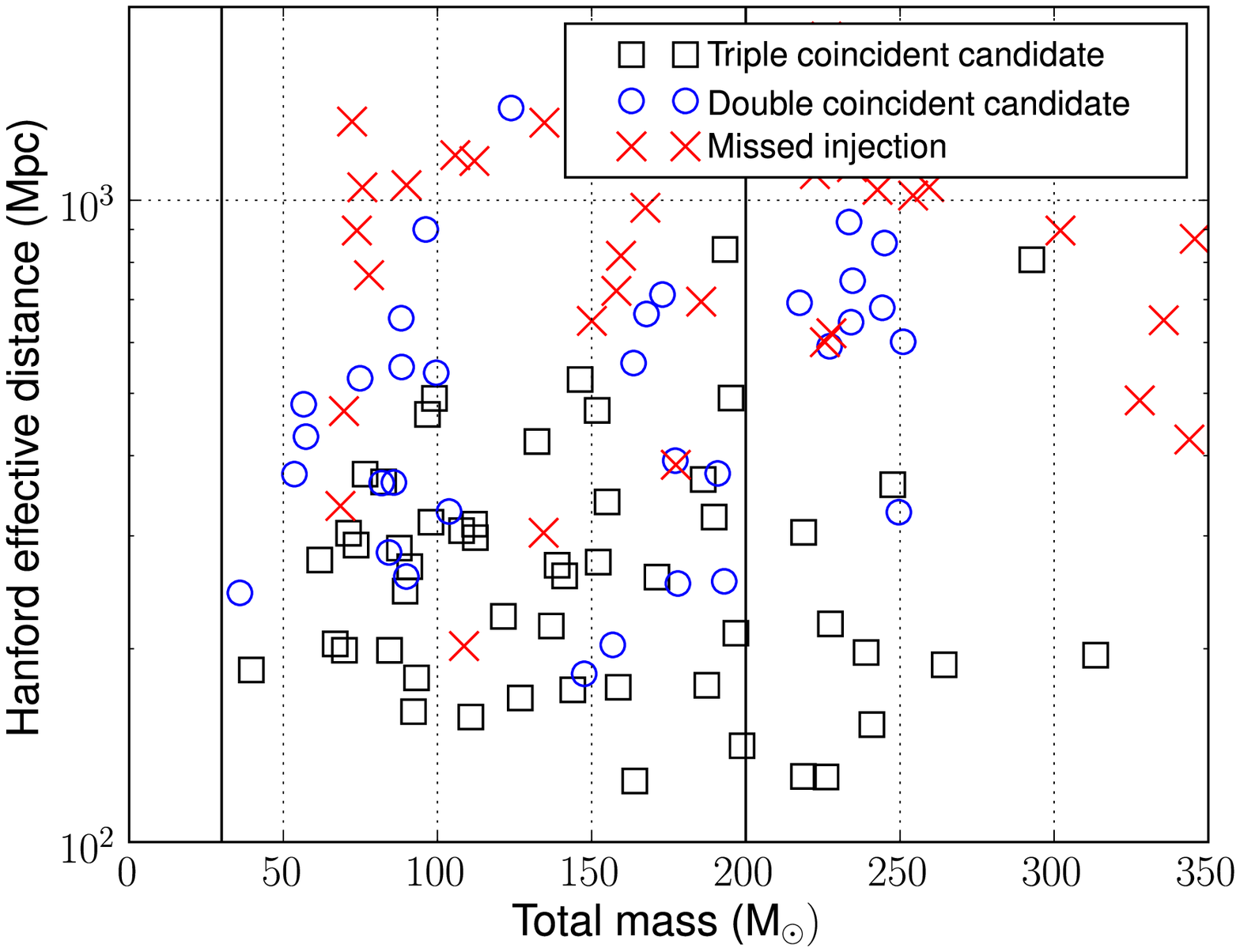}
}
\subfigure[Search with phenomenological templates.]
{
\label{fig:PhenomFoundMissed}
\includegraphics[width=0.5\textwidth]{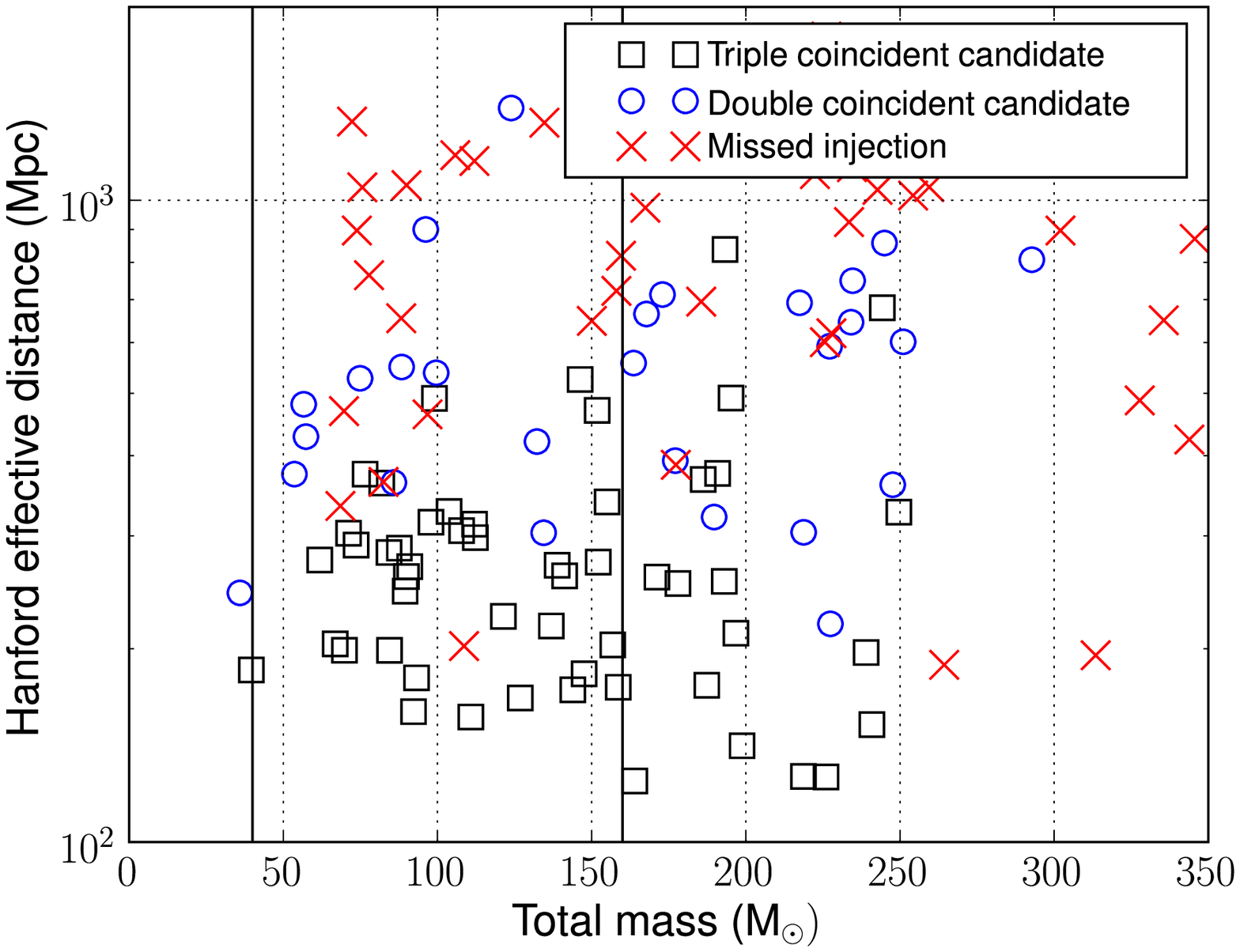}
}
\caption{\textbf{Found and missed injections for the EOBNR and
Phenomenological templates.}  The figures shows found and missed
injections as a function of the injected effective distance in Hanford
and the total mass.  \textbf{Left:} Results for the EOBNR search.
\textbf{Right:} Results for the search with phenomenological waveforms.
The vertical bars mark the limits of the template bank used in the
search.}
\label{fig:IMRfoundMissed}
\end{figure}

\begin{figure}
\includegraphics[width=0.5\textwidth]{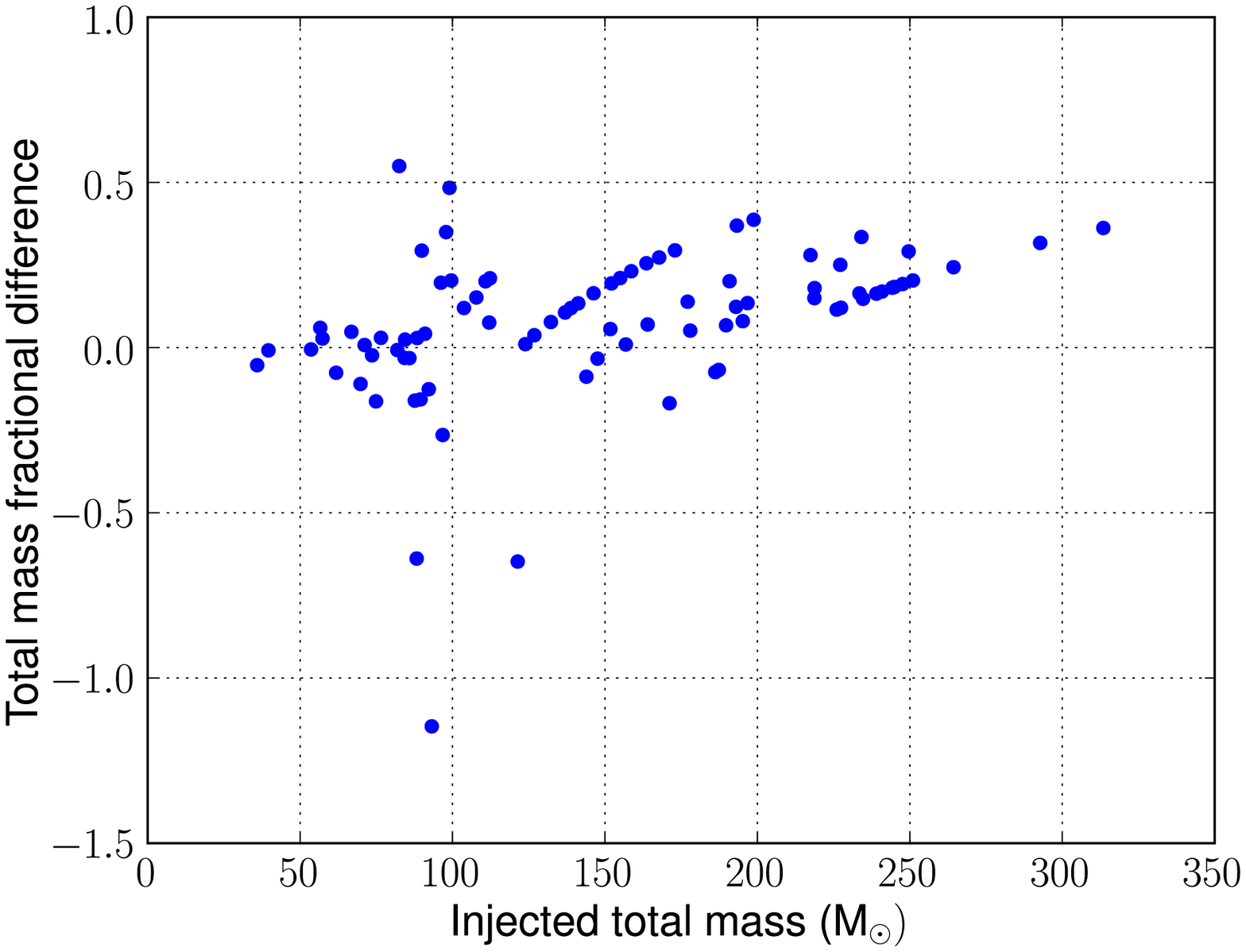}
\includegraphics[width=0.5\textwidth]{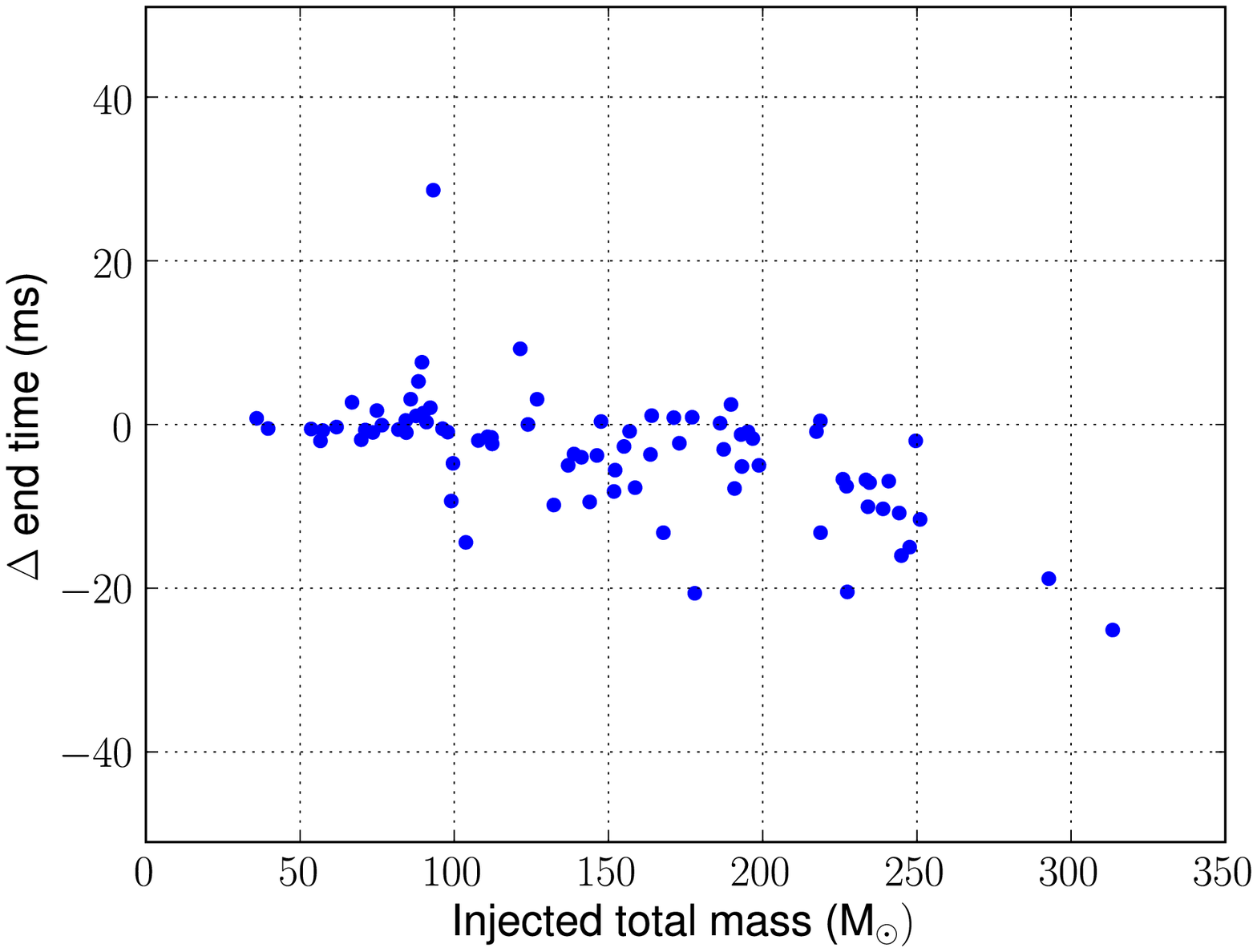}
\caption{\textbf{Parameter accuracy for EOBNR templates.} \textbf{Left:}
Accuracy with which the total mass is recovered. 
The
template bank covers the region $30 M_\odot \le M \le 200 M_\odot$, hence
the mass of injections with $M > 200 M_\odot$ are always underestimated.
Most of the injections
with total mass less than 200 $M_\odot$ were recovered with a mass
accurate within to a few tens of percent, demonstrating that the EOBNR
templates are more faithful to the injected signal than the TaylorF2 templates shown in
Figure~\ref{fig:3_5pn_params}.  Higher mass injections are
necessarily recovered with underestimated total mass, because the
template bank did not cover the entire simulation region.
\textbf{Right:} Accuracy of determining the coalescence time of the
injections.  The end time for injections with total mass less than 200
$M_\odot$ was typically recovered to within a few milliseconds.  The end
time for injections with total mass above 200 $M_\odot$ (outside the
range of the template bank) was typically recovered to within 10 or 20
milliseconds.}  
\label {fig:EOBNRParam}
\end{figure}

\begin{figure}
\includegraphics[width=0.5\textwidth]{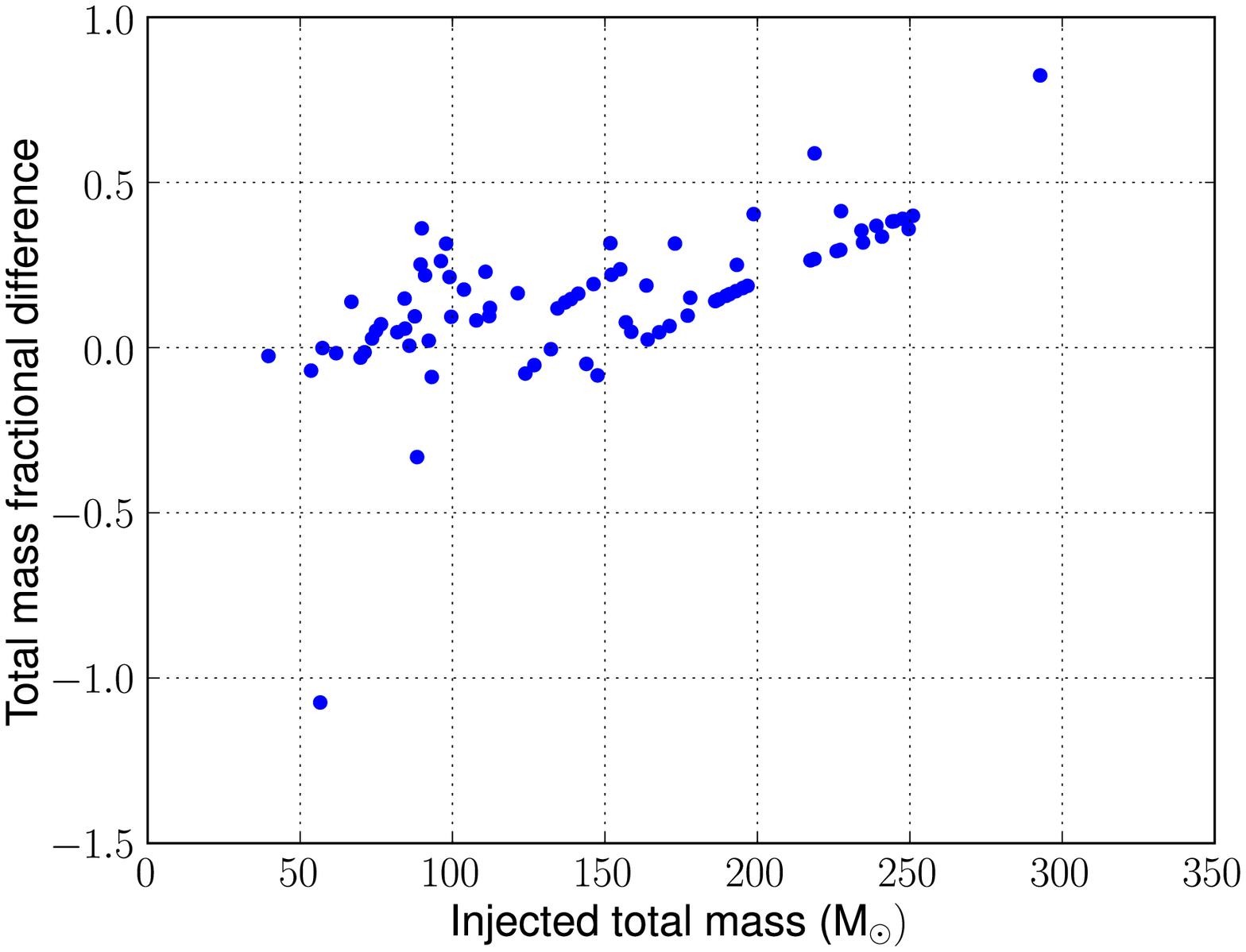}
\includegraphics[width=0.5\textwidth]{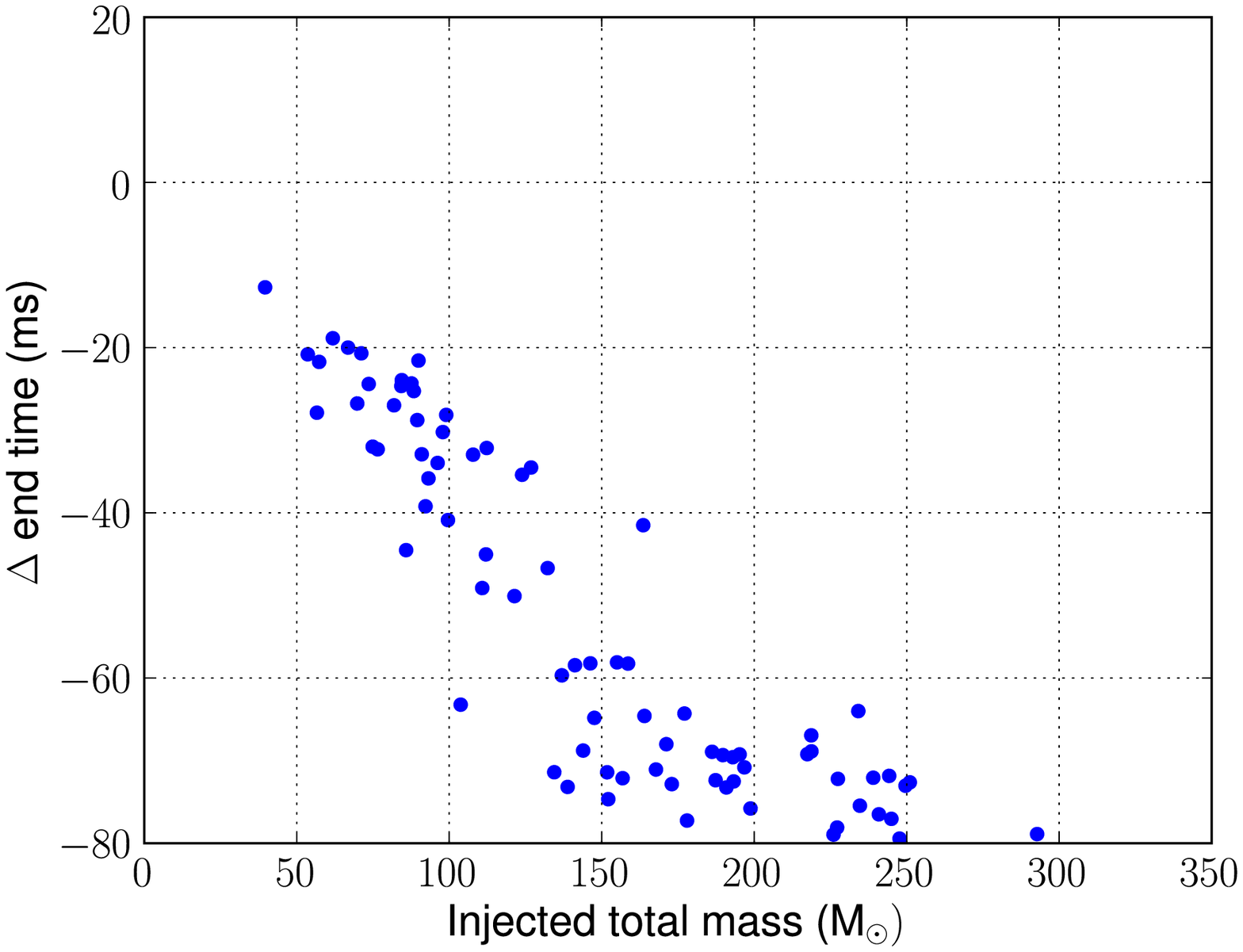}
\caption{\textbf{Parameter accuracy for phenomenological templates.}
\textbf{Left:} Accuracy with which the total mass is recovered. The
total mass is typically recovered within 20\%, for signals within the
template space.  For higher mass injections, there is an inevitable
underestimation of the mass due to the limited reach of the template
bank.  \textbf{Right:} Accuracy of determining the coalescence time of
the injections.  The timing plot shows the systematic offset discussed
in the text.}
\label{fig:PhenomParam} 
\end{figure}

Plots of found and missed injections for the searches are shown in
Figure \ref{fig:IMRfoundMissed}.  For the most part, simulated signals
in the mass range covered by the template banks are well recovered.
Some of the missed signals at lower distance correspond to waveforms
from simulations of spinning black holes.  Since all searches make use
of non-spinning waveforms this drop is expected.  
Finally, we turn to parameter estimation.  Figures \ref{fig:EOBNRParam}
and \ref{fig:PhenomParam} show the parameter recovery accuracies for the
EOBNR and phenomenological searches respectively.  In both cases, the
accuracy of recovering the total mass of the simulations is greatly
improved over TaylorF2 waveforms shown in Figure \ref{fig:3_5pn_params}.
This is likely related to the increased mass range of the searches, as
well as the use of full waveforms.  The timing accuracy for EOBNR is
comparable with the TaylorF2 results, while for the phenomenological
waveforms, the known timing bias affects the results.

Both the EOBNR and phenomenological models will be improved in the
future.  Further accurate EOBNR models have already appeared in the
literature~\cite{Buonanno:2007pf,Damour:2007yf,Damour:2007vq,%
Damour:2008te,Boyle:2008ge} since the time the EOBNR model used in this
analysis was implemented, and extensions to include spin and
eccentricity are under development.  There are a number of obvious
improvements in the phenomenological waveforms that can be made: 
Calculating the parameter space metric for the
phenomenological waveforms would enable the use of an optimal
template bank and allow for improved coincidence 
algorithms. The construction of the phenomenological waveform
model can itself be significantly improved by extending the fitting to
higher mass ratios and spins, quantifying the error on the
phenomenological parameters, matching to post-Newtonian theory as early
as possible and including higher order modes in the waveform.  The results of
the NINJA analysis also demonstrate a clear need to improve accuracy in
measuring the end time of the signal. This is not
straightforward, however, since there is no clear definition of the time of merger
for the phenomenological waveforms or the numerical signals~\cite{Hannam:2009hh}.  Work on the improvements to both
the EOBNR and phenomenological searches are being made, and will be
applied in and guided by future NINJA projects.

\subsubsection{Ringdown Templates}
\label{sssec:ring}
As described in section~\ref{ssec:modeled}, ringdown templates can be computed
using black hole perturbation theory and so matched filtering can be used to
search for these signals. Ringdown templates are exponentially damped
sinusoids parametrised by the ringdown frequency $f$ and quality factor $Q$.
The LSC ringdown search pipeline~\cite{Goggin:2008} has been used 
to filter the NINJA data against a bank of ringdown templates with frequencies
between 50 Hz and 2 kHz, and quality factors between 2 and 20.  The bank had a
maximum mismatch of 3\% and contained 583 templates.
A lower-frequency cutoff of 45 Hz was applied when filtering the NINJA data
generated with the LIGO noise curves and 35 Hz for data with the Virgo noise
curve.
The goals of these analyses were to ascertain the detectability of the
injected numerical waveforms using ringdown templates at single and coincident
detector levels and the accuracy with which the final black hole parameters
can be estimated. The current searches use single-mode templates.
The waveforms described in this paper are known to contain higher
order multipoles. The potential effects of ignoring these in the
search are discussed in Ref.~\cite{Berti:2007zu} (see in particular
Fig.~8 in there).

\begin{figure}
\begin{center}
\includegraphics[width=0.47\textwidth]{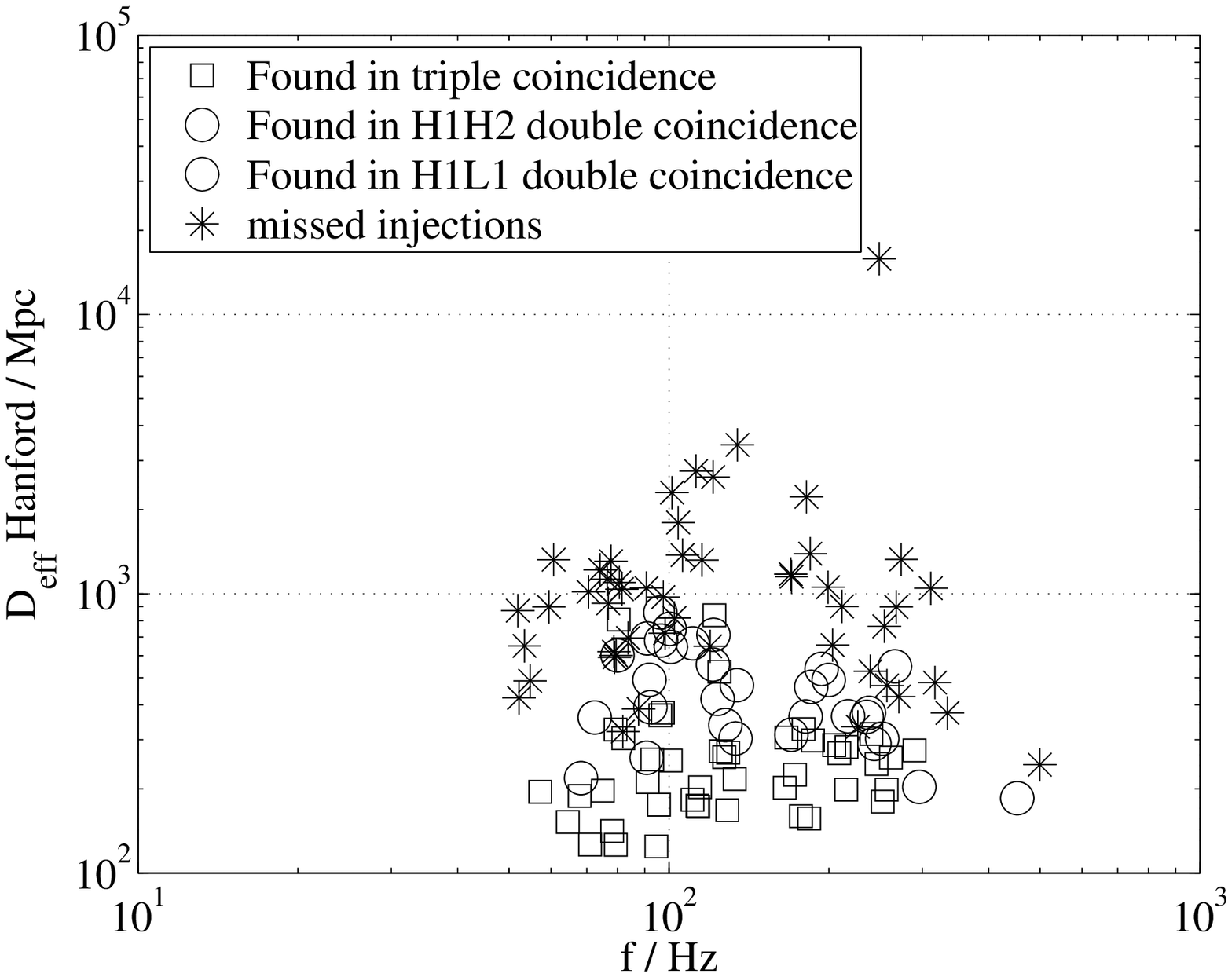}
\includegraphics[width=0.47\textwidth]{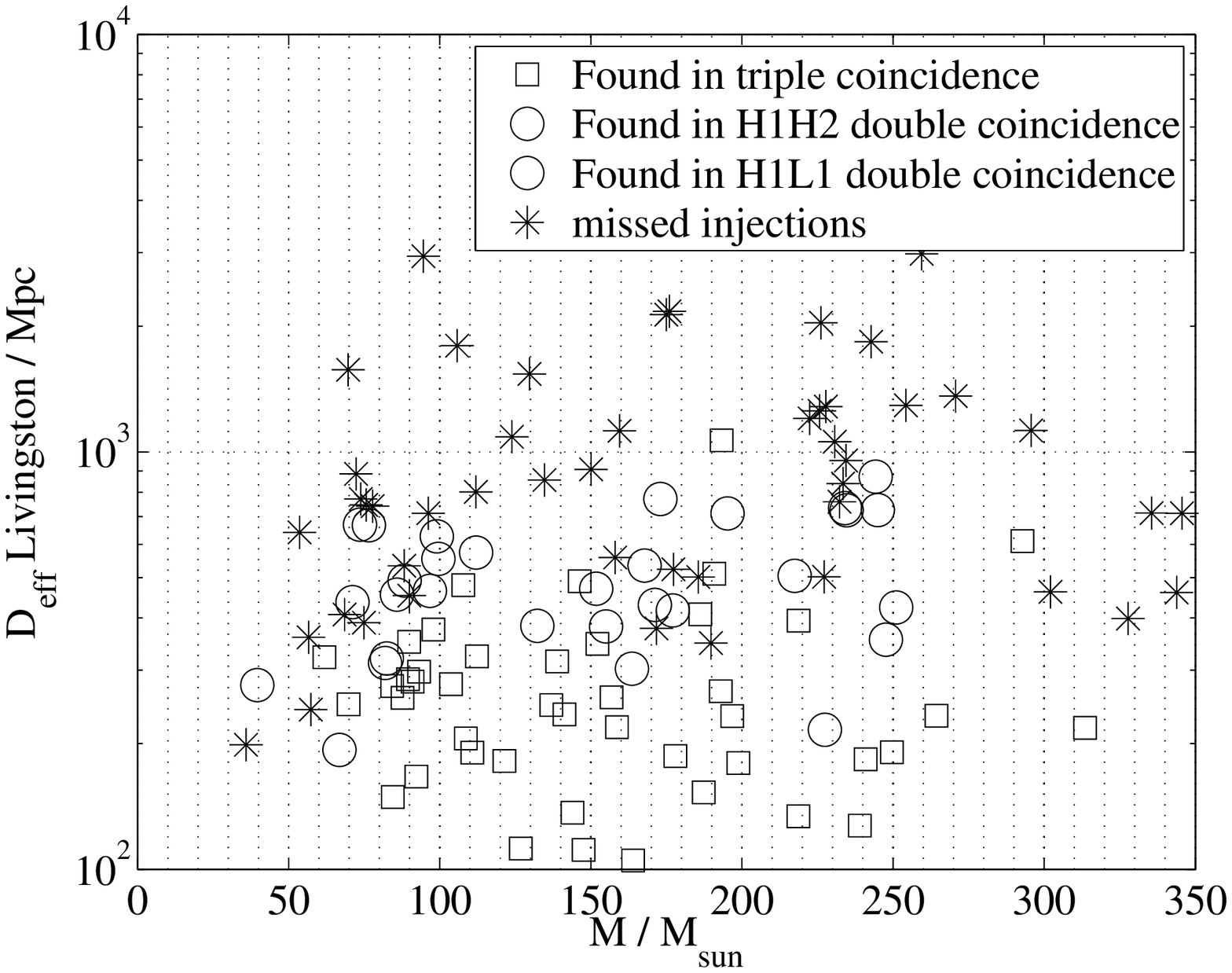}
\end{center}
\caption{{\bf Distribution of injections found and missed by the ringdown
pipeline.} The left figure shows the effective distance
of the injected signal in the LIGO Hanford
Observatory as a function of the predicted ringdown frequency. The right
figure shows the effective distance of the injected signal in the LIGO
Livingston Observatory 
as a function of the
total initial mass of the signal. The figures show signals found in triple
coincidence (blue crosses), in double coincidence in H1H2 (green stars), in
double coincidence in H1L1 (cyan stars), and missed (red circles). }
\label{fig:mfall}
\end{figure}

An injection is defined as found if a set of coincident triggers lies within
10~ms of the peak time of the injection (as specified in the contributed
numerical data). If more than one set of coincident triggers satisfies this criterion,
that with the largest value of $\sum_i \rho_i^2$ is selected, where $\rho_i$
is the signal to noise ratio in the $i^{\textrm{th}}$ detector. Of the 126
injections made into the three simulated LIGO detectors, $45$ were found in triple
coincidence, $24$ in H1 and L1 (only), and $7$ in H1 and H2 (only).
Figure~\ref{fig:mfall} shows the distribution of found and missed injections
for this analysis. The ringdown frequency and quality is computed via the 
Echeverria formulae~\cite{Echeverria:1989hg}:
\begin{eqnarray}
f&=&\frac{1}{2 \pi} \frac{c^3}{GM}\left[ 1-0.63\left( 1-a\right)
^{\frac{3}{10}}\right]  \label{eqn:Echeverria_fofMa} \\
Q&=&2\left( 1-a\right) ^{-\frac{9}{20}}.
\label{eqn:Echeverria_Qofa}
\end{eqnarray}
More recent and accurate
fits for a variety of modes are listed in the Appendices of
Ref.~\cite{Berti:2005ys}.
The final black hole mass $M$ and spin $a$ can be computed from the component
masses and spins of the numerical simulation, as described in \cite{Buonanno:2007pf} and 
\cite{Rezzolla:2007rz}, respectively. See also
Refs.~\cite{Berti:2007fi,Berti:2007dg} for a discussion and comparison
of different numerical techniques to perform the necessary fits.


As expected, we see that in general, the closest injections (measured by
effective distance $D_\mathrm{eff})$, defined in equation~\ref{eq:effdist}) were found in triple coincidence, those
with a large Livingston effective distance were found in H1 and H2 only, while
those with a large Hanford effective distance were not found in H2, and the
furthest injections were missed in at least two detectors. The plots show that
there are three missed injections which, given their frequencies and effective
distances, we would have expected to find. However, all three of these are
(non-spinning) injections with mass ratio of 4:1, and thus the energy emitted 
in the ringdown is less than would be emitted by a binary of the same total 
mass but with a mass ratio of 1 \cite{Berti:2007fi}.
This is not taken into account in the
calculation of effective distance. 

\begin{figure}
\begin{center}
\includegraphics[width=0.47\textwidth]{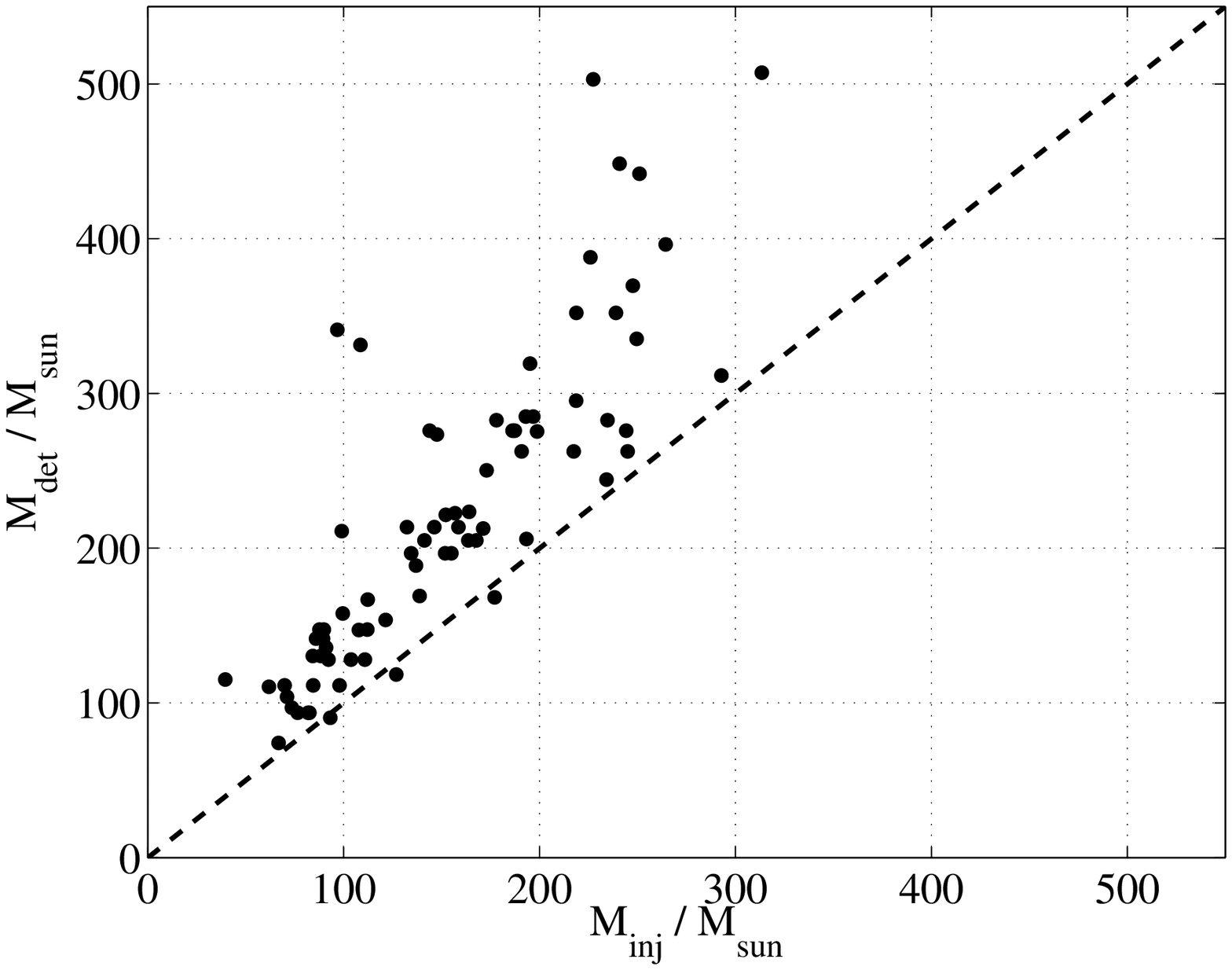}
\includegraphics[width=0.47\textwidth]{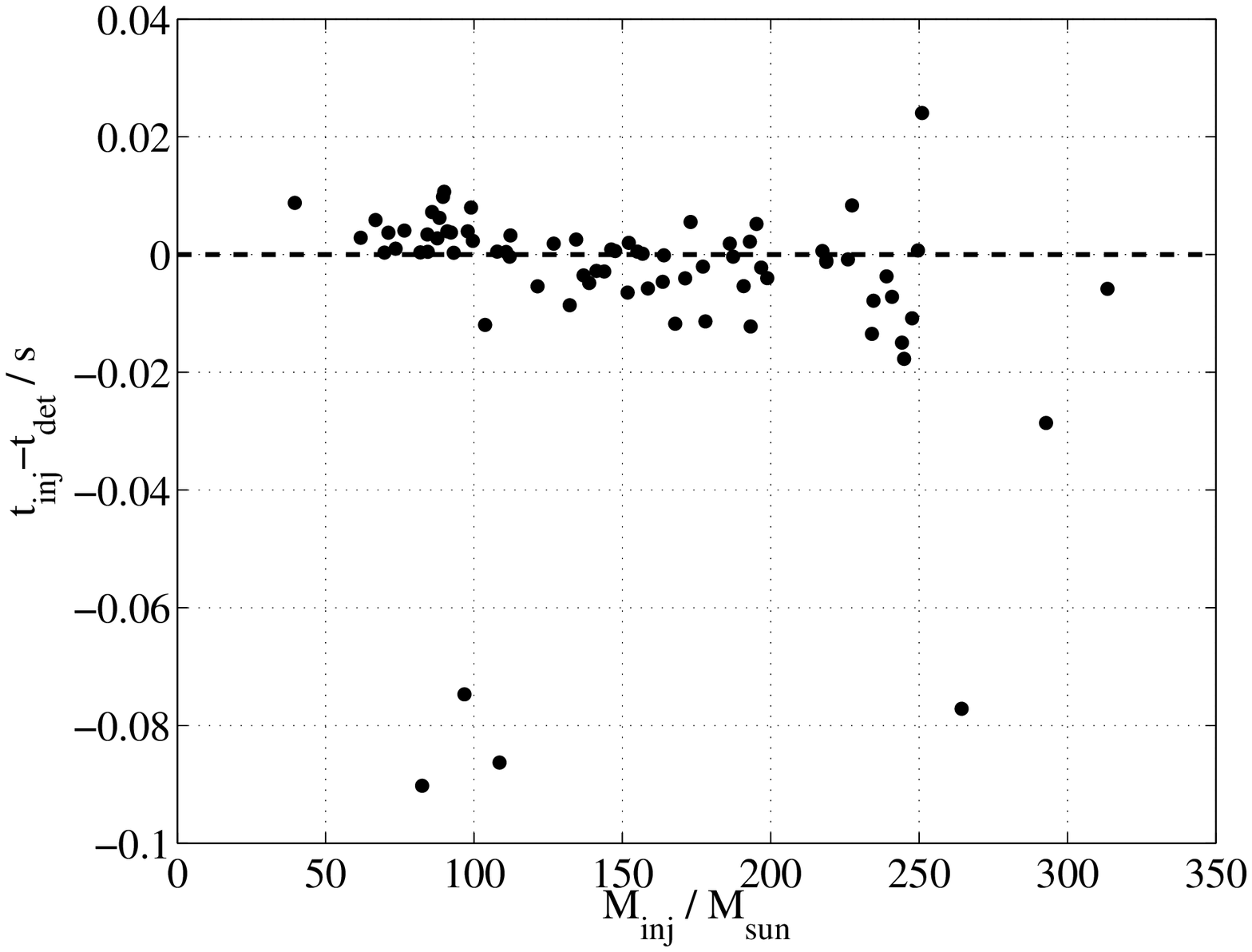}
\end{center}
\caption{{\bf Accuracy of measuring the ringdown parameters.} The left figure
shows the detected ringdown mass versus total injected mass for all found
injections. The right figure shows the difference between the time of injected waveform
peak amplitude and the start time of the ringdown as found by the search.}
\label{f:ringacc}
\end{figure}

Equations~(\ref{eqn:Echeverria_fofMa}) and (\ref{eqn:Echeverria_Qofa}) can be
inverted to calculate $M$ and $a$ from the template parameters $f$ and $Q$ of a given
gravitational-wave candidate. Figure~\ref{f:ringacc} shows the accuracy with
which the ringdown search measures the mass and peak time of the injected
signals. Given that mass is radiated during the ringdown phase (the exact
amount depends on the initial mass ratio) one would expect the measured mass
to underestimate the mass of the signal, and hence the data points would
lie below the diagonal.  However, the recovered frequency is systematically
underestimated due to the presence of the preceding inspiral, leading to an
overestimation of the mass. The peak time of the signal is measured with
similar accuracies to the coalescence time measured by the TaylorF2 templates
described in Section~\ref{sssec:insp}.  The three data points with a large time
difference and masses lying in the range 80 and 110 $M_\odot$ are part of the
\verb|PU_T52W| non-spinning, equal mass group where the peak amplitude
occurred early in the waveform (i.e~prior to the merger).  

\subsection{Search pipelines to detect un-modelled waveforms}
\label{ssec:unmodeled}
Several algorithms exist to detect gravitational wave transients with minimal
assumptions on their origin and waveform; these techniques are often referred
to as \emph{burst searches}. Burst searches do not use templates and
instead target excesses of power in the time-frequency plane. The LSC and
Virgo collaborations have developed several burst search algorithms
which use different transformations for the generation of time-frequency data
maps. The identification of coherent signatures across multiple detectors has
proven to be very effective at suppressing false alarms.

Since they do not assume a template and they target short transients burst
searches are suited for the detection of the merger phase of the coalescence.
They have the potential to probe a large parameter space, inclusive of spin and
ellipticity, at no additional computational cost. For this reason, the NINJA
data was analysed by two burst algorithms: Q-pipeline and HHT.

The Q-pipeline~\cite{Chatterji:2005,Chatterji:2004qg} is one of the algorithms
used in to search for burst sources in LIGO's fifth science run~\cite{LSCburstS5y1}.
It is a multi-resolution time-frequency search for statistically significant
excess signal energy, equivalent to a templated matched filter search for
sinusoidal Gaussians in whitened data. The template bank is constructed to
cover a finite region in central time, central frequency, and quality factor
such that the mismatch between any sinusoidal Gaussian in this signal space and
the nearest basis function does not exceed a maximum mismatch of 20\% in
energy. For the purpose of the NINJA analysis, and to explore detectability
and parameter estimation, the Q-pipeline analysis was focused on the detection
efficiency at the single detector, for all four detectors, using a nominal SNR
threshold comparable to the one used in the matched filter searches.

The Hilbert-Huang Transform (HHT)~\cite{Huang:1998,Camp:2007ee} is an adaptive
algorithm that decomposes the data into Intrinsic Mode Functions (IMFs), each
representing a unique locally monochromatic frequency scale of the data. The
original data is recovered by constructing a sum over all IMFs.  The Hilbert
transform as applied to each IMF unveils instantaneous frequencies and
amplitudes as a function of time, thus providing high time-frequency
resolution to detected signals without the usual time-frequency-uncertainty as
found in basis set methods like the Fourier transform. 

In this section we briefly describe how the algorithms were applied and
highlight their performance, while Section~\ref{ssec:comparison} compares the
performance of the burst searches to the matched filtering algorithms.

\subsubsection{Q-pipeline}
\label{ssec:q_pipeline}
The simulated LIGO and Virgo data streams were filtered by the Q-pipeline~\cite{Cadonati:2009jg} with
the same configuration used in the LSC S5 burst analysis~\cite{LSCburstS5y1}.
Data is processed in 64 s analysis blocks with frequency range 48--2048 Hz and
Q range 3.3--100.  The resulting triggers, once clustered, indicate a
time-frequency interval and a significance of the excess power in that
time-frequency tile. This significance can be easily converted into the
signal-to-noise ratio of a matched filter with sine-Gaussian templates.

\begin{figure}
   \begin{center}
   \includegraphics[width=0.49\linewidth]{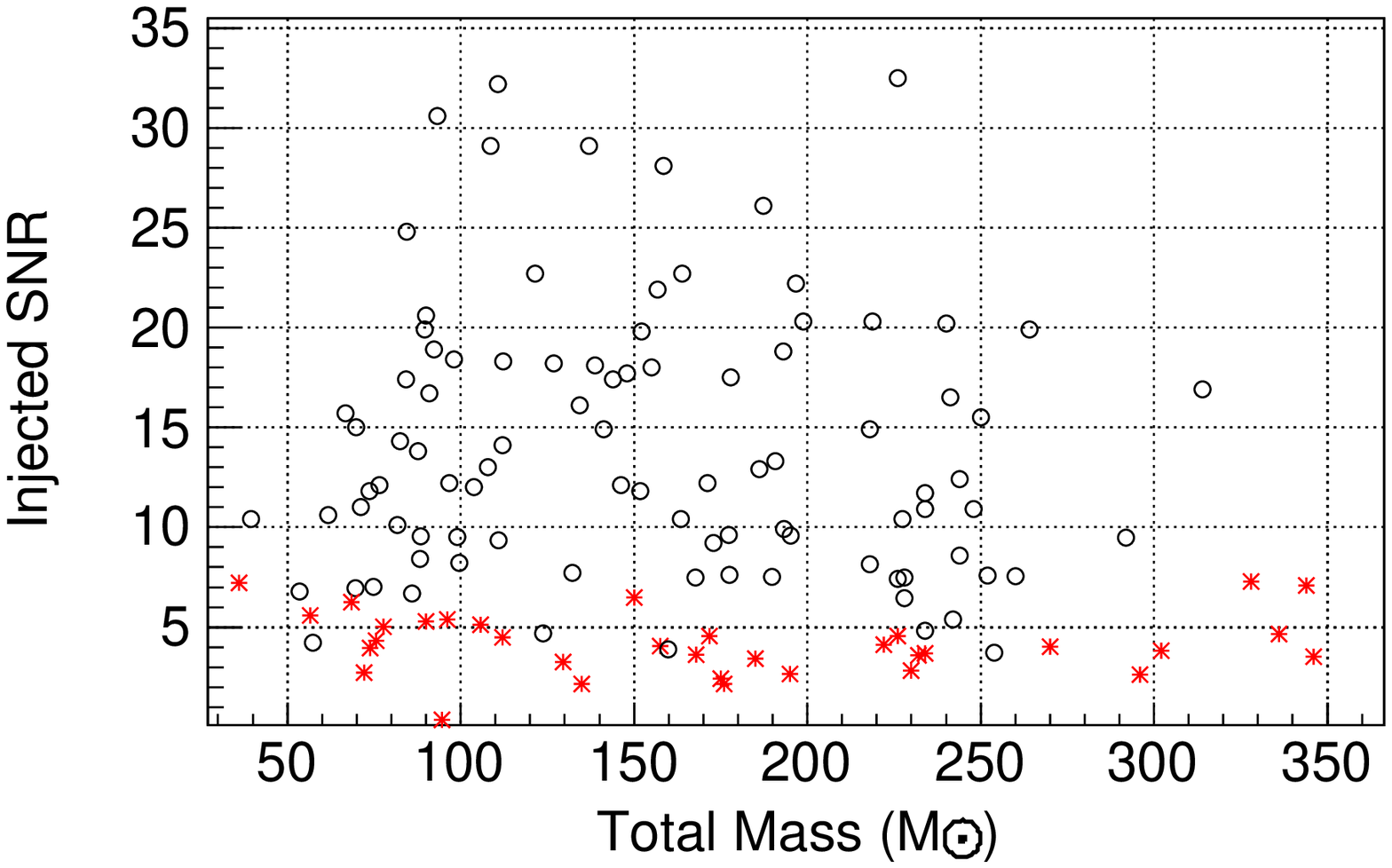}
   \includegraphics[width=0.49\linewidth]{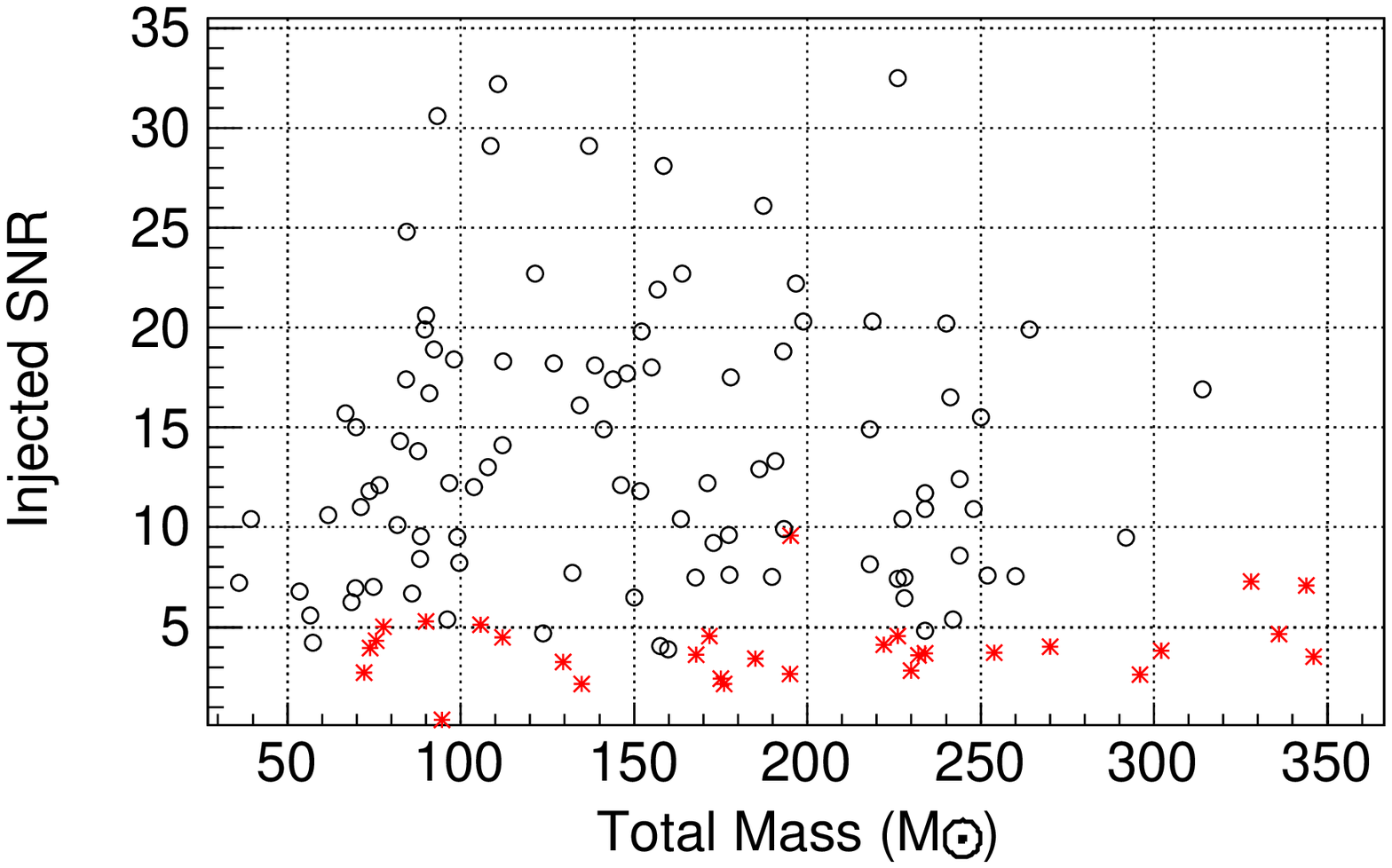}
   \end{center}
   \caption{{\bf Distribution of injections found by the Q-pipeline (left) in 
   the LIGO Hanford 4~km detector, to be compared to the distribution for the 
   EOBNR matched filtering search.} The total mass of the injected signal is 
   shown on th $x$-axis and the optimal matched filer signal-to-noise ratio of 
   the injection is shown on the $y$-axis. Circles show found injections and
   crosses show missed injections. At the single-detector level, with the same 
   SNR threshold, Q-pipeline and the EOBNR search have comparable performances. } 
   \label{f:Qresults}
\end{figure}

\begin{figure}
   \begin{center}
   \includegraphics[width=0.49\textwidth]{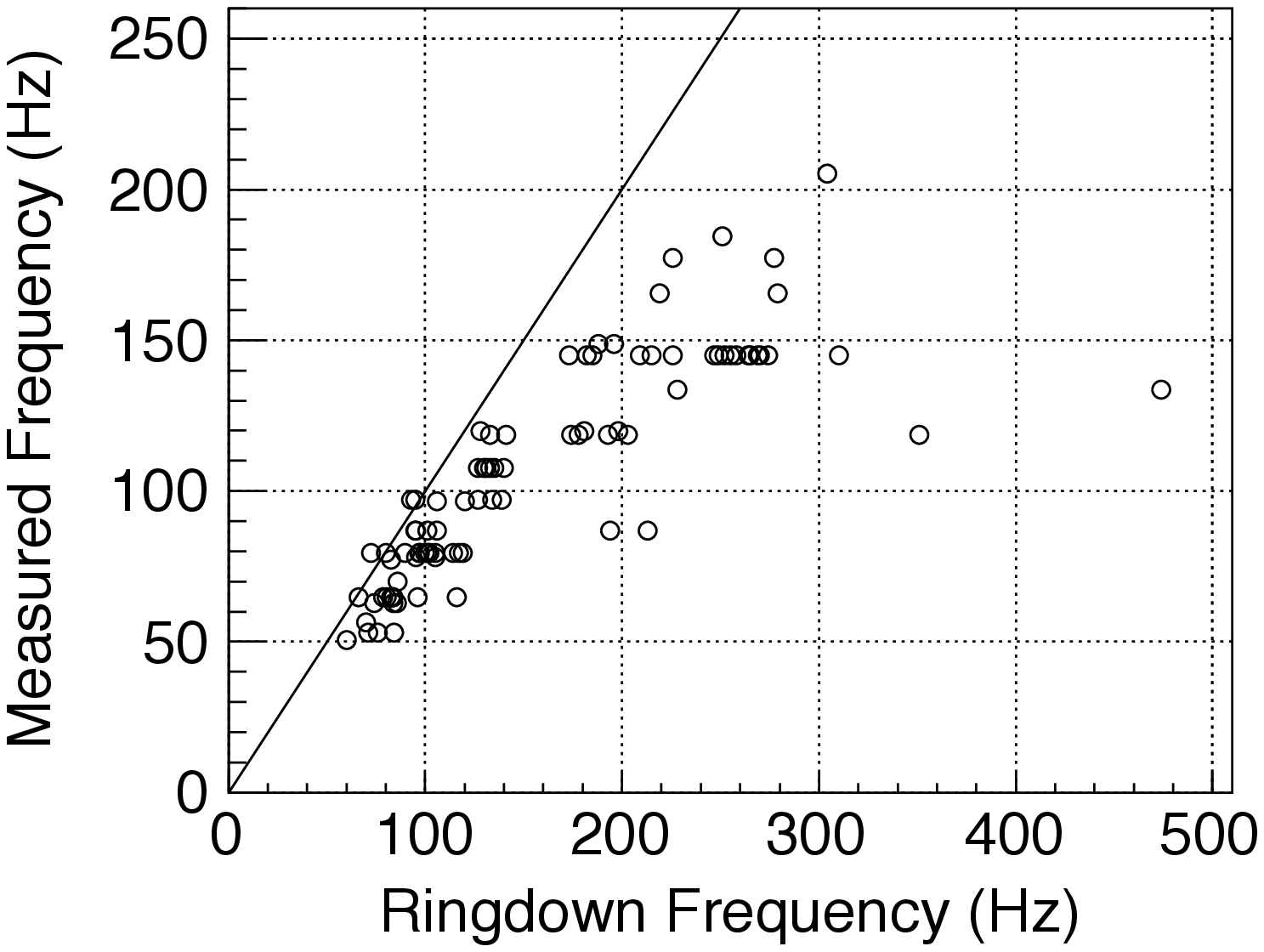}
   \includegraphics[width=0.49\textwidth]{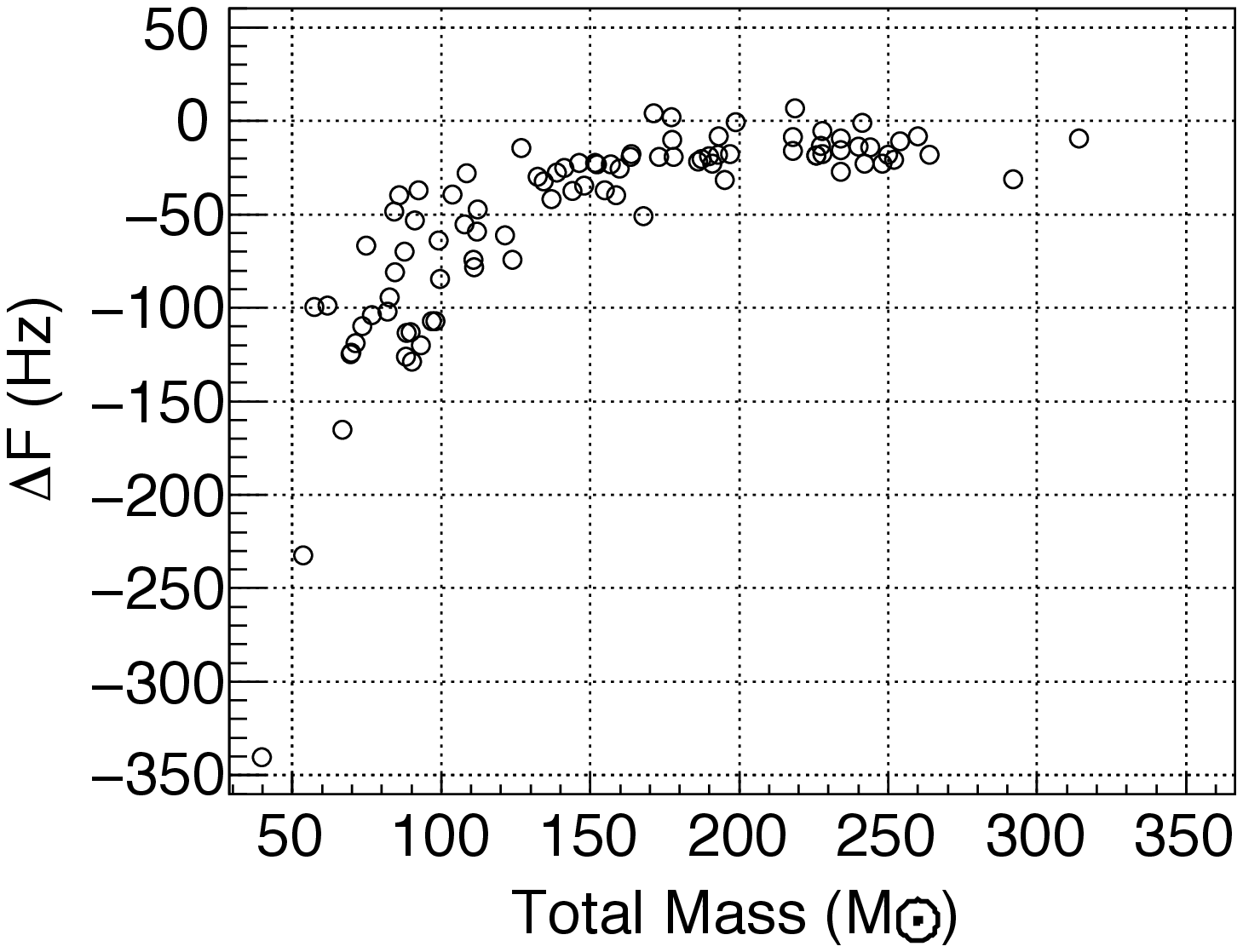}
   \end{center}
   \caption{{\bf Parameter measurement of signals using the Q-pipeline.} The
   left plot shows the measured frequency of the gravitational-wave candidate
   versus the injected ringdown frequency $f_\mathrm{ring}$ computed by
   equation~(\ref{eqn:Echeverria_fofMa}). The right plot shows the  difference
   between Q-pipeline measured frequency and $f_\mathrm{ring}$, against the
   total mass of the injection. These plots show the algorithm preferentially triggers on the portion of the signal that is in the most sensitive band of the detector (50-200 Hz). This is consistent with the behavior of the ringdown search; see figure~\ref{f:qringcomp} for an event-by-event comparison.}
   \label{f:Qring}
\end{figure}

Figure~\ref{f:Qresults} shows the distribution of found and missed injections
in the H1 detector as a function of the total mass and the matched filter SNR
of the injected waveforms for Q-pipeline (left) and the EOBNR matched filtering search described in section~\ref{sssec:imr}, where both detection thresholds are set at $\textrm{SNR}=5.5$. At the single detector level, the two algorithms have comparable performance. 
Figure~\ref{f:Qring} shows the the central frequency of
the most significant tile reported by Q-pipeline versus the ringdown frequency
computed from equation~(\ref{eqn:Echeverria_fofMa}) and the difference between
these two frequencies as a function of mass.  These results demonstrate that
the Q-pipeline preferentially detects the portion of the signal that is in the
most sensitive frequency band of the detector (50-200Hz): the ringdown for
higher masses, or the inspiral for lower masses.


\subsubsection{Hilbert-Huang Transform}
\label{sssec:hht}
An automatic, two-stage HHT pipeline was applied to the NINJA data to detect
and characterise the injected signals. Since the HHT
pipeline~\cite{Stroeer:2008} is a new development its application to NINJA
data was restricted to the simulated 4~km LIGO detectors H1 and L1.  The data
was pre-whitened and a $1000$~Hz low pass zero-phase finite impulse response
filter was applied prior to use of the HHT.  In the first \emph{detection
stage}, the instantaneous amplitudes from each detector in turn are divided
into blocks with similar statistical properties according to the Bayesian
Block algorithm~\cite{McNabb:2004bu}.  These blocks are then scanned for
excess power, with triggered blocks yielding start and end times, thus
coincidences between detectors, the maximum frequency, and the signal-to-noise
ratio (SNR) of the signal. The second \emph{characterisation stage} computes
the instantaneous frequencies, detailed time-frequency-power maps and
time-frequency-power cluster-enhanced maps for the region of data containing
the signal identified in the first stage.  The overlap of the individual
cluster-enhanced maps is used to estimate the time lag between the signals in
the detectors and to construct a coherent addition of the two detector data
streams used in a final characterisation of the signal. 

\begin{figure}
\begin{center}
\includegraphics[width=0.48\linewidth]{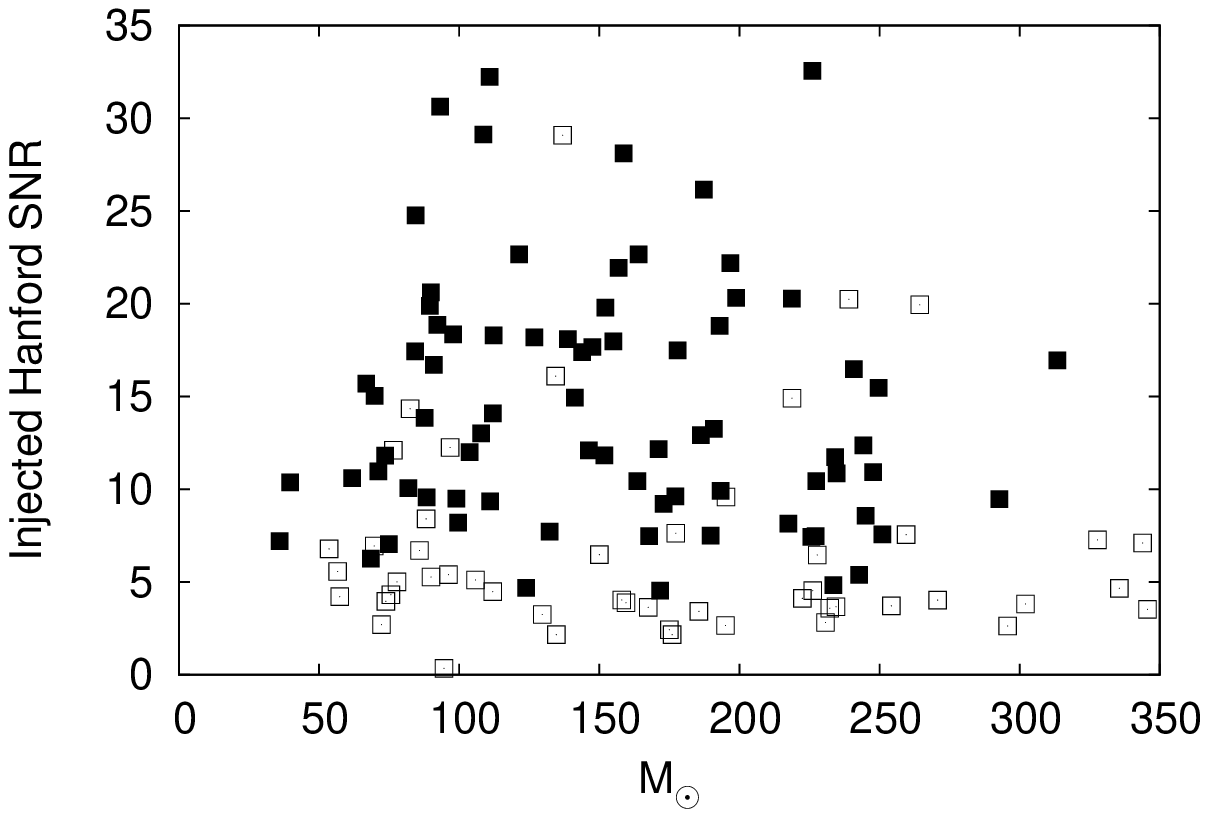}
\includegraphics[width=0.48\linewidth]{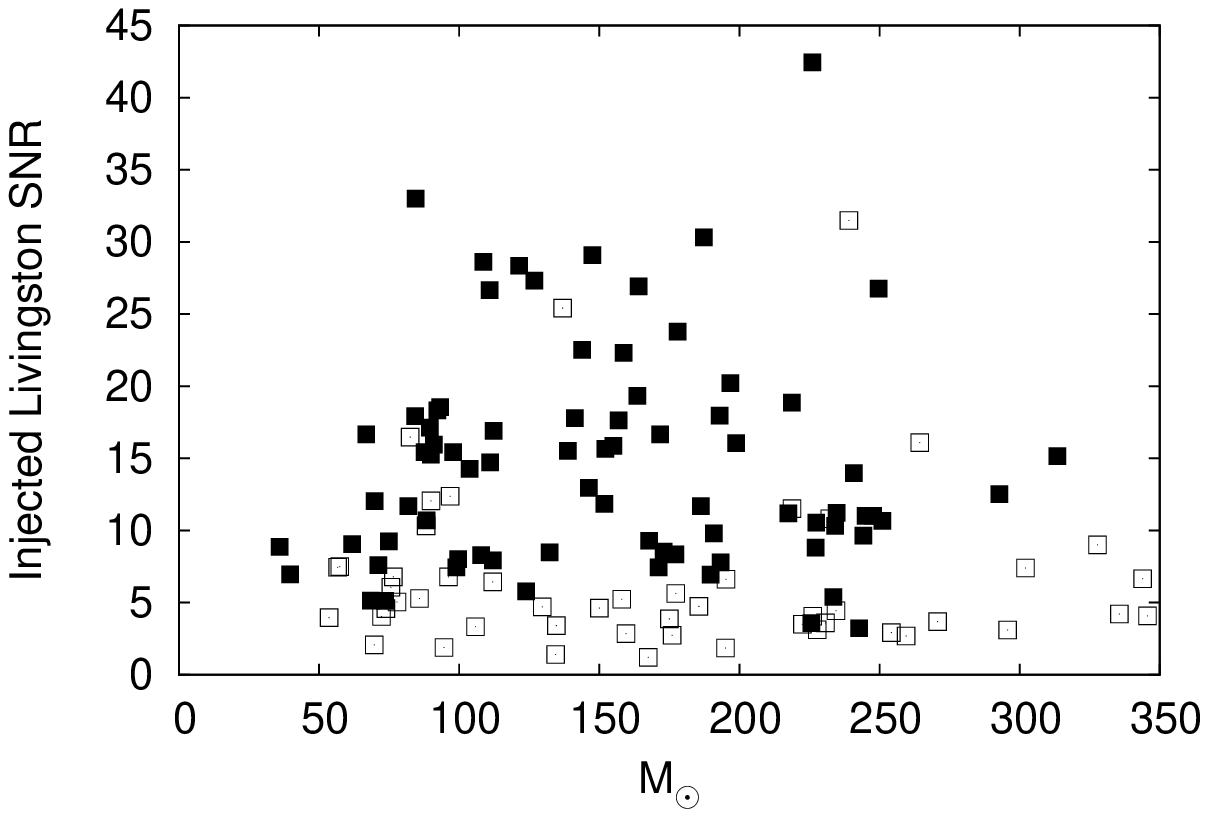}
\end{center}
\caption{{\bf Distribution of found and missed injections for the
Hilbert--Huang pipeline.} The left figure shows the results of the search on
the simulated LIGO Hanford 4~km detector, and the right figure shows the
result for the simulated LIGO Livingston 4~km detector. Detected signals are
shown in black and missed signals in white, as functions of the injected
matched filter signal-to-noise ratio and total mass.}
\label{Fig:hht_F0}
\end{figure}

\begin{figure}\begin{center}
\includegraphics[width=0.48\linewidth]{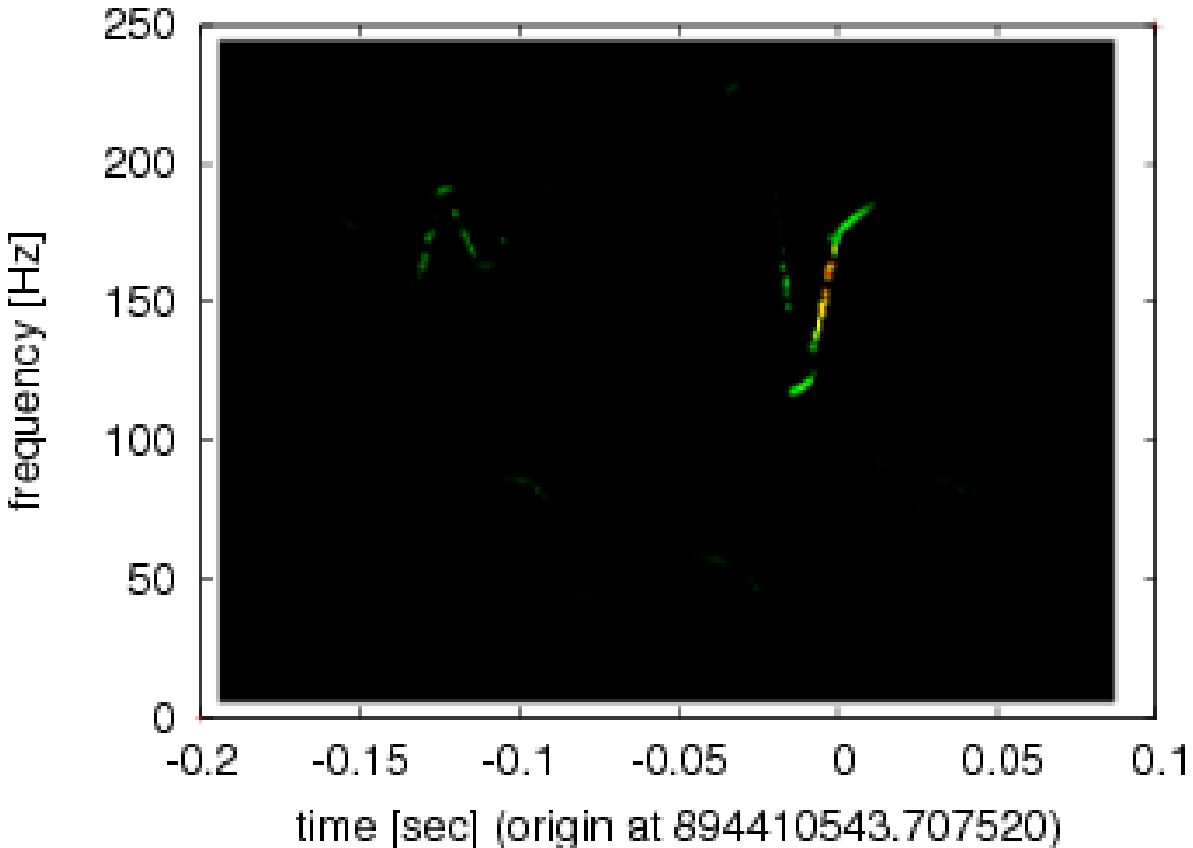}
\includegraphics[width=0.48\linewidth]{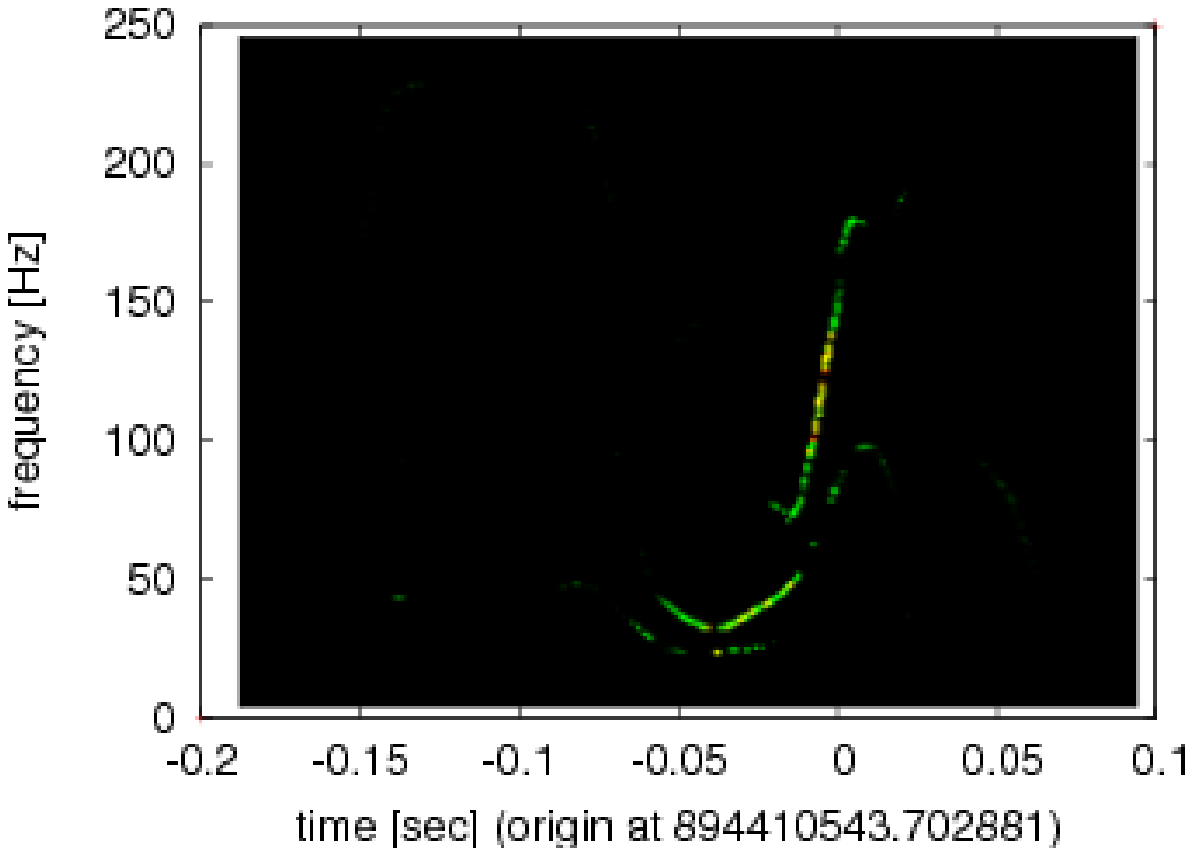}
\end{center}
\caption{{\bf The cluster-enhanced time-frequency maps of the BAM HHB S00 signal injected with a total mass of $69.8 M_\odot$.} H1 is shown in the left panel and L1 in the
right panel. We clearly see the burst part of the signal, thus the actual
merger. The ringdown and inspiral may be obscured by noise at low SNR.}
\label{Fig:hht_F1}
\end{figure}

The excellent resolution of the HHT in time and frequency was used to reject
false events due to overly short triggers, failed coincidences or mismatched
time-frequency ranges~\cite{Stroeer:2008}. We identified the latter as a
powerful veto tool which could be used to improve the sensitivity of
gravitational-wave data analysis pipelines.  We were ultimately able to
identify 80 events in coincidence, as shown in Figure~\ref{Fig:hht_F0}. Three
of these candidate events were determined to be false positives when the
candidate events were compared to the list of injected signals. Out of the 50
missed events, 39 have injected $\textrm{SNR}<10$, five have injected
$\textrm{SNR}<10$ in one detector and $\textrm{SNR}>10$ in the other, and six
had injected $\textrm{SNR}>10$ in both detectors.  We therefore reason that
most of the of missed events are low SNR cases in which no blocks were
triggered.  Most of the cluster-enhanced maps show the time-frequency
evolution of the signal with high precision (see Figure~\ref{Fig:hht_F1} for
one particular example). Time lag estimates and coherent additions show strong
potential and can be seen as proof of principle, but need further refinement
to work reliably in an automatic pipeline. SNR estimates are difficult since
only the burst region or diverse fragments of the signal are visible in our
search.  We refer to~\cite{hht:astroeer2008} for further details of this
analysis.

\subsection{Comparison of Inspiral-Burst-Ringdown Results}
\label{ssec:comparison}
In this section we consider comparisons between several of the search
pipelines described in the preceding sections.  The performance of a pipeline
depends on many parameters, such as the signal-to-noise thresholds, the
trigger coincidence and coherence tests, signal-based vetoes tests, allowed
false alarm rates, etc.  Search pipelines are tuned to suppress false alarms
while preserving detection efficiency and tuning decisions can dramatically
affect the relative performance of one pipeline versus another. Given the
limited scope of the NINJA data set, a comprehensive comparison of pipelines
was not possible. However, to make a first-order comparison between search
pipelines applied to the NINJA data, disentangled from pipeline tuning
decisions, we compared the number of injections found \emph{in a single
detector} at a fixed matched-filter signal-to-noise threshold.

Table~\ref{t:found} reports the number of injections found in single
interferometers and in multi-interferometer networks, using triggers from the
Q-pipeline burst algorithm (which targets the merger by match filtering to
sine-Gaussian templates), matched filter to ringdown templates, matched filter
to inspiral templates and matched filter to non-spinning, full coalescence
EOBNR waveforms.
For all algorithms the same nominal threshold of $\textrm{SNR} \ge 5.5$ was
imposed.  In addition, the number of detected injections in AND (injections
detected simultaneously from multiple algorithms) and OR (injections detected
by at least one algorithm) in an inspiral-merger-ringdown analysis is
reported.

The statistics of this sample is too small to make inferences on which
pipeline performs better in which parameter region; a more systematic study is
needed. Furthermore, since the NINJA data contains only Gaussian noise and
signals and so this comparison does not take into account noise transients
which may cause false detections.  Nevertheless, we have an indication that
all pipelines have comparable chances to find these injections. The
differences between pipelines are in the accuracy with which they can measure
the parameters of the signal.

\begin{table}
\begin{center}
\begin{tabular}{r|llllllll}
\hline
          & H1 & H2 & L1 & V1 & H1L1 & H1H2L1 & H1L1V1 & H1H2L1V1\\
\hline
$\textrm{SNR}_\mathrm{injected}\ge 5.5$ & 94 & 60   & 93   & 105  &  84   & 58 & 68 & 48 \\
\hline
Q-pipeline (M)           & 88 & 55   & 85   & 92   &  77   & 51 & 57 & 40 \\
Ringdown (R) & 88 & 56   & 83   & 93   &  76   & 52 & 56 & 40 \\
\hline
M-R AND            & 87 & 55   & 82   & 91   &  76   & 51 & 56 & 40 \\M-R OR             & 89 & 56   & 86   & 94   &  77   & 52 & 57 & 40 \\\hline
TaylorF2 (I)  & 85 & 43   & 82   & --   &  75   & 43 & -- & -- \\
I-M-R AND  & 82 & 42   & 77   & --   &  72   & 41 & -- & -- \\
I-M-R OR   & 91 & 57   & 88   & --   &  80   & 55 & -- & -- \\\hline
EOBNR (E)           & 90 & 56   & 88   & 100  &  79   & 54 & 64 & 45 \\
E-M-R AND        & 86 & 53   & 81   & 88   &  75   & 49 & 56 & 39 \\
E-M-R OR         & 92 & 58   & 91   & 104  &  81   & 56 & 65 & 47 \\\hline
\end{tabular}\end{center}
\caption{%
\label{t:found}
{\bf Number of injections found} with $\textrm{SNR} \ge 5.5$. This table
takes into account for each detector only signals with injected $\mathrm{SNR}
\ge 5.5$. If the same injection is found in more than one detector or
algorithm, it is counted as coincidence. The inspiral triggers are from the
2~PN TaylorF2 templates, terminated at $f_\mathrm{ISCO}$ (total mass in
$20-90\,M_{\odot}$)  and from the EOBNR waveforms.  Note that, in an actual
search, different thresholds may be needed for different pipelines, depending
on the false alarm rate and the morphology of noise transients, so this is
strictly a comparison between search techniques on the limited NINJA data set,
not of the performance of full pipelines on actual data.}
\label{tab:n_inj_S5.5}
\end{table}

\begin{figure}
   \begin{center}
   \includegraphics[width=0.8\linewidth]{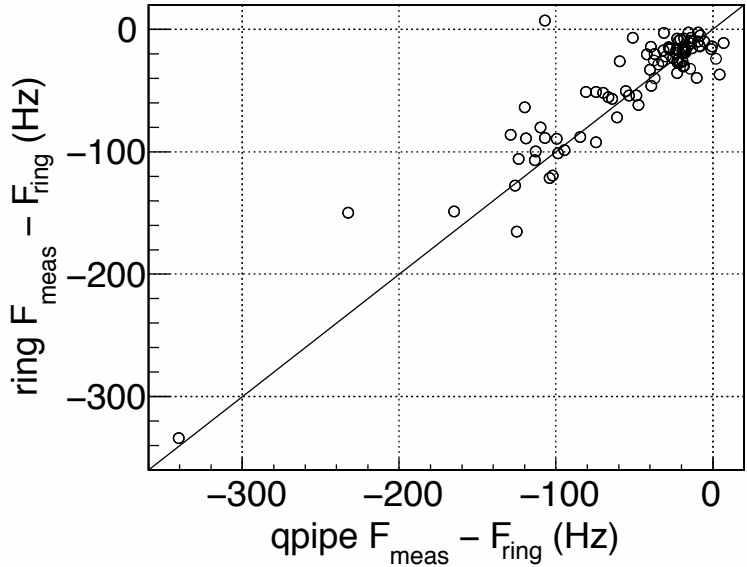}
   \end{center}
   \caption{{\bf Comparison of the frequencies measured by the Q-pipeline and
   ringdown searches.} The $x$-axis shows the difference between the central
   frequency reported by the Q-pipeline and the frequency of the injected
   ringdown $f_\mathrm{ring}$. The $y$-axis shows the difference between the
   frequency measured by the ringdown search and the injected ringdown
   frequency.}
   \label{f:qringcomp}
\end{figure}

\begin{figure}
   \begin{center}
   \includegraphics[width=0.49\textwidth]{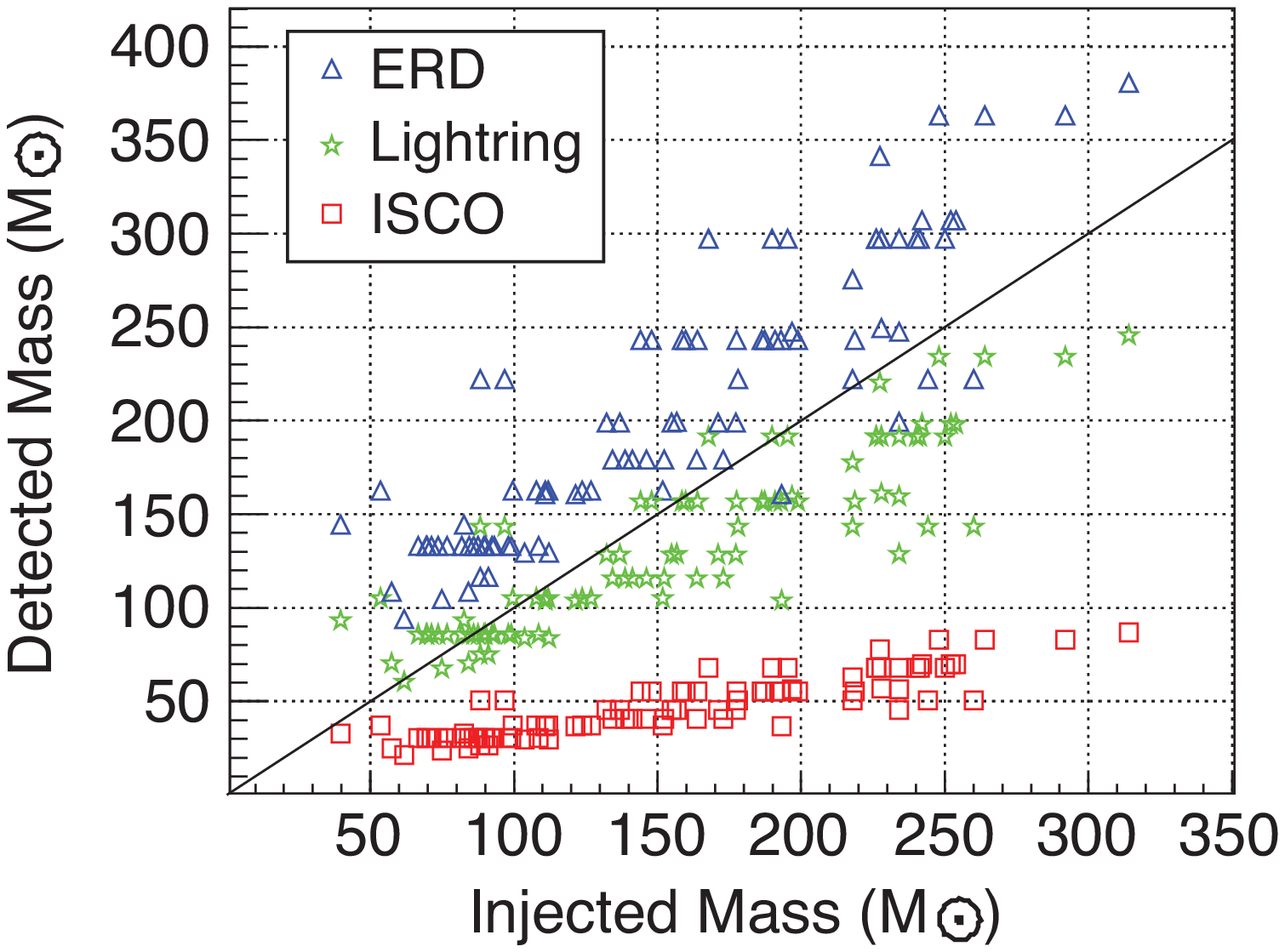}
   \includegraphics[width=0.49\textwidth]{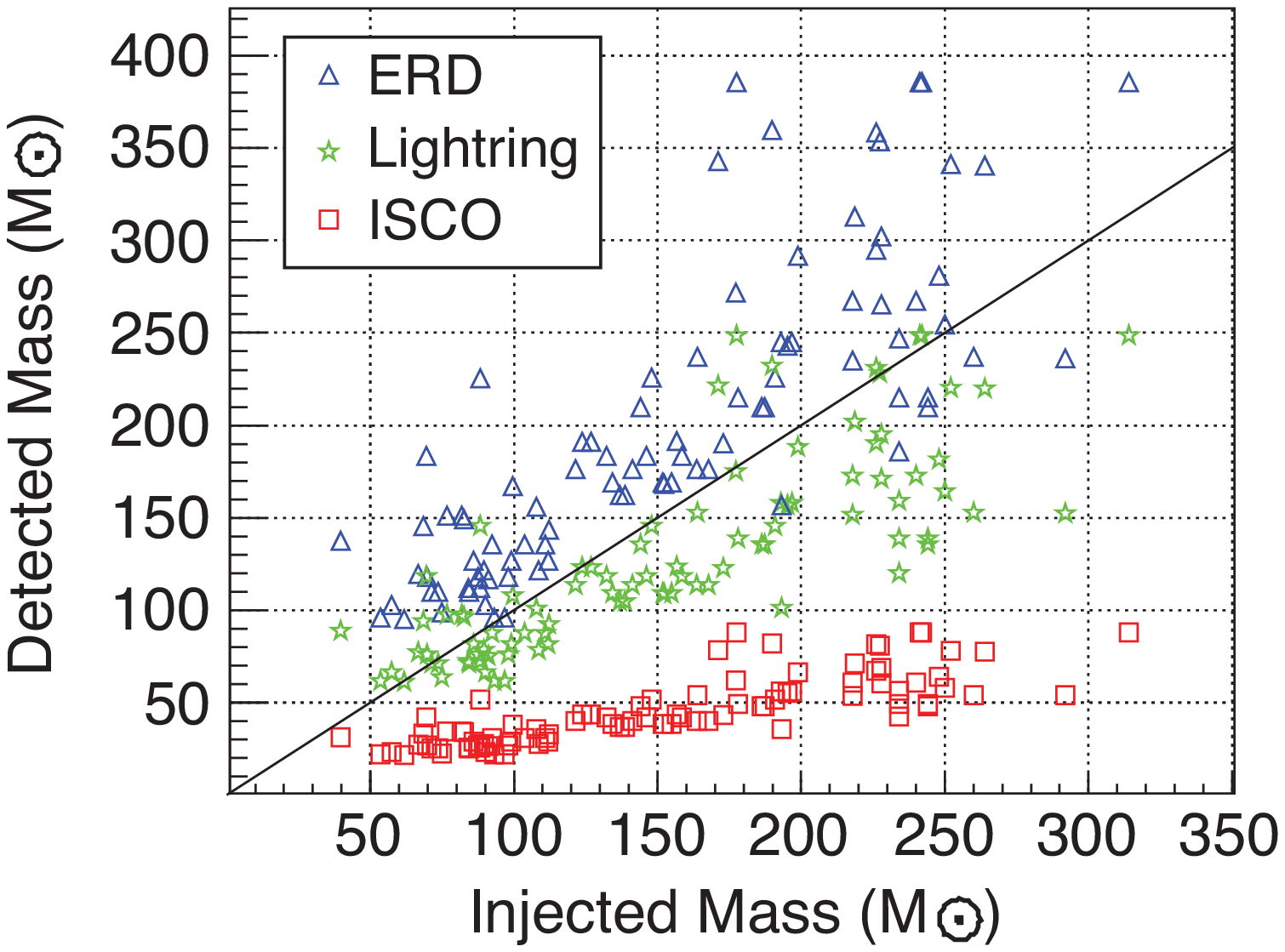}
   \end{center}
   \caption{{\bf Comparison of the injected and recovered masses for the
   Q-pipeline and ringdown searches.} For both searches we start from the
   frequency reported by the algorithm and compute the corresponding mass,
   using equation~(\ref{e:freqs}), assuming that the measured frequency is
   $f_\mathrm{ISCO}$, $f_\mathrm{LR}$, and $f_\mathrm{ERD}$. Both algorithms
   measure a frequency somewhere between the light-ring and ringdown
   frequencies.} 
   \label{f:MvM1}
\end{figure}

\begin{figure}
   \centering
   \includegraphics[width=0.49\textwidth]{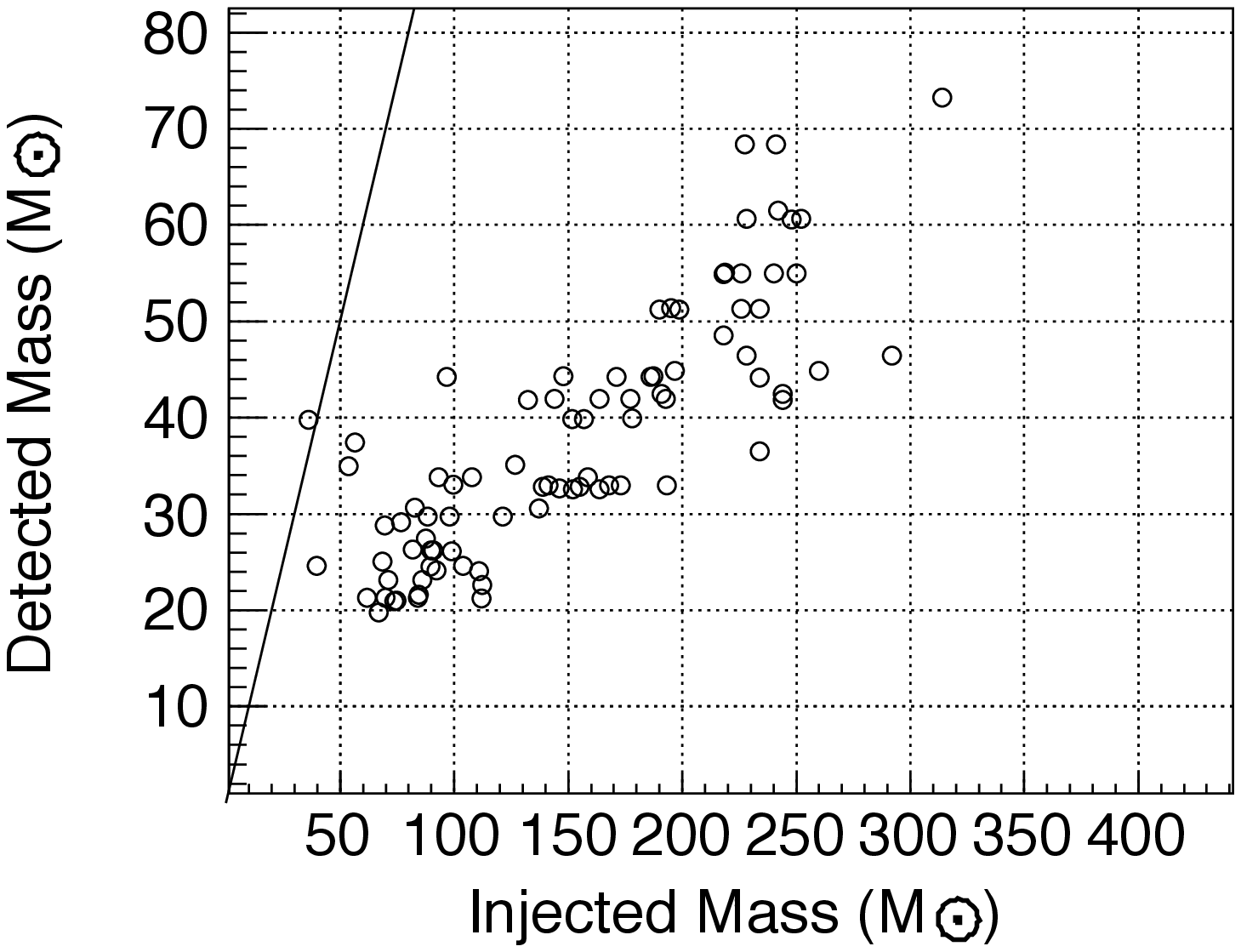}
   \includegraphics[width=0.49\textwidth]{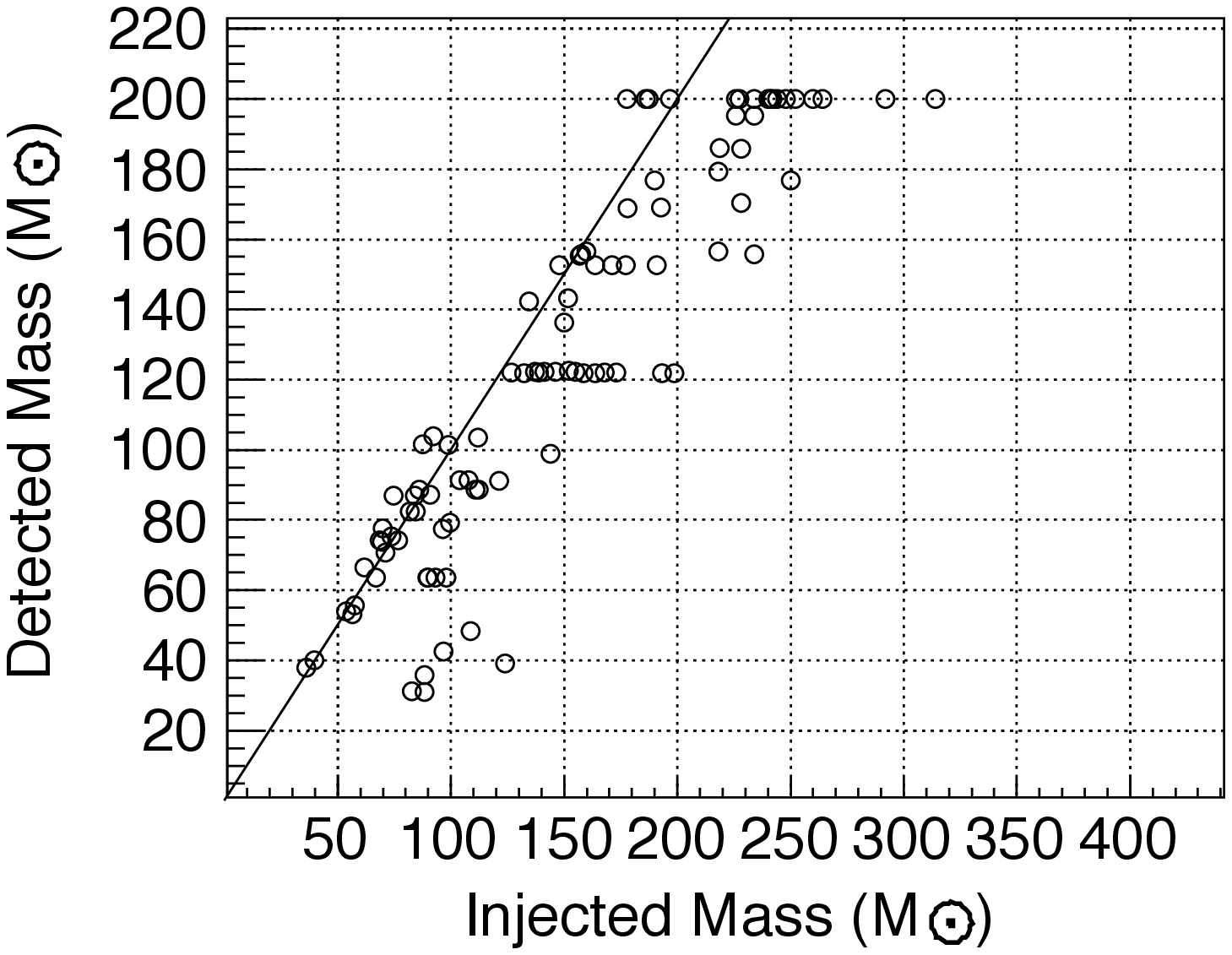}
   \caption{{\bf Comparison of the injected and recovered masses for the
   TaylorF2 and EOBNR searches.} The left plot shows the comparison for 2.0
   order post-Newtonian TaylorF2 templates terminated at $f_\mathrm{ISCO}$ and
   the right plot shows the comparison for EOBNR templates.}
   \label{f:MvM2}
\end{figure}

Figure~\ref{f:qringcomp} compares the accuracy with the Q-pipeline and
ringdown searches measure the frequency of the signal and Figures~\ref{f:MvM1}
compares the accuracy with these pipelines measure the total mass of the
injected signal. Both the ringdown and Q-pipeline searches report a frequency
not a mass, so to compare to the injected and detected total masses we must
calculate a mass from the frequency of the candidate. There is no unique way
to do this, and but we can calculate the total mass under the assumption the
algorithm is detecting a given portion of the waveform. Figure~\ref{f:MvM1}
shows the result of using the formulae for ISCO, light-ring and effective
ringdown frequencies to compute the total
mass~\cite{Pan:2007nw,Buonanno:2000ef}:
\begin{equation}
f_\mathrm{ISCO}=\frac{c^3}{6\sqrt{6}\pi GM}, \quad
f_\mathrm{LR}=\frac{c^3}{3\sqrt{3}\pi GM}, \quad
f_\mathrm{ERD}=\frac{0.5967 c^3}{2\pi GM}.
\label{e:freqs}
\end{equation}
Both burst and ringdown code preferentially detect the portion of the signal
that is in the detectors' most sensitive band.  Figure~\ref{f:MvM1} shows that
for both algorithms, at the lowest injected masses this correspond to
$f_{ISCO}$, but as the injected mass increases, the algorithms trigger between
light ring and ringdown, as expected. 

This comparison is more straightforward for matched filtering codes that use a
template with a given mass. Figure~\ref{f:MvM2} shows the detected total mass
against the injected total mass for (i) 2.0 post-Newtonian order TaylorF2
templates terminated at $f_\mathrm{ISCO}$, with templates in the mass range
20--90 M$_{\odot}$ and (ii) EOBNR templates with masses in the range to
30--200~M$_{\odot}$. The TaylorF2 templates significantly underestimate the
masses of the injected signals, due to the fact that most of the injected
signals lie outside the template bank. The EOBNR search, with its larger mass
range and inspiral-merger-ringdown waveforms, measures the masses of the
injected signals with better accuracy.

\subsection{Bayesian pipelines}
\label{ssec:param_est}
Bayesian inference~\cite{Gregory:2005} is a powerful means of extracting information from
observational data based on the calculation of posterior probabilities and
probability density functions. Computation of these quantities is expensive and
so these algorithms are not typically used to search for candidate events in
detector data. They are, however, useful in the closer study of candidates
identified by the search pipelines described in Sections~\ref{ssec:modeled} and
\ref{ssec:unmodeled}.  This section describes the results of applying two
different Bayesian inference algorithm to the NINJA data. The
first is designed to estimate the parameters of the signal assuming a 
gravitational wave is 
present in the data. The second calculates the confidence in the presence of
the signal, quantified by the {\it odds ratio} between the signal and noise
models of the data.

Both approaches require the calculation of the posterior probability-density
function (PDF)  on the parameter space of the signal, given the data $d$,
which is
\begin{eqnarray}
p(\vec{\theta}|d)&=&\frac{p(\vec{\theta})p(d|\vec{\theta})}{p(d)} \nonumber \\
&\propto&p(\vec{\theta}) 
\exp \left( -2 \int_0^\infty  \frac {\left|\tilde{d}(f)
    - \tilde{h}(f; \vec{\theta}) \right|^2}{S_\mathrm{n}(f)}\, df  \right)
\end{eqnarray}
in the presence of Gaussian noise with power spectral density
$S_\mathrm{n}(f)$, where $p(\vec{\theta})$ is the prior probability density of
the parameters $\vec{\theta}$ and $h(\vec{\theta})$ is the model used to
describe the signal~\cite{Rover:2006ni}.

A Markov-Chain Monte-Carlo approach~\cite{Gilks:1996} 
was used to coherently analyse data from
multiple detectors in order to evaluate the posterior PDFs.  This technique
stochastically samples the parameter space in a search for the parameters that
best match the observed data; it does so by attempting a random jump from the
current set of parameter values to a new one, then deciding whether the jump
should be taken by comparing the likelihoods of the old and new locations in
parameter space.  In this way,  one simultaneously searches for the set of
parameters that yield the best fit to the data, and determines the accuracy of
the parameter estimation.

Bayesian model selection, based on a different Monte-Carlo technique 
known as nested sampling \cite{Skilling:2004}, was employed as a tool to 
measure the confidence of a detection using different waveform families. 
This approach requires the calculation of the {\it marginal likelihood} 
of the signal and noise models, obtained by computing the integral of 
the posterior PDF $\int{}p(\vec{\theta})p(d|\vec{\theta})d\vec{\theta}$ 
to find the total probability of the model. It was possible to calculate 
this integral for the nine-parameter model of a non-spinning binary 
coalescence signal described coherently in multiple interferometers. The 
ratio of likelihoods of the signal and noise models is known as the 
``Bayes factor,'' and is used to multiply prior odds, giving the 
posterior odds ratio between the two models, taking into account the 
observational data; in turn, the value of this Bayes factor corresponds 
to the level of confidence in the detection. As a straightforward 
by-product of the nested-sampling algorithm, it is also possible to 
infer the maximum-likelihood values of the parameters of the detected 
signals; this was used to obtain further information on the ability 
of different waveform approximants to recover the source parameters.

\subsubsection{Parameter Estimation Using Markov Chain Monte Carlo}
\label{ssec:mcmc}
A selection of injected numerical signals were analysed with a 
Markov-Chain Monte Carlo (MCMC) code~\cite{vanderSluys:2007st,vanderSluys:2008qx}.  The signal model was based on
waveforms with phase evolution at 1.5 post-Newtonian order and
leading-order amplitude evolution.  Parameter estimation
was successful on NINJA injections with relatively low total 
mass in which the inspiral contained a significant fraction of the 
total signal-to-noise ratio.  For high-mass injections, the algorithm 
attempted to match the merger and ringdown portions of the injected 
signal to inspiral templates, resulting in poor parameter estimation.

The post-Newtonian waveforms used in this analysis include the spin of the
larger body $m_1$, allowing us to use the analytical simple-precession
waveform~\cite{Apostolatos:1994mx}.  The parameter space thus consists of
twelve independent parameters:
\begin{equation}
\vec{\theta} = \{\mathcal{M},\eta,\alpha,\delta,\psi,\iota,d,a_\mathrm{spin},\kappa,\phi_\mathrm{c},\alpha_\mathrm{c},t_\mathrm{c}\},
\end{equation}
where $\mathcal{M}$ and $\eta$ are the chirp mass and symmetric mass ratio,
respectively; $\alpha$ (right ascension) and $\delta$ (declination) identify
the source position in the sky; the angles $\psi$ and $\iota$ identify the
direction of the total angular momentum of the binary; $d$ is the luminosity
distance to the source; $0 \le a_\mathrm{spin} \equiv S_1/m_1^2 \le 1$ is the
dimensionless spin magnitude; $\kappa$ is the cosine of the angle between the
spin and the orbital angular momentum; and $\phi_\mathrm{c}$ and $\alpha_\mathrm{c}$ are
integration phase constants that specify the gravitational-wave phase and the
location of the spin vector on the precession cone, respectively, at the time
of coalescence $t_\mathrm{c}$.

The MCMC algorithm used for the NINJA analysis was optimised by including a 
variety of features to efficiently 
sample the parameter space, such as parallel tempering \cite{vanderSluys:2008qx}.
This MCMC implementation can be run on a data set from a single detector, or on data sets 
from multiple detectors. In the latter case, a coherent search among all detectors significantly 
improves the determination of source position and orientation \cite{vanderSluys:2007st,vanderSluys:2008qx}. 

The MCMC code was run on a selection of injected signals in the NINJA data.  
It was found that although the MCMC runs are clearly able to detect a signal 
whenever the inspiral contains a sufficient signal-to-noise ratio (SNR), 
they were unable to correctly determine the signal parameters for many injections.  
For the high masses typical of most NINJA injections, the SNR is dominated by the 
merger and ringdown, so that the inspiral-only templates tried to match the merger and 
ringdown portions of the injected waveform.  Typically, it is found that in such cases 
the time of coalescence is overestimated since the injected ringdown is matched to 
an inspiral; the chirp mass is underestimated since the merger/ringdown frequency 
is higher than the inspiral frequency for a given mass, so that matching them to an 
inspiral requires the mass to be lower; the mass ratio is underestimated, which 
allows the waveform to contain more energy in the narrow frequency band corresponding 
to quasi-normal ringing; and the spin rails against the upper prior of $1$ since the 
innermost stable circular orbit frequency is highest for an inspiral into a 
maximally-spinning Kerr black hole.  We tried to circumvent the problem of matching 
to the merger and ringdown by introducing more-restrictive priors on spin and/or $\eta$.  
These efforts still failed when the total masses were too high, but were successful in 
the case of lower total masses and longer inspiral signals. 

\begin{figure}[htb]
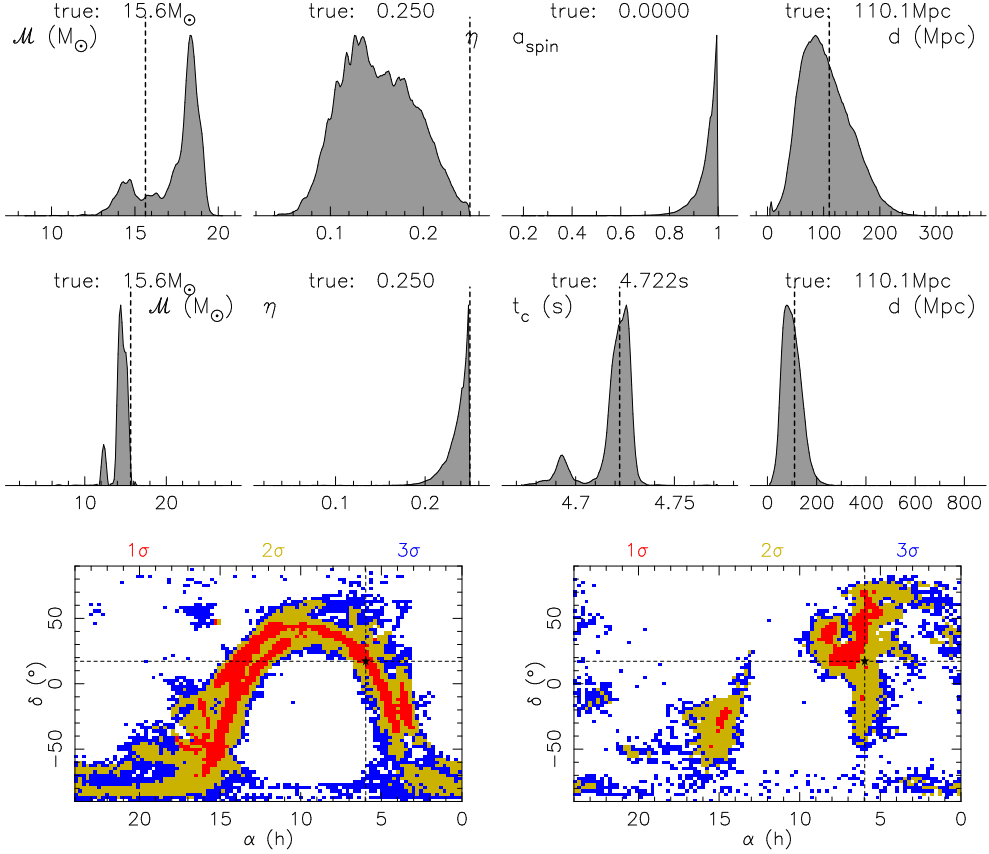

\begin{center}
\includegraphics[keepaspectratio=true, width=1.0\textwidth]{figures/mcmc_1a}
\vskip0.15in

\includegraphics[keepaspectratio=true, width=1.0\textwidth]{figures/mcmc_2a}

\vskip0.15in
\includegraphics[keepaspectratio=true, width=0.45\textwidth]{figures/mcmc_1b}
\hskip0.25in
\includegraphics[keepaspectratio=true, width=0.45\textwidth]{figures/mcmc_2b}

\end{center}
\caption{{\bf Results of the MCMC analysis for one of the injected NINJA
waveforms.} The top row shows the marginalised PDFs for $\Mc$, $\eta$,
$a_\mathrm{spin}$, and $d$ produced by a two-detector MCMC analysis on an injected non-spinning
equal-mass {\tt SpEC} Cornell/Caltech waveform (true values $\Mc=15.6\,M_\odot$, 
$\eta=0.25$, $t_\mathrm{c}=4.7223$\,s, and $a_\mathrm{spin}=0$).  Middle row: the same PDFs
(but with $a_\mathrm{spin}$ replaced by $t_\mathrm{c}$) for a three-detector run with constrained spin on
the same injection.  Bottom row: two-dimensional PDFs for the sky position
with the 2-detector run on the left and the three-detector MCMC run on the right; 
the 1-$\sigma$, 2-$\sigma$ and 3-$\sigma$ probability areas are displayed in 
different colours, as indicated in the top of each panel.
Dashed lines denote the true values of injected parameters.}
\label{MCMCfigure}
\end{figure}

Figure~\ref{MCMCfigure} shows the PDFs produced by runs on an injected equal-mass non-spinning 
{\tt SpEC} Cornell/Caltech waveform with $\Mc=15.6\,M_\odot$.  This particular injection was chosen 
because it had the lowest total mass, and {\tt SpEC} waveforms typically have more inspiral cycles; 
runs on other injections show comparable results, with the general trend that the higher the total 
mass (and, thus, the lower the relative fraction of the SNR contributed by the inspiral), the poorer 
the parameter estimation becomes.

Data from two detectors, H1 and L1, were used to compute the PDFs shown in the top row of Figure~\ref{MCMCfigure}.  
We used wide, flat priors for intrinsic parameters (e.g., $\Mc \in [2\,M_\odot, 100\,M_\odot]$, 
$\eta \in [0.03, 0.25]$, $a_\mathrm{spin} \in [0,1]$).  We find that the values of the intrinsic 
parameters are not determined very accurately.  In particular, the spin rails against the upper bound 
of $1$ while $\eta$ is underestimated, as expected.  We find that the sky location is nevertheless 
constrained to an arc of a ring containing the true value; the 2-$\sigma$ ($\sim 95$\%) sky area of 
the ring shown in the bottom left of Fig.~\ref{MCMCfigure} is $\sim 10000$ square degrees.

In the middle row of Fig.~\ref{MCMCfigure}, we plot the PDFs based on data from three detectors: 
H1, L1, and V1.  The spin parameter was constrained to its true value $a_\mathrm{spin}=0$ for this run.  
This had the effect of forcing the MCMC to match the inspiral only, significantly improving the resolution 
of other parameters: for example, the PDF of $\eta$ now rails against $0.25$, which is its true value.  
The chirp mass is still somewhat underestimated; 
a higher SNR may be necessary to improve the mass determination.  Promisingly, it was found that the sky location 
is constrained to a smaller patch on the sky: the 2-$\sigma$ sky area in the bottom right of Fig.~\ref{MCMCfigure} 
is  $6300$ square degrees. In fact, the sky localisation is even better when the spin parameter is allowed 
to vary, allowing the MCMC to use the SNR contributed by the ringdown; removing the spin-parameter constraint 
reduces the 2-$\sigma$ sky area to $2750$ square degrees.  This ability to determine the source position will 
be significant in any future searches for electromagnetic counterparts of gravitational-wave triggers. 

We hope that in the future it will be possible to test MCMC codes on numerical signals in a lower mass range, 
where the inspiral portion would dominate the SNR, so that inspiral-only templates are not at a significant 
disadvantage.  Meanwhile, we have recently implemented templates at 3.5~PN order in phase 
that include the spin for both members of the binary, thus improving the accuracy 
of parameter estimation and increasing the range of applicability of our code.

\subsubsection{Bayesian Model Selection Pipeline}
\label{ssec:bayesian}
The primary goal of this analysis was to investigate the performance of different template families on the confidence of detection of the injections contained in the NINJA data set. The approach to this problem used a method described in~\cite{Veitch:2008ur,Veitch:2008wd}, which can be summarised as follows. Two hypotheses are under consideration: (i)
$H_N$---The data $\{\tilde d_k\}$ are described by (Gaussian and stationary) noise only: $\tilde d_k= \tilde n_k$, 
and (ii) 
$H_S$---The data $\{\tilde d_k\}$ are described by (Gaussian and stationary) noise $\{\tilde n_k\}$ {\em and} a gravitational wave signal $\{\tilde h^{(a)}_k(\vec{\theta})\}$, according to a given approximant $a$, where $\vec{\theta}$ represents the vector of the (unknown) signal parameters: $\tilde d_k= \tilde n_k + \tilde h^{(a)}_k(\vec{\theta})$. 
The marginal likelihood of $H_S$ is calculated by performing the integral 
\begin{equation}
p(\{\tilde{d_k}\}|H_S,a) = \int p(\vec{\theta})p(\{\tilde{ d_k}\}|H_S,a,\vec{\theta})d\vec{\theta}\,.
\end{equation}
The ratio of probabilities or ``odds ratio'' of the two models is
\begin{eqnarray}
{\cal O}_{SN,a} &=& \frac{P(H_S|\{\tilde d_k\},a)}{P(H_N|\{\tilde d_k\})}
\nonumber \\
&=& \left[\frac{P(H_S)}{P(H_N)}\right]\,\left[\frac{P(\{\tilde d_k\}|H_S,a)}{P(\{\tilde d_k\}|H_N)}\right] \nonumber \\
&=& {\cal P}\,B_{SN}^{(a)},
\end{eqnarray}
where ${\cal P}$ is the \emph{prior odds ratio} and $B_{SN}^{(a)}$ is the \emph{Bayes factor}.

In this analysis the model includes the response of all available simulated detectors (L1, H1, H2 and V1) coherently, and the gravitational waveforms are calculated using function in the LAL library~\cite{lal}. For the gravitational waveform, two different approximants were considered: the standard (2~PN) TaylorF2 waveform family, with inspiral truncated at $f_\mathrm{ISCO}$, and the phenomenological inspiral-merger-ringdown IMRPhenA templates described in~\cite{Ajith:2007kx}. In each case the waveforms were truncated at low frequency at 30 Hz.


\begin{figure}[htbp] 
   \centering
   \includegraphics[width=5in]{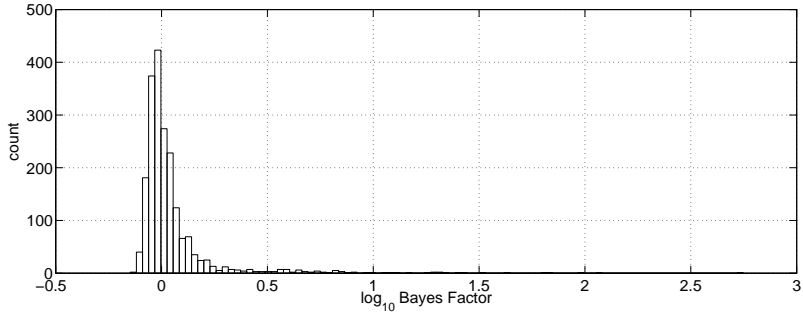}
\caption{{\bf The distribution of Bayes factors obtained by running the algorithm with TaylorF2 approximants on signal-free data sets.} The pipeline was run coherently on the four detectors, and the same was done for the results reported in this figure. The total number of trials is 2000 and the highest value of the Bayes factor is $\log_{10} B = 2.77$.
}
\label{fig:TF2_fa} 
\end{figure}

The choice of priors in the analysis was as follows.  For the TaylorF2 approximant, the prior on $\vec{\theta}$ was uniform on chirp mass, symmetric mass ratio and distance, within the following limits: time of coalescence in a window $\pm 0.5$\,s around the actual $t_c$ of each injection, symmetric mass ratio $\eta$ in the range $0.01 \le \eta \le 0.25$, chirp mass within the bounds set by $\eta$ and the total mass in the range $50 \Ms \le M \le 150 \Ms$, and distance $1\le D \le 500$ Mpc. The parameters for orientation and position on the sky of the binary were allowed to vary over their entire angular ranges. For the IMRPhenA approximant, the limits were identical on all the parameters, with the exception of the total mass whose upper boundary was set to $475 \Ms$. We also calculated the Bayes factors for data segments containing no signal in order to estimate the background distribution of Bayes factors. Figure~\ref{fig:TF2_fa} shows the distribution of Bayes factors (using TaylorF2 approximants) when running the analysis algorithm on portions of data without injections: 2000 trials were carried out, with a maximum value of $\log_{10}B = 2.77$. If interpreted as a threshold value on the Bayes factors to decide whether a signal has been detected or not, it corresponds to a false alarm of $0.05 \%$. The distribution obtained with the IMRPhenA approximant is very similar. In the analysis, a range of ``detection thresholds'' on $\log_{10}B$ was considered, in order to explore how the two different approximants (and the algorithm) respond to different numerical relativity injections. 

\begin{figure}[htbp] 
   \centering
   \includegraphics[width=2.5in]{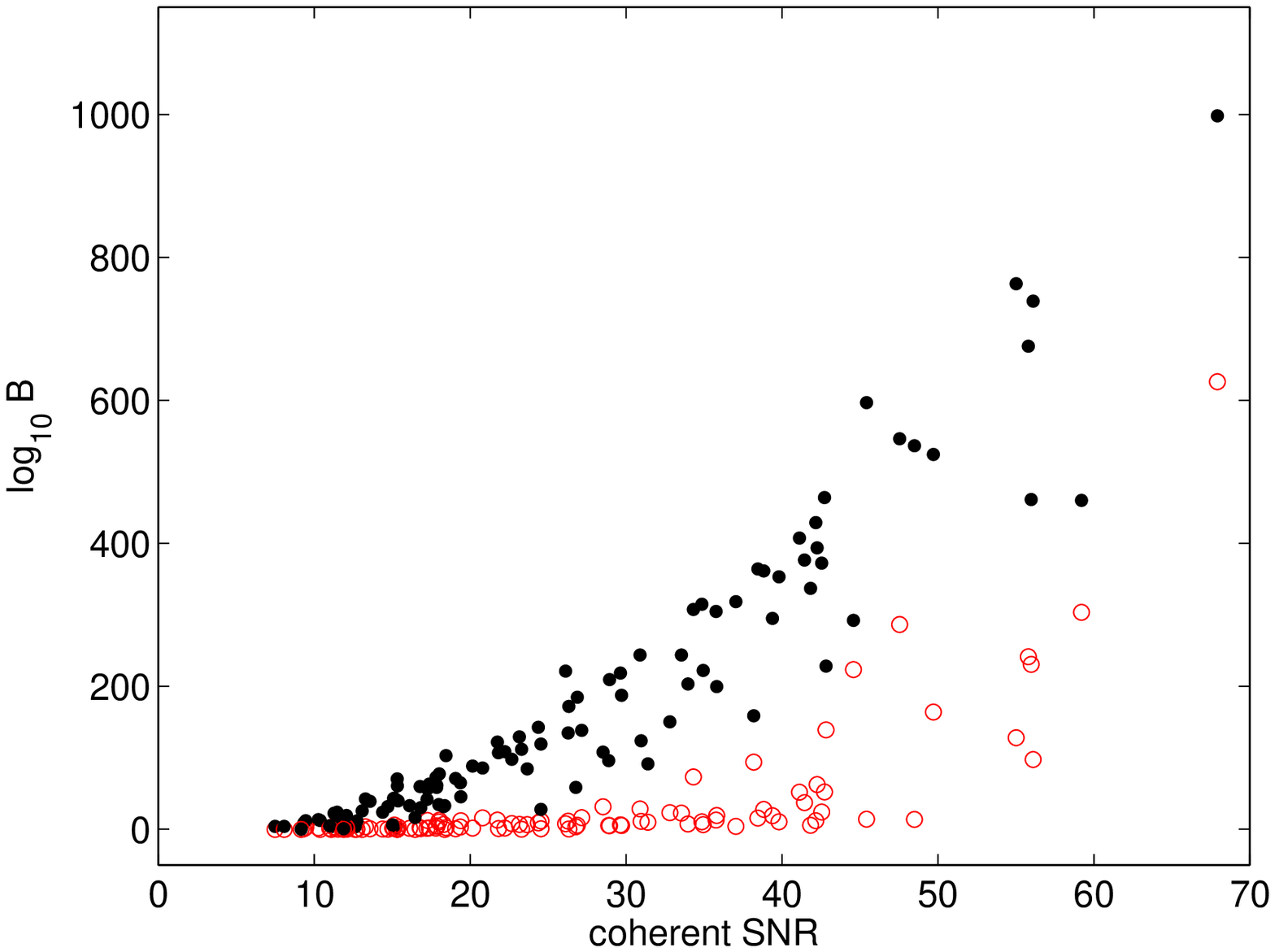}
   \includegraphics[width=2.5in]{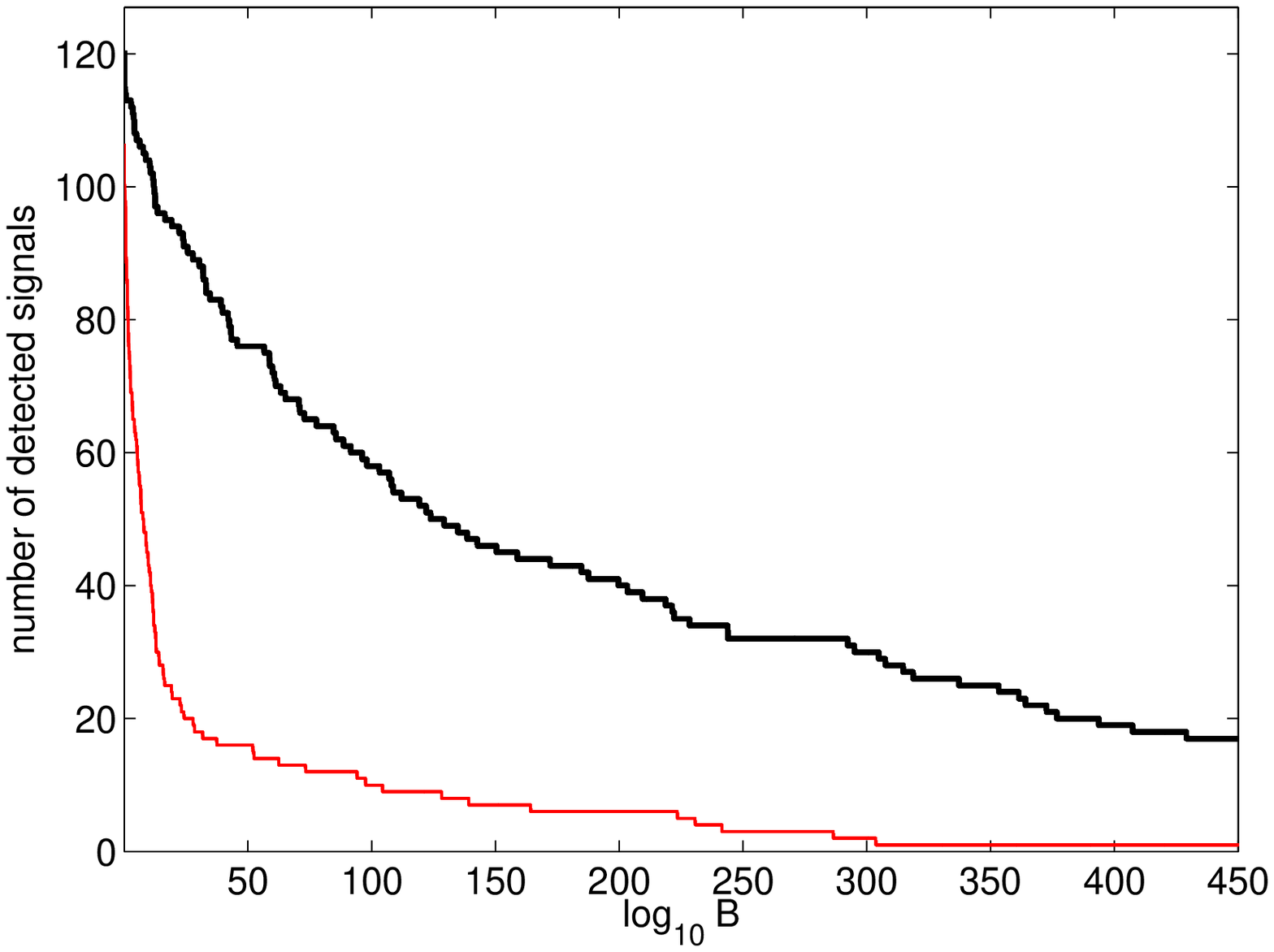}
   \caption{{\bf Comparison of the Bayes factors for TaylorF2 and IMRPhenA approximants.} {\bf Left:} The values of the Bayes factors obtained in the analysis of the NINJA data set as a function of the optimal coherent (L1-H1-H2-V1) signal-to-noise ratio at which the signals were injected into instrument noise. The solid (black) dots and the open (red) circles refer to the Bayes factors obtained by using the IMRPhenA and TaylorF2 approximants, respectively. {\bf Right:} The cumulative number of injections recovered as a function of the Bayes factor. The thin (red) solid line corresponds to the results obtained by using the TaylorF2 approximant, whereas the thick (black) solid line refers to the IMRPhenA approximant. A threshold of $\log_{10} B_{SN} = 3$ has been used.}
   \label{fig:summary_B}
\end{figure}

Figure~\ref{fig:summary_B} and Table~\ref{tab:results_bayes}  summarise the main results. The left panel of Figure~\ref{fig:summary_B}  shows the value of the Bayes factor computed for the two approximants as a function of the coherent four-detector signal-to-noise ratio at which the waveforms were injected. For \emph{all} the injections, IMRPhenA approximants return a Bayes factor which is (significantly) larger than TaylorF2 approximants. This is not surprising, as the TaylorF2 waveforms do not contain the merger and ring-down portion of the coalescence and are truncated at $f_\mathrm{ISCO}$. Figure~\ref{fig:summary_B}  (right panel) shows the number of injections that are recovered at a given Bayes factor (or above). Once more the effectiveness of IMRPhenA approximants is striking compared to the TaylorF2 waveform family. These results are broadly in agreement with the matched-filter analysis carried out with inspiral-merger-ringdown waveforms described in Section~\ref{sssec:imr}.

\begin{table}
\centering
\begin{tabular}{c|cc}
\hline
Threshold 	& \multicolumn{2}{c}{Number of detected injections}	 \\
  $(\log_{10} B$)& TaylorF2	&	IMRPhenA\\
\hline
3	& 69		&	112	\\
5	& 61		&	107	\\
10	& 43		&	104	\\
30	& 28		&	89	\\
100	& 10		&	58	\\
\hline
\end{tabular}
\caption{
{\bf The number of detections in the NINJA data with model-selection pipeline
as a function of the Bayes-factor threshold using
TaylorF2 and IMRPhenA approximants.}
The total number of injections was 126. 
See also Figure~\ref{fig:summary_B}.
} 
\label{tab:results_bayes} 
\end{table}

\begin{figure}[htbp] 
   \centering
   \includegraphics[width=2.5in]{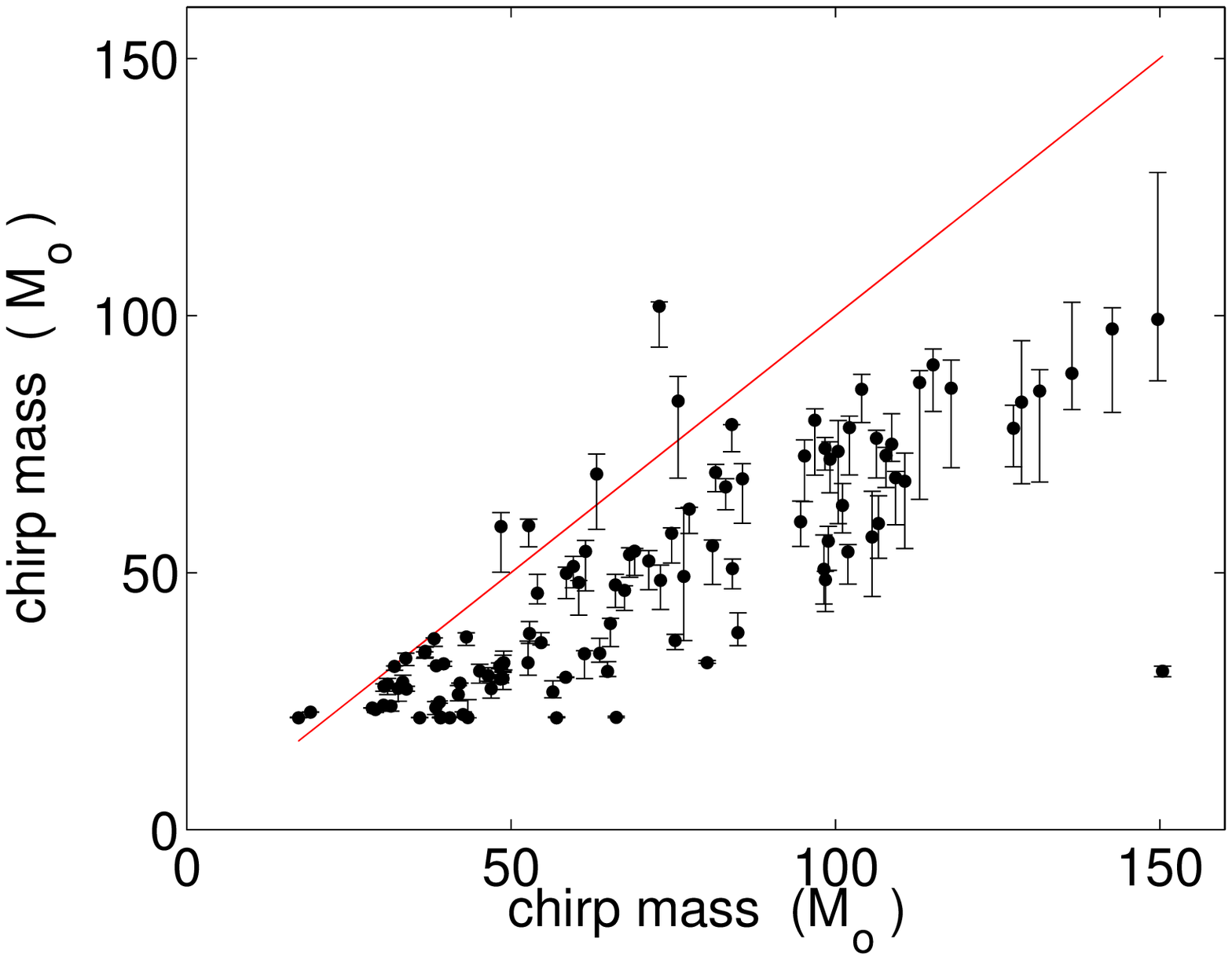}
   \includegraphics[width=2.5in]{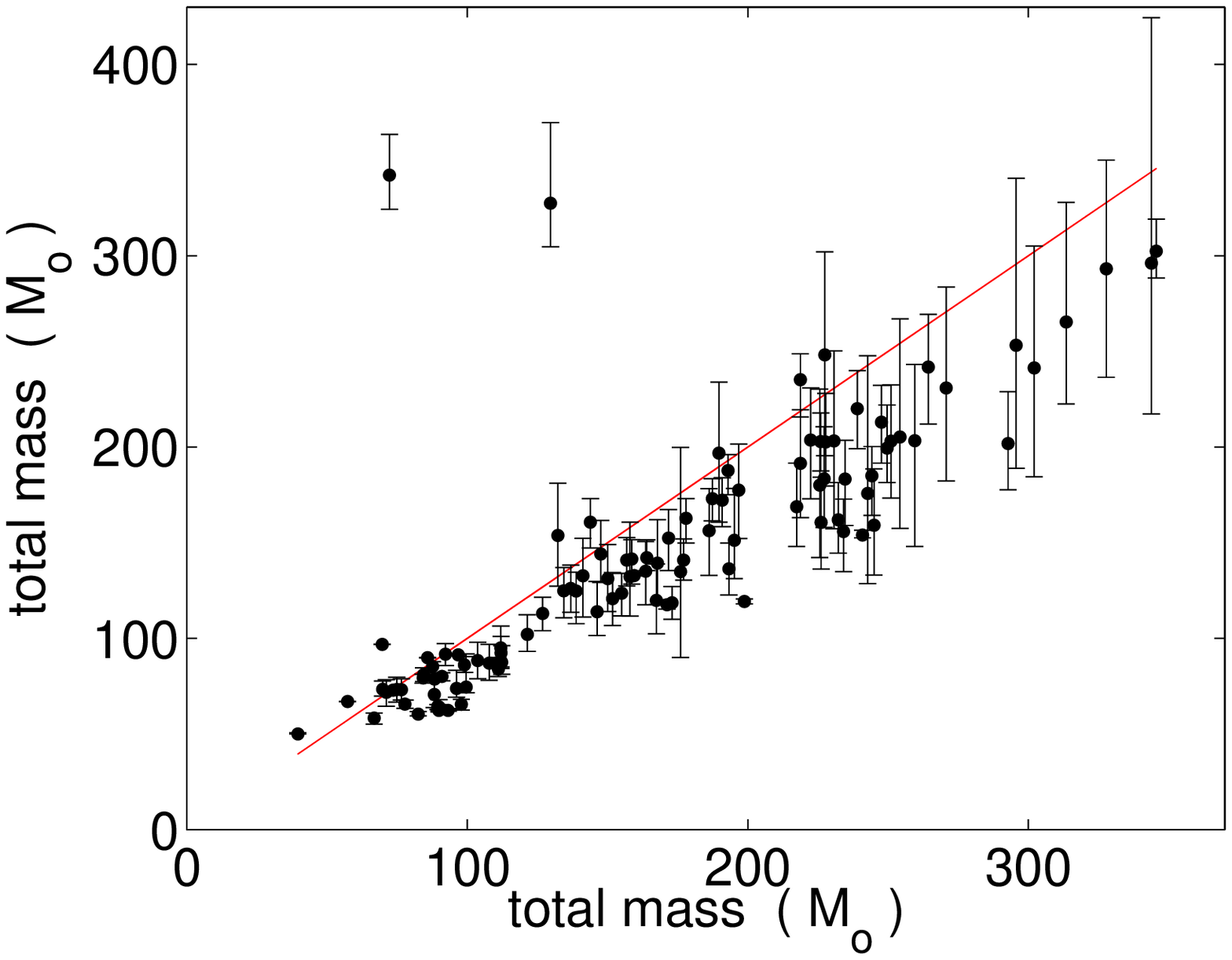}
   \caption{{\bf Comparison of the recovered mass parameters for the IMRPhenA approximant.} {\bf Left:} The recovered (maximum likelihood) values of the chirp mass as a function of the injected values. {\bf Right:} The recovered (maximum likelihood) values of the total mass as a function of the injected values.
The IMRPhenA approximant was used with a threshold of $\log_{10} B_{SN} = 3$.
}
   \label{fig:error_nest_m}
\end{figure}

\begin{figure}[htbp] 
   \centering
   \includegraphics[width=2.5in]{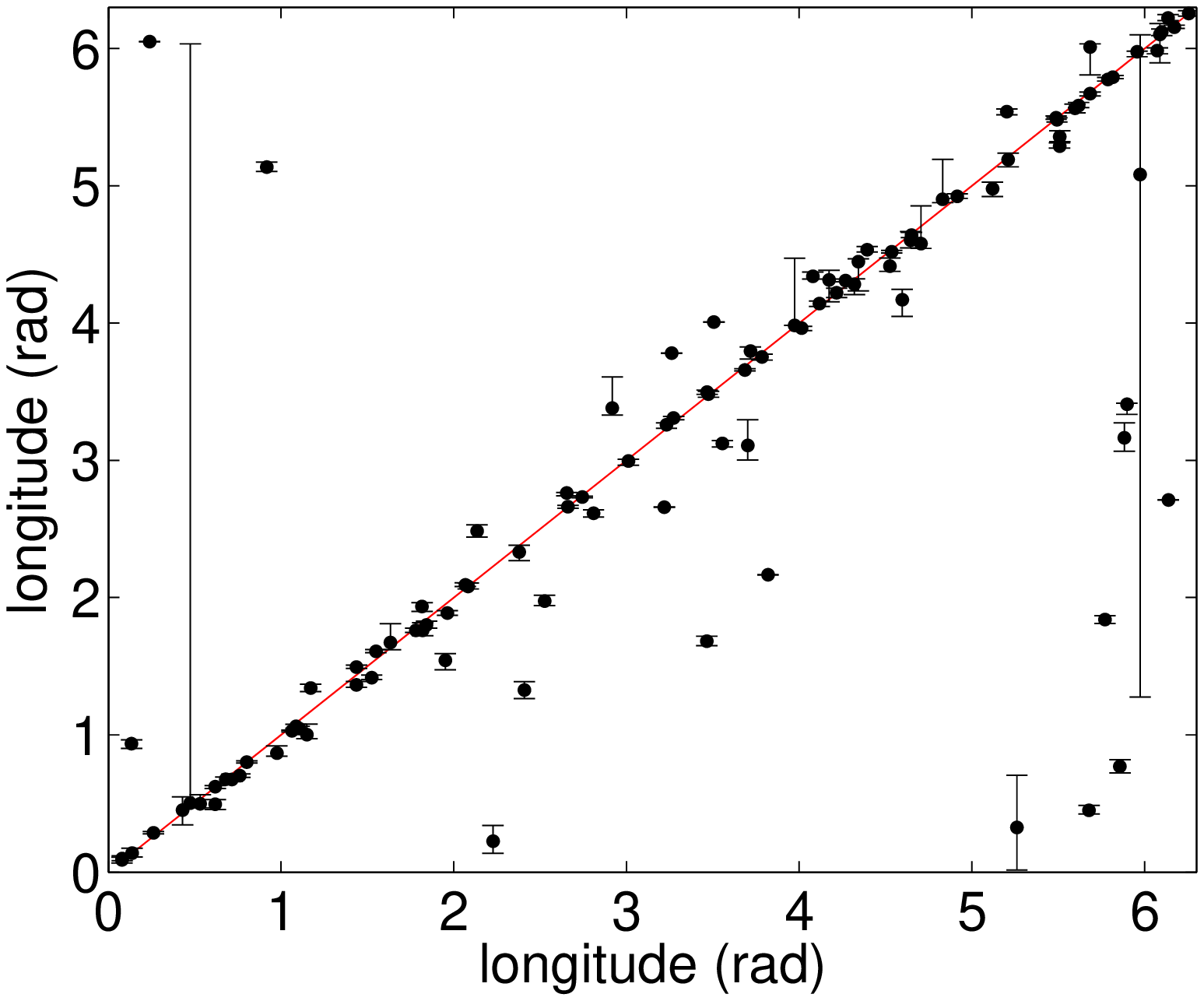}
   \includegraphics[width=2.5in]{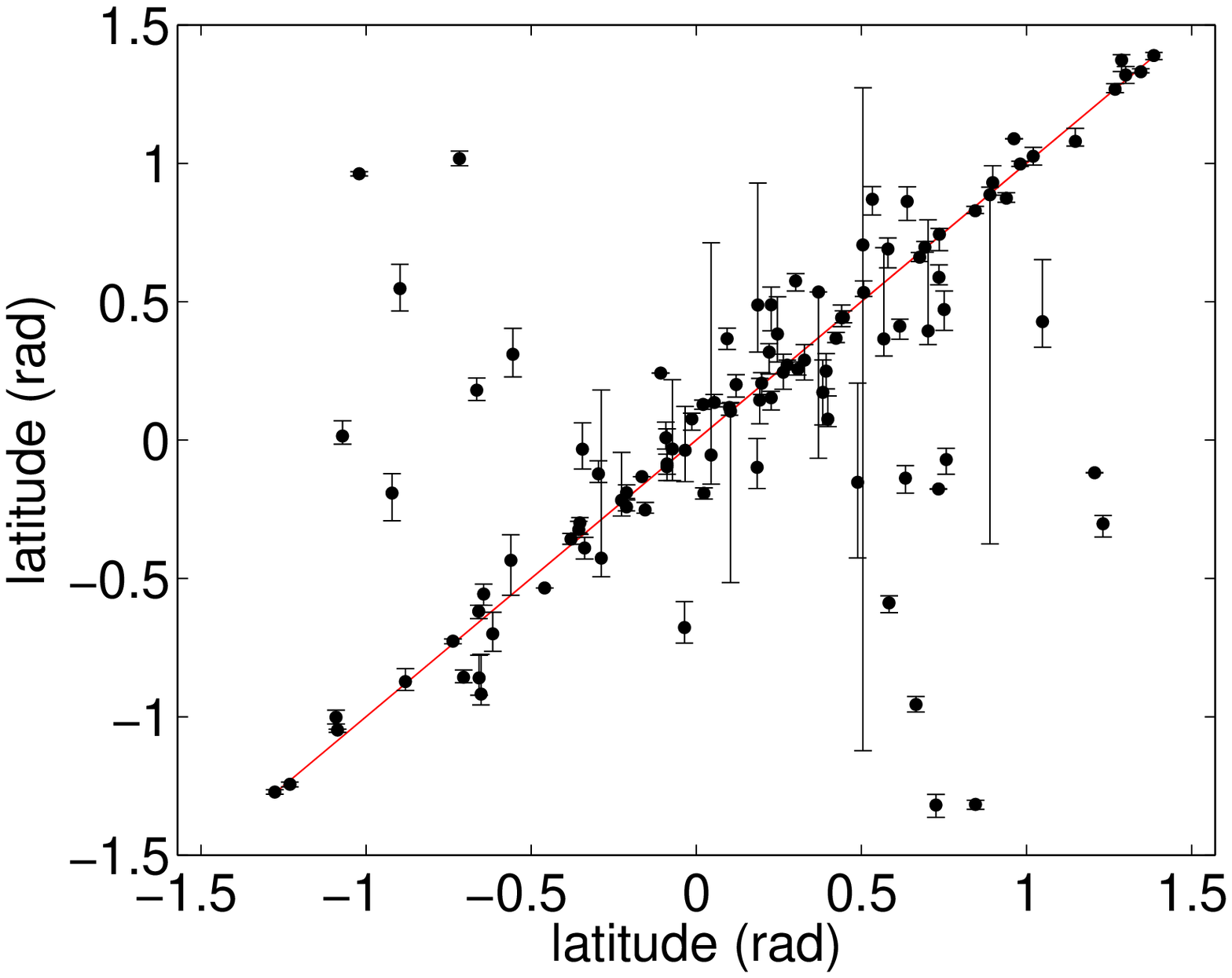}
   \caption{{\bf The recovered source-location parameters for the IMRPhenA approximant.} {\bf Left:} The recovered (maximum likelihood) values of the source longitude as a function of the injected values. {\bf Right:} The recovered (maximum likelihood) values of the source latitude as a function of the injected values.
The IMRPhenA approximant was used with a threshold of $\log_{10} B_{SN} = 3$.
}
   \label{fig:error_nest_longlat}
\end{figure}

The nested-sampling algorithm used for model-selection can also be used for parameter estimation. In particular, one can generate in a straightforward way the maximum likelihood estimate of the recovered parameters, and have an indication of the statistical errors on such values. For simplicity, in this analysis we identified the statistical errors (the error bars in Figures~\ref{fig:error_nest_m} and~\ref{fig:error_nest_longlat})  with the region of parameter space in which the likelihood values were not lower than a factor $e$ with respect to the maximum likelihood. The results for chirp mass, total mass and the two coordinates of the source in the sky -- latitude and longitude -- are shown in Figures~\ref{fig:error_nest_m} and~\ref{fig:error_nest_longlat}; here we restrict to only the IMRPhenA approximant and to all the signals that yielded $\log_{10} B \ge 3$. The results for the masses show a behaviour that is qualitatively consistent with the results obtained using a matched-filtering analysis, see \emph{e.g.} Figures~\ref{fig:EOBNRParam}, \ref{fig:PhenomParam}, and~\ref{f:MvM2}. The total mass is (in most of the cases) systematically underestimated, although for 34 injections the recovered values were consistent with the injected total mass. These injections correspond in all cases to waveforms with (near) zero eccentricity and in 21 (out of 34) instances to non-spinning waveforms. We have also checked that the errors on the masses do not show any significant correlation with the value of the Bayes factor at which the injections were recovered or the injected signal-to-noise ratio. However, despite the systematic errors on the physical parameters, the sky location is on average fairly well determined. This is most likely due to the fact that there is enough information in the (source-location dependent) time of arrival of the signals at different instrument sites to recover meaningful information about the position of the source in the celestial sphere. This is currently under careful investigation and more details about this and other  aspects of the analysis can be found in Ref.~\cite{AylottVeitchVecchio:2009}.

\section{Conclusion}
\label{sec:discussion}
The NINJA project was conceived as a first step towards a long-term
collaboration between numerical relativists and data analysts with the
goal of using
numerical waveforms to enhance searches for gravitational waves. NINJA is
unique in that it focused on running existing gravitational-wave search
algorithms on data containing waveforms obtained from numerical simulations.
Since this constitutes the first such analysis, the scope of the project was
deliberately kept somewhat modest: restrictions were placed on the number of
waveforms to be submitted by each numerical group, no attempt was made to
include transient noise sources in the data and only a limited number of
simulated signals were produced for the data analysis. This helped to encourage
significant involvement from both the numerical relativity and data analysis
communities, with ten numerical relativity groups providing waveforms and 
data-analysis contributions from nine different groups.

Communication between the data analysis and numerical communities has been
smooth and fluent during the course of the NINJA project. The format
described in \cite{Brown:2007jx} provided
a good starting point from which to interchange data between the communities.
As the project was being developed, several improvements were made to the
format, which we expect will continue evolving as more experience is
gained with a broader
family of waveforms, including those containing matter.

The limited number of signals in the NINJA data makes it dangerous to draw
strong conclusions from the comparison of different waveform families and
different search methods.  Overall, it is clear that many of the data analysis
methods were capable of detecting a significant fraction of the simulated
waveforms.  This is immediately significant as several of the analyses
performed are routinely used in searches of the LIGO and Virgo data.  However,
since the NINJA data set did not include the type of non-Gaussian transients
seen in real gravitational-wave detector data, it is difficult to translate
the efficiencies observed here into statements about LIGO or Virgo
sensitivity.

NINJA has demonstrated that more work is required to measure the parameters of
signals in detector data. Parameter estimation is poor for most pipelines, and
several methods tend to associate a candidate event to that part of the
waveform which lies in the most sensitive band of the detector.  For example,
in a search with inspiral only templates, the ringdown of a high mass black
hole which occurs at around 100 Hz might be picked up.  This will lead to poor
estimation of both the binary's mass and coalescence time.  Similarly, the
un-modelled burst searches will correctly identify the signal but, without
knowing which part of the coalescence it corresponds to, have difficulty
providing accurate parameters.  There is some evidence that using full
inspiral-merger-ringdown waveform templates alleviates this problem, as well
as evidence that estimation of the sky location of the signal is largely
independent of the mismatches between simulated and template waveform.  These
are all issues which warrant further investigation. 

We hope that this work will provide a foundation for future analyses, and plans
are envisioned to continue and extend the NINJA project. Several suggestions
have been made to broaden this work and make it more systematic: in addition to
expanding the parameter space explored by numerical simulations, two crucial
steps will be to construct hybrid analytic-numerical waveforms (which will
allow a lower range of masses to be injected) and to consider data containing
non-stationary noise.  It would also be natural to include other waveform
families, such as supernovae or binary mergers comprising one or two neutron
stars.  Subsequent NINJA projects could provide a noise-free data set for
tuning parameter estimation and measurement pipelines and release ``training''
and ``challenge'' data sets, as has proven successful in the Mock LISA
Data Challenges ~\cite{Arnaud:2007vr,Babak:2007zd}, in which the parameters of
the waveforms are known and unknown to the analysts, respectively.
The numerical data sets may also be useful for efforts aimed at
using the best-available waveforms to explore and develop LISA data analysis
approaches and in evaluating parameter
estimation accuracy for LISA. These efforts, as carried out by
the Mock LISA Data Challenge Task Force and the LISA Parameter Estimation Task Force, are summarised in
Ref.~\cite{Babak:2008sn,Arun:2008zn}.

However
future analyses progress, it is clear that a significant amount remains to be
learned from collaborations between the numerical relativity and
gravitational-wave data analysis communities.

\section*{Acknowledgements}

We thank Alan Weinstein for helpful comments on this paper and the Kavli
Institute for Theoretical Physics (KITP) Santa Barbara for hospitality during
the workshop ``Interplay between Numerical Relativity and Data Analysis,''
where the NINJA project was initiated. The Kavli Institute is supported by
National Science Foundation grant PHY-0551164.

This project was supported in part by DFG grant SFB/Transregio~7
``Gravitational Wave Astronomy'' (BB, MH, SH, DP, LR, US); 
%
%
by National Science Foundation grants
PHY-0114375 (CGWP),
PHY-0205155 (UIUC),
PHY-0354842 (RM),
DMS-0553302 (MB, LB, TC, KM, HP, MS),
DMS-0553677 (LK, AM),
PHY-0553422 (NC),
PHY-0555436 (PL),
PHY-0600953 (PB, LG, RAM, RV),
PHY-0601459 (MB, LB, TC, KM, HP, MS),
PHY-0603762 (AB, EO, YP),
PHY-0649224 (BF),
PHY-0650377 (UIUC),
PHY-0652874 (FAU),
PHY-0652929 (LK, AM),
PHY-0652952 (LK, AM),
PHY-0652995 (MB, LB, TC, KM, HP, MS),
PHY-0653303 (PL, DS, MC, CL),
PHY-0653321 (VK, IM, VR, MvdS),
PHY-0653443 (DS),
PHY-0653550 (LC),
PHY-0701566 (ES),
PHY-0701817 (PB, LG, RAM, RV),
PHY-0714388 (MC, CL, YZ),
OCI-0721915 (ES),
PHY-0722315 (MC, CL),
PHY-0745779 (FP),
PHY-0801213 (FH),
PHY-0838740 (BF, LS, VR) 
and NSF-0847611 (DB, LP);
%
%
by NASA grants 
HST-AR-11763 (CL, MC, JF, YZ),
NNG-04GK54G (UIUC),
NNG-04GL37G (RM),
NNG-05GG51G (LK, AM),
NNG-05GG52G (MB, LB, TC, KM, HP, MS),
05-BEFS-05-0044 (GSFC),
06-BEFS06-19 (GSFC),
07-ATFP07-0158 (MC, CL, YZ),
and NNX-07AG96G (UIUC),
%
%
and by NSF cooperative agreement PHY-0107417 (DK, SC). 

%
%
BA was supported by a Vacation Bursary of the UK Engineering and Physical
Sciences Research Council. AV, JV and BS acknowledge support by the UK Science
and Technology Facilities Council.  SF acknowledges the support of the Royal
Society.  MH was supported by SFI grant 07/RFP/PHYF148.  FP 
acknowledges support from the Alfred P. Sloan Foundation. SH acknowledges
support from  DAAD grant D/07/13385, grant FPA-2007-60220 from the Spanish
Ministry of Science and Education and VESF.
MB, LB, TC, KM, HP and MS acknowledge support from the
Sherman Fairchild Foundation and the Brinson Foundation.  LK and AM
acknowledge support from the Fairchild Foundation.
BK was supported by the NASA Postdoctoral Program at the Oak Ridge Associated
Universities. SM was supported in part by the Leon A. Herreid Graduate 
Fellowship. RS was supported by an EGO sponsored fellowship, EGO-DIR-105-2007.

%
Computations were carried out under LRAC allocations 
MCA08X009 (PL, DS),
TG-MCA08X010 (FAU),
TG-MCA02N014 (LSU),
TG-MCA99S008 (UIUC),
TG-PHY990002 (MB, LB, TC, LK, KM, AM, HP, MS),
on LONI systems (LSU),
and on the clusters at the AEI, Cardiff
University, Northwestern University (NSF MRI grant PHY-0619274 to VK),
the LIGO Laboratory, 
NASA Advanced Supercomputing Division (Ames Research Center),
Syracuse University, 
LRZ Munich (BB, MH, SH, US), 
the University of Birmingham,
and the RIT NewHorizons cluster.

\appendix
\section{Glossary of terms}



\begin{tabular}{ll}
\hline
Term & Meaning \\

\hline

{\bf ADM} &Arnowitt-Deser-Misner. \\

{\bf ASCII} &American Standard Code for Information Interchange.\\

{\bf \texttt{BAM}} & Bifunctional Adaptive Mesh code developed at University of Jena.\\

{\bf BBH} &Binary Black Hole. \\

{\bf BSSN} &Baumgarte-Shapiro-Shibata-Nakamura 3+1 formulation of \\
& Einstein's equations. \\

{\bf CBC} &Compact Binary Coalescence. \\

{\bf \texttt{CCATIE}} &AEI/LSU numerical relativity code. \\

{\bf EOB} &Effective One Body. \\

{\bf EOBNR} &Effective-one-body waveforms calibrated to numerical data. \\

{\bf ERD} &Effective Ringdown. \\

{\bf GH} &Generalized harmonic formulation of Einstein's equations. \\

{\bf \texttt{Hahndol}} &Numerical-relativity code developed at NASA-Goddard. \\

{\bf HHT} &Hilbert-Huang Transform. \\

{\bf IMF} &Intrinsic Mode Functions. \\

{\bf IMR} &Inspiral-Merger-Ringdown.  \\

{\bf ISCO} &Innermost Stable Circular Orbit. \\

{\bf L1, H1, H2, V1} &LIGO Livingston 4~km, Hanford 4~km, Hanford 2~km \\
& and Virgo 3~km gravitational-wave detectors. \\

{\bf LAL} &LSC Algorithm Library.  \\

{\bf \texttt{Lean}} &Numerical-relativity code developed by Ulrich Sperhake. \\

{\bf \texttt{LazEv}} &Brownsville/RIT numerical relativity code. \\

{\bf LSC} &LIGO Scientific Collaboration. \\

{\bf \texttt{MayaKranc}} &Numerical-relativity code developed at Penn State \\
                         &using the Kranc code-generation package developed at AEI,\\
                         &Southampton and Penn State.\\

{\bf MCMC} &Markov-Chain Monte-Carlo. \\

{\bf NINJA} &Numerical INJection Analysis. \\

{\bf NR} &Numerical Relativity. \\

{\bf PDF} &Probability-density Function. \\

{\bf PN} &Post-Newtonian. \\

{\bf \texttt{PU}} &Numerical-relativity code developed by Frans Pretorius. \\

{\bf S5} & Fifth LIGO science run.\\

{\bf SNR} &Signal-to-noise ratio. \\

{\bf SPA} &Stationary Phase Approximation. \\

{\bf \texttt{SpEC}} &Spectral Einstein Code developed at Caltech and Cornell.\\

{\bf TT} &Transverse-Traceless. \\

{\bf \texttt{UIUC}} &University of Illinois at Urbana-Champagn numerical \\
&relativity code. \\

{\bf VSR1} &Virgo science run 1. \\

{\bf WRD} &Weighted ringdown. \\

\hline

\end{tabular}

\section*{References}
\bibliography{bibliography/ninja}

\providecommand{\newblock}{}

\end{document}